\documentclass[11pt]{article}

\usepackage{enumerate}

\usepackage{amsmath}
\usepackage{amssymb}
\usepackage{latexsym}
\usepackage{graphicx}
\usepackage{enumerate}
\usepackage{colortbl}
\usepackage{amsthm} 
\usepackage{bm}
\usepackage{mathtools}
\usepackage{caption}
\usepackage{subcaption}
\usepackage{natbib}
\usepackage{url}
\mathtoolsset{showonlyrefs=true}
\usepackage[table]{xcolor}

\usepackage[utf8]{inputenc}
\usepackage[T1]{fontenc}

\usepackage{hyperref}

\usepackage{caption}
\graphicspath{
	{./Figures/}
}

\theoremstyle{thmstyleone}%
\newtheorem{theorem}{Theorem}%
\newtheorem{proposition}[theorem]{Proposition}%
\newtheorem{definition}{Definition}
\usepackage{booktabs}

\begin{document}

\title{\textbf{\Large Regression for spherical responses with linear and spherical covariates using a scaled link function}}

\author{{\scshape Shogo\ Kato}\,$^{a}$, \quad {\scshape Kassel\ L.\ Hingee}\,$^{b}$, \vspace{0.1cm}\\
{\scshape Janice\ L.\ Scealy}\,$^{b}$, \quad {\scshape and} \quad {\scshape Andrew\ T.A.\ Wood}\,$^{b}$ \vspace{0.5cm}\\
\textit{\(^{a}\) Institute of Statistical Mathematics, Tokyo, Japan} \vspace{0.1cm}\\
\textit{\(^{b}\) Australian National University, Canberra, Australia} \vspace{0.5cm}\\ 
}
\maketitle

\abstract{We propose a regression model in which the responses are spherical variables and the covariates include  linear and/or spherical variables.
A novel link function is introduced by extending the M\"obius transformation on the sphere.
This link function is an anisotropic mapping that enables scale control along each axis of the spherical covariates and for each linear covariate.
It generalizes several well-known link functions for circular or linear covariates.
Each parameter of the link function is clearly interpretable.
For the error distribution, we consider a general class of elliptically symmetric distributions, which includes the Kent distribution, the elliptically symmetric angular Gaussian distribution, and the scaled von Mises--Fisher distribution.
Axes of symmetry of the error distribution are determined using a method involving parallel transport.
Maximum likelihood estimation is feasible via reparameterization of the proposed model.
Moreover, the parameters of the link function and the shape/scale parameters of the error distribution are orthogonal in the sense of the Fisher information matrix.
The proposed regression model is illustrated using two real datasets.
An R software package accompanies this article.
} \vspace{0.5cm}

\noindent%

\maketitle

\section{Introduction}
The analysis of directional data is a subfield of statistics that has made great strides in recent decades.  See, for example, the book by \citet{mardia2000di} and the review article by \citet{Pewsey2021}.  However, despite some notable progress, an area where statistical methodology is still rather under-developed is spherical regression, i.e. regression models for a unit vector response.  There are numerous scientific fields where spherical regression is potentially useful and important.  For example, \citet[pp.~258-262]{mardia2000di} mention five  examples where some form of spherical regression is needed: the orientation of crystals in crystallography; the orientation of satellites in space science; ``geometric integrity'' in industrial quality control; applications of machine vision; and the rotation of tectonic plates in geophysics.  The two real-data applications we consider below are to oceanography and to moment tensors in seismology.

The main contributions of this paper are the following. First, we present a new, flexible link function for the mean direction, and discuss how it relates to certain simpler link functions in the literature for the mean direction on the circle and the sphere.  Second, we develop a framework which allows one or both of a Euclidean covariate vector and a spherical (i.e. unit vector ) covariate vector.
Third, a reparameterization of our link function simplifies maximum likelihood estimation.
Fourth, rather than tie our development to a particular error distribution, such as the \citet{Kent1982} distribution, the scaled von Mises--Fisher distribution proposed by \citet{Scealy2019} or the elliptically symmetric angular Gaussian distribution  \citet{Paine2020}, our development applies to a large class of elliptically symmetric distributions which includes all three families mentioned above.  Fifth, we develop a parallel transport approach for determining the axes of symmetry of the relevant error distributions, which side-steps the need to introduce a large number of parameters (especially in more than 3 dimensions) required to determine the orientation of the error distribution as a function of the covariate vector.  Sixth, in our applications, the assumption of a rotationally symmetric error distribution is shown to be inadequate and it is demonstrated that the flexibility of elliptical symmetry is needed.  Moreover, we would expect this finding to hold for most applications of spherical regression.

Well-behaved link functions for circular or spherical responses can be difficult to design, and care is needed to ensure parameter identifiability and avoid highly multimodal likelihood functions \citep[e.g.~helical regression functions, p257]{mardia2000di}.
\citet{Fisher1992} proposed a circular regression model with linear covariates.
A regression model with a circular covariate was developed by \citet{Rivest1997} for predicting a circular response with a rotation of the decentered circular covariate.
\citet{Downs2002} proposed another circular regression model that uses a link function expressed in the form of a M\"obius transformation.
Regression models related to that of \citet{Downs2002} have been discussed, for example, by \citet{Kato2008bb,Rueda2016,Rueda2019,Jha2021}.

Previous work on spherical regression includes \citet{Scealy2011}, who use the \citet{Kent1982} distribution as an error distribution, \citet{Cornea2017}, who propose a method for regression on Riemannian symmetric spaces, \citet{Scealy2019} and \citet{Paine2020}.  From the perspective of this paper, the limitation of the spherical regression models of \citet{Scealy2011} and \citet{Scealy2019} are that the link functions are defined for response data on the positive orthant, corresponding to compositional response data after applying the square-root transformation. \citet{Cornea2017} develop method-of-moment estimation for regressions with responses on Riemannian symmetric manifolds with the sphere and stereographic projection one of their important applications.  However, \citet{Cornea2017} incorporate neither a spherical covariate nor an anisotropic error distribution, both of which are important components of our contribution here.  Finally, in \citet{Paine2020}, the link functions are quite different to the link function proposed here and require more parameters (especially in higher dimensions). In one of our real data examples we compare the performance of the different link functions.

In the case of a single spherical covariate (i.e. unit vector) of the same dimension as the spherical response,  \citet{Chang1986} considers a rotational link function where the parameter matrix is a rotation matrix, whereas \citet{Rosenthal2014} consider a generalization of \citet{Chang1986} where the link function has a parameter matrix that is non-singular with determinant 1.  In the online supplementary material, we discuss the similarities and differences between the \citet{Rosenthal2014} link function and our link function in the case of a single spherical covariate.

As already mentioned, in this paper we propose a regression model in which responses take values in the sphere and covariates consist of both linear and spherical variables.
Throughout the document, we assume that $S^{p-1} = \{ x \in \mathbb{R}^p : \|x\| = 1 \}$ denotes the unit sphere in $\mathbb{R}^p$, $\|\cdot\|$ represents the Euclidean norm, $D^p = \{ x \in \mathbb{R}^p : \|x\| \leq 1 \}$ denotes the closed unit disc in $\mathbb{R}^p$, $\overline{\mathbb{R}}^p = \mathbb{R}^p \cup \{ \infty \}$ is the compactified Euclidean space, $e_j$ is the $p$-dimensional unit vector with its $j$-th element equal to one, $0_p$ denotes the $p$-dimensional zero vector, $I_p$ is the $p \times p$ identity matrix, $O_{p \times q}$ is a $p \times q$ zero matrix.
Define $\infty + c = \infty$ for $c \in \overline{\mathbb{R}}^p$, and $B \infty = \infty$ if $B \neq O_{p \times q}$ and $B \infty = 0$ if $B= O_{p \times q}$.
    
In the next section (Section \ref{sec:regression}) we introduce our link function and relevant error distributions. In Section \ref{sec:link} we investigate properties of our link function, including special cases and relationships with other well-known link functions.
We then (Section \ref{sec:estimators}) describe maximum likelihood estimation of our regression models and specify how the orientation of our error distribution may be conveniently  determined by parallel transport.
In Section \ref{sec:applications} we apply our regression methods to locations on $S^2$ (midatlantic ridge and continental locations) and $S^4$ (earthquake moment tensors). We conclude in Section \ref{sec:conclusion}.
All proofs are given in the supplementary material and a software package written in R accompanies this article.

\section{Regression models for spherical data} \label{sec:regression}
Section 2.1 presents a new type of link function for spherical data.  In Section 2.2, a class of spherical distributions with elliptical symmetry is specified. This class includes the Kent (1982) family and also more recent proposals such as the elliptically symmetric angular Gaussian family, the scaled von Mises--Fisher family and the tangent elliptical distributions proposed by, respectively, \citet{Paine2018}, \citet{Scealy2019} and \cite{Garcia-Portugues2020}.  The proposed link function can be used with any family of distributions within this class.
\subsection{The link function}
\begin{definition}[Link function] \label{def:link}
Let \( x = (x_e^\top, x_s^\top)^\top \) be a single case of the covariates, where \( x_e \in \mathbb{R}^{q_e} \) is a Euclidean variable with \( q_e \geq 1 \), and \( x_s \in S^{q_s - 1} \) is a spherical variable with \( q_s \geq 2 \).
Then the link function for an \( S^{p-1} \)-valued response, where \( p \leq \min(q_e + 1, q_s) \), is defined by
\begin{align} 
\mu(x) & = B_0 \mathcal{S}^{[-1]} \left(  B_{s} \mathcal{S} (R_{s}^\top x_{s}) + B_e  R_e^\top x_e \right), \quad x_e \in \mathbb{R}^{q_e}, \ x_{s} \in S^{q_{s}-1}, \label{eq:link}
\end{align}
where \( B_0 \) is a \( p \times p \) rotation matrix, \( B_{s} = \mbox{diag}( \beta_{s2} , \ldots, \beta_{sp}) \), \( B_{e} = \mbox{diag}( \beta_{e2} , \ldots, \beta_{ep}) \), $\beta_{sj},\beta_{ej} \geq 0$, $j=2,\ldots,p$, $\beta_{sp}^2+\beta_{ep}^2 \leq \cdots \leq \beta_{s2}^2 + \beta_{e2}^2$, \( R_{s} \) is a \( q_{s} \times p \) matrix satisfying $R_{s}^\top R_{s} = I_p$, and \( R_e \) is a \( q_e \times (p-1) \) matrix with $R_e^\top R_e= I_{p-1} $.
Here $\mathcal{S}$ denotes the stereographic projection extended to be defined on the unit disc
\begin{equation}
\mathcal{S}(x) = \frac{1}{1+x_1} (x_2,\ldots,x_p)^\top, \quad x=(x_1,\ldots,x_p)^\top \in D^{p} \setminus \{-e_1\}, \label{eq:stereo}
\end{equation}
and $\mathcal{S}^{[-1]}$ is the inverse stereographic projection
\begin{equation}
\mathcal{S}^{[-1]}(y) = \frac{1}{1+\| y \|^2} (1-\| y \|^2, 2 y_1, \ldots, 2 y_{p-1})^\top, \quad y =( y_1,\ldots, y_{p-1})^\top \in \mathbb{R}^{p-1}. \label{eq:inverse}
\end{equation}
Also $\mathcal{S}(-e_1) = \infty$ and $\mathcal{S}^{[-1]}(\infty) = -e_1$.
\end{definition}

While the detailed properties of this link function, including a simpler form, will be discussed in Section \ref{sec:link}, we briefly outline its intuition here.
Note that the equation \eqref{eq:link} leads to the expression
\begin{equation}
\mathcal{S} (B_0^\top \mu (x) ) = B_{s} \mathcal{S} (R_{s}^\top x_{s}) + B_e R_e^\top x_e . \label{eq:link_linear}
\end{equation}
Since the stereographic projection $\mathcal{S}$ maps the sphere onto $\overline{\mathbb{R}}^{p-1}$, the left-hand side, \( \mathcal{S} (B_0^\top \mu(x)) \), and each term on the right-hand side of this equation take values in \( \overline{\mathbb{R}}^{p-1} \).
This implies that the link function \eqref{eq:link} assumes a `linear' relationship between the linearized spherical variable \( \mathcal{S} (R_{s}^\top x_{s}) \) and the Euclidean variable \( x_e \).

Each parameter of this link function can be interpreted as follows.
The parameter \( B_0 \) is a rotation parameter, functioning similarly to an intercept in a linear regression model.
The \((i,i)\)-elements of \( B_{s} \) and \( B_e \) control the scale of the \( i \)-th elements of the linearized spherical variable \( \mathcal{S} (R_{s}^\top x_{s}) \) and the transformed Euclidean variable \( R_e^\top x_e \), respectively, playing roles analogous to slope parameters in linear regression.
The parameters \( R_{s} \) and \( R_e \) determine the orthonormal bases onto which \( x_{s} \) and \( x_e \) are projected, respectively.  
These parameters do not have direct counterparts in linear regression; however, \( R_{s} \) appears in the link function for a model with a single circular covariate (\( p = q_s = 2 \)) in the model of \citet{Downs2002}, where it serves as the parameter controlling the location of the covariate $x_{s}$ and the positive or negative relationship between two variables.
Proposition \ref{prop:link2}, below, provides further insights into the interpretation of the link function \eqref{eq:link}. 

As in the multivariate linear regression model, it is straightforward to incorporate a term that shifts the linear variables by setting one of the elements of $x_e$ to one, for example, by letting $x_e = (1, x_{e2}, \ldots, x_{e q_e})^\top$.

\subsection{A class of spherical distributions with elliptical symmetry}
Before introducing a broad class of spherical distributions with elliptical symmetry, we give the probability density, with respect to induced Lebesgue measure on $S^{p-1}$, of the scaled von Mises--Fisher distribution \citep{Scealy2019} to be used later:
\begin{equation}
\label{eq:log_lik} 
f_Y(y)=\{c_p(\kappa)a_1\}^{-1}J^{-(p-1)/2} \exp \left \{ \frac{\kappa y^\top \mu/a_1}{(y^\top \mu/a_1)^2 + \sum_{j=2}^p (y^\top \gamma_j/a_j)^2}  \right \},
\end{equation}
where $J=(y^\top \mu/a_1)^2 +\sum_{j=2}^p (y^\top \gamma_j/a_j)^2$.  When $a_1=\cdots = a_p$, the standard von Mises--Fisher distribution is recovered.  Given data $(y_i, x_i)$, $i=1, \ldots , n$, where $y_i$ is a unit vector response and $x_i$ denotes a spherical covariate, a Euclidean vector covariate or both, $\mu(\cdot)$ is the link function (\ref{def:link}), the mean direction of observation $i$ is $\mu(x_i)$, and $\mu(x_i), \gamma_2(x_i), \ldots , \gamma_p(x_i)$ is an orthonormal set of vectors.

We now specify a broad class of distributions with ellipse-like symmetry.  Let $g: \mathbb{R}^2 \rightarrow \mathbb{R}_+=\{z \in \mathbb{R}: z \geq 0\}$ be any non-negative function for which the right-hand side of \eqref{class} below integrates to 1.
Suppose $\mu =\gamma_1\in \mathbb{S}^{p-1} \subset \mathbb{R}^p$ is the unit vector mean direction and suppose that $\gamma_2, \ldots , \gamma_p$ are unit vectors such that $\gamma_1, \ldots , \gamma_p$ is an orthonormal basis of $\mathbb{R}^p$. Given $g$, the pdf of the model  of interest is given by
\begin{equation}
f(y\vert \mu, \gamma_2, \ldots , \gamma_p,\kappa, \lambda) =c_{g,p}(\kappa, \lambda)^{-1} g \left (\kappa \mu^\top y, \lambda_1 (y^\top \mu)^2+\sum_{j=1}^{p} \lambda_j (y^\top \gamma_j)^2 
\right ),   \label{class}
\end{equation}
where $\lambda=(\lambda_1, \ldots ,\lambda_{p})^\top$ and $c_{g,p}(\kappa, \lambda)$ is a normalizing constant.  The shape and concentration of the distribution are determined by $\kappa$ and $\lambda$ while $\mu=\gamma_1, \gamma_2, \ldots , \gamma_p$ determine the orientation of the distribution; specifically, these vectors determine the axes of symmetry of the distribution.  A simple form of the distribution arises when $\mu$
and the $\gamma_j$, $j=2, \ldots , p$, correspond to the coordinate axes.  E.g. when $\mu=e_1$ and $\gamma_j=e_j$, $j=2, \ldots p$,  it is seen that
\[
f(y\vert e_1, e_2 \cdots e_p, \kappa, \lambda)
=c_{g,p}(\kappa, \lambda)^{-1} g \left (\kappa y_1, \sum_{j=1}^{p} \lambda_j y_{j}^2 \right ).
\]

From symmetry considerations it is easy to see  that the mean direction of a distribution of the form \eqref{class} is $\mu \in \mathbb{S}^{p-1}$ provided $\kappa>0$ and that $\mathbb{E}[y y^\top]$ has eigenvectors $\mu, \gamma_2, \ldots , \gamma_p$.  In fact, $\kappa$, along with $\lambda_1, \ldots , \lambda_p$, determines the concentration of the distribution about the mean direction $\mu$. Moreover, in the unimodal case, the contours of constant density have ellipse-like symmetry.

It turns out that there is some indeterminacy in the parametrization (\ref{class}).  To see this, we consider three  known families of distributions which have ellipse-like symmetry and are of the form (\ref{class}).  

First, we consider the well-known \citet{Kent1982} distribution. This corresponds to taking $g(u,v)=\exp(u+v)$, $\lambda_1=0$ and $\sum_{j=2}^p \lambda_j=0$ in (\ref{class}).  In the case $p=3$, $\lambda_2 = -\lambda_3$.  Note that the von Mises--Fisher distribution \citep[see][]{mardia2000di}, a rotationally symmetric distribution, is recovered when $\lambda_2=\cdots =\lambda_p$.

A second family is the elliptically symmetric angular Gaussian (ESAG) distribution, proposed in \citet{Paine2018}, and discussed in a spherical regression context in \citet{Paine2020}.  Here, 
$g(u,v)=v^{-p/2} \exp  (u^2/(2v))
M_{p-1}(u/v^{1/2})$,
where
\[
M_{p-1}(\alpha)=\int_0^\infty t^{p-1} \frac{1}{\sqrt{2 \pi}}\exp\{ -(t-\alpha)^2 /2\}dt,
\]
$\lambda_1=1$ and $\prod_{j=2}^p \lambda_j=1$.  In \citet{Paine2018}, ESAG was parametrized by $\mu \in \mathbb{R}^p$ and the symmetric positive-definite $p \times p$ matrix $V$.  The connection between the original ESAG parameters (on the left) and the parameters of (\ref{class}) (on the right) is
\[
\mu \mapsto \kappa \mu \hskip 0.2truein \textrm{and} \hskip 0.2truein V^{-1} \mapsto \sum_{j=1}^p \lambda_j \gamma_j \gamma_j^\top.
\]

A third family is the scaled von Mises--Fisher distribution, introduced by \citet{Scealy2019}.  In this case, we have
$g(u,v)=v^{-(p-1)/2} \exp(-u/v^{1/2})$
in (\ref{class}), where $\lambda_1=1$ and $\prod_{j=2}^p \lambda_j=1$.  Note that, when $a_1, \ldots , a_p$ are the positive scaling factors in \citet{Scealy2019}, $\lambda_j=a_j^{-1}$, $j=1, \ldots , p$ and the normalising constant is $c_{g,p}(\kappa, \lambda) = c_p(\kappa)a_1 = c_p(\kappa)/\lambda_1$, where $c_p(\kappa)$ is the normalising constant of the von Mises--Fisher distribution.
The tangent elliptical distributions of \citet[Theorem 2.1]{Garcia-Portugues2020} also fall into the class given by \eqref{class}.

In all of the above families, $\lambda_1$ is set to a fixed value because it is difficult or impossible to estimate both $\kappa$ and $\lambda_1$ simultaneously.  The second constraint, either $\sum_{j=2}^p \lambda_j=0$ in the first case or $\prod_{j=2}^p \lambda_j=1$ in the second and third cases, is needed to remove the indeterminancy due to the constraint $y^\top y=1$.

The key point to note is that we can use the link function proposed in Section 2 to model the mean direction for any family of distributions of the form (\ref{class}).   If $x$ is a covariate vector, the unit vector $\mu(x)$ determines the mean direction via the link function specified in Section 2.1, while the orthonormal set $\{\mu(x), \gamma_2(x), \ldots  , \gamma_p(x)\}$ 
determines the axes of symmetry of  the distribution. 
The selection of the vectors $\gamma_2(x), \ldots, \gamma_p(x)$ will be discussed in Section \ref{sec:gamma}.

\section{Properties of the proposed link function} \label{sec:link}

In this section we investigate some properties of the proposed link function \eqref{eq:link} and its special cases.
For convenience, let $b_{0j}$, $r_{sj}$, and $r_{ej}$ denote the $j$-th columns of $B_0$, $R_s$, and $R_e$, respectively.
Let $A_{-1}$ represent the submatrix of a matrix $A$ with its first column removed.

\subsection{General properties} \label{sec:general}
Our link function (\ref{eq:link}) can be expressed in a simple and closed form as follows.

\begin{proposition} \label{prop:link2}
The link function \eqref{eq:link} can be expressed as
\begin{equation} 
\mu(x) 
= \frac{(1-\|t(x)\|^2) b_{01} + 2 \sum_{j=2}^p t_j(x) b_{0j} }{1+\|t(x)\|^2}, \quad x_e \in \mathbb{R}^{q_e}, \ x_s \in S^{q_s-1} \setminus \{ - r_{s1} \}, \label{eq:link2}
\end{equation}
where  
$$
t(x) = \frac{B_s R_{s,-1}^\top x_s}{1+r_{s1}^\top x_s} + B_e R_e^\top x_e ,
$$
and $t_j(x)$ denotes the $(j-1)$-th element of $t(x)$.
Also, if $x_s=-r_{s1}$, then $\mu(x) = -b_{01} $ for $B_s \neq O_{(p-1) \times (p-1)}$ and $\mu(x) = b_{01} $ for $B_s = O_{(p-1) \times (p-1)}$.
\end{proposition}

%
This expression implies that $\mu(x)$ is a linear combination of $b_{01}, \ldots, b_{0p}$. 
Here $b_{01}$ is referred to as the reference direction, which mainly governs the direction in which the transformed covariate $t(x)$ is attracted or repelled.
The scales of \( B_s \) and \( B_e \) determine the closeness to the reference direction \( b_{01} \).
For \( \beta \geq 0\), as \( \beta \) decreases, the transformed point \( \mu(x; \beta B_s, \beta B_e ) \) moves closer to the direction \( b_{01} \) than \( \mu(x; B_s, B_e ) \).  
Conversely, as \( \beta \) increases, the transformed point \( \mu(x; \beta B_s, \beta B_e ) \) approaches the direction \( -b_{01} \) compared to \( \mu(x; B_s, B_e) \).
See Proposition \ref{prop:beta} in  the online supplementary material for details.

Note that $t_j(x)$ may be written simply as
$$
t_j(x) = \frac{\beta_{sj} r_{sj}^\top x_s}{1+r_{s1}^\top x_s} + \beta_{ej} r_{e,j-1}^\top x_e .
$$
Then it follows that $b_{0j}$ determines the axis for the scale change of the $j$-th elements of $R_s^\top x_s$ and $R_e^\top x_e$.

	\begin{proposition} \label{prop:mu_dim}
	The following properties hold for the proposed link function \eqref{eq:link}, where the domain of $x_e$ is extended to $\overline{\mathbb{R}}^{q_e}$.
    
	\begin{enumerate}[(i)\ ]
	\item If $ \beta_{ep}^2+\beta_{sp}^2 > 0$, then $\mu( \overline{\mathbb{R}}^{q_e} \times  S^{q_s-1} ) = S^{p-1}$.
	
	\item 
	Let $\beta_{ek}^2+\beta_{sk}^2 > 0$ and $\beta_{e,k+1}= \beta_{s,k+1} =0$ for $  2 \leq k \leq p-1 $.
	Then $\mu$ maps $\overline{\mathbb{R}}^{q_e} \times S^{q_s-1}$ onto the $(k-1)$-dimensional unit sphere $\tilde{S}^{k-1}, $ embedded in $S^{p-1}$, given by
    \begin{equation}
    \tilde{S}^{k-1} = \left\{ x \in S^{p-1} : x = u_1 b_{01} + \cdots + u_k b_{0k} , \ (u_1,\ldots,u_k) \in S^{k-1} \right\}.
    \end{equation}

	\item 
	If $\beta_{e2} = \beta_{s2} = 0$, then $\mu( \overline{\mathbb{R}}^{q_e} \times  S^{q_s-1} ) = b_{01}$, so that $\mu(x)$ does not depend on $x$.
	\end{enumerate}
\end{proposition}

Proposition \ref{prop:mu_dim} implies that the number of indices \( j \) for which either \( \beta_{ej} \) or \( \beta_{sj} \) equals zero corresponds to the reduction in the number of dimensions of the transformed covariate, \( t(x) \), caused by the transformation \( \mu(x) \).

Note that this proposition holds except for an exceptional case even when the domain of $x_e$ remains $\mathbb{R}^{q_e}$.
Clearly, Proposition \ref{prop:mu_dim}(iii) also holds in this situation.
Propositions \ref{prop:mu_dim}(i) and \ref{prop:mu_dim}(ii) hold as well, provided the additional assumption $\beta_{s2} > 0$ is satisfied.
The only exception is the case $\beta_{s2} = 0$, where $-b_{01}$ is excluded from $\mu(\mathbb{R}^{q_e} \times S^{q_s-1})$ in Propositions \ref{prop:mu_dim}(i) and \ref{prop:mu_dim}(ii).

If $p=q_s$, then $\mathcal{S}$ reduces to the stereographic projection on $S^{p-1}$ and $\mathcal{S}^{[-1]}$ becomes its inverse, namely, $ \mathcal{S}^{[-1]} = \mathcal{S}^{-1}$.
The stereographic projection $\mathcal{S}$ with $p=q_s$ is a bijective mapping which maps the unit sphere $S^{p-1}$ onto  $\overline{\mathbb{R}}^{p-1}$ \citep[see, for example, ][Section 4.2]{Ratcliffe2019}.
In this case the projected point $\mathcal{S}(x_s)$ corresponds to the point at the intersection of the embedded Euclidean space $\mathbb{R}^{p-1} \times \{0\}$ and the line connecting $x$ and the south pole $-e_1$.

\subsection{A special case: a spherical covariate only} \label{sec:special_sphere}

Throughout this subsection, we discuss a special case of the link function with only a spherical covariate, namely, $q_e=0$.
In this case the link function \eqref{eq:link} reduces to
\begin{align} 
\begin{split} \label{eq:link_s}
\mu_s(x_s) & = B_0 \mathcal{S}^{[-1]} \left(  B_{s} \mathcal{S} (R_{s}^\top x_{s}) \right) \\
 & = \frac{(1-\|t_s(x_s)\|^2) b_{01} + 2 \sum_{j=2}^p t_{sj}(x_s) b_{0j} }{1+\|t_s(x_s)\|^2}, \quad x_s \in S^{q_s-1} \setminus \{ - r_{s1} \},
 \end{split}
\end{align}
where $ t_s (x_s) = (B_s R_{s,-1}^\top x_s)/(1+r_{s1}^\top x_s) $
and $t_{sj}(x_s)$ denotes the $(j-1)$-th element of $t_s(x_s)$.
Also $\mu_s( - r_{s1}) = -b_{01}$.
This subclass includes some existing link functions and transformations for spherical data.
Proposition \ref{prop:mobius_2}, below, for $p=q_s=2$, implies that the link functions, or related functions, of the statistical models proposed by \citet{Kato2008bb}, \citet{Kato2010a}, \citet{Rueda2016}, \citet{Rueda2019} and \citet{Jha2021}---which apply the link function of \citet{Downs2002}--correspond to a special case of the link function~\eqref{eq:link_s}.
When $p=3$, \eqref{eq:link_s} contains the link function of \citet{Downs2003} (see Appendix \ref{sec:downs} of the supplementary material) and Proposition \ref{prop:mobius_p}, below, considers general $p$.
While all the special cases of the link functions discussed in this subsection have simple forms, and some possess the appealing mathematical property of conformality, none allow for scale changes with different strengths along each axis.
For statistical analysis, such axis-specific scale changes are important to enhance the flexibility of the link function.

\begin{proposition} \label{prop:mobius_2}
Consider the proposed link function \eqref{eq:link_s} with $p=q_s=2$.
Let 
$$
x_s = \begin{pmatrix} \cos \theta_{x} \\ \sin \theta_{x} \end{pmatrix}, \quad
\mu_s(x_s) = \begin{pmatrix} \cos ( \mu_{DM}(\theta_x)) \\ \sin (\mu_{DM} (\theta_x)) \end{pmatrix},
$$
$$
B_0 = \begin{pmatrix} \cos \beta_0 & -\sin \beta_0 \\ \sin \beta_0 & \cos \beta_0  
\end{pmatrix}, \quad
R_s = 
\begin{pmatrix} \cos \eta & -\sin \eta \\ \sin \eta & \cos \eta
\end{pmatrix}
\begin{pmatrix} 1 & 0 \\ 0 & \delta
\end{pmatrix},
$$
where $\theta_{x}, \beta_0, \eta \in [-\pi,\pi)$, $\delta \in \{-1,1\}$.
Then the link function \eqref{eq:link_s} can be expressed as the link function of \citet{Downs2002} which is of the form
\begin{equation}
\mu_{DM}(\theta_x) = \beta_0 + 2 \arctan \left\{ \delta \beta_{s2} \tan \left( \frac{\theta_{x}-\eta}{2} \right) \right\}. \label{eq:mobius_2}
\end{equation}
\end{proposition}

\begin{figure}
	\begin{center}
       \makebox[15cm][l]{ \hspace{1.5cm} \makebox[6cm][l]{(a)} \hspace{0.3cm} \makebox[6cm][l]{(b)} } \\
       \includegraphics[width=6cm,height=6cm]{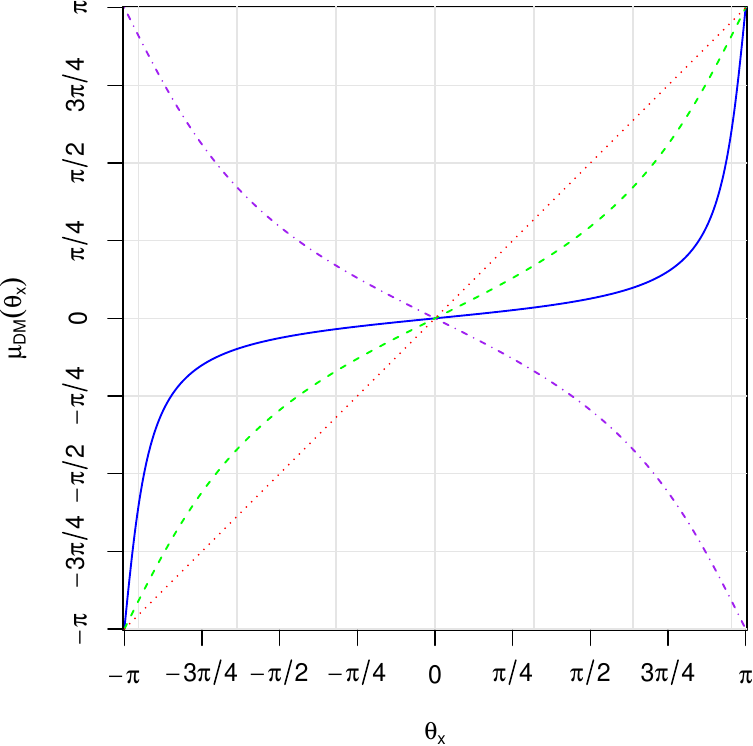}
        \quad \includegraphics[width=6cm,height=6cm]{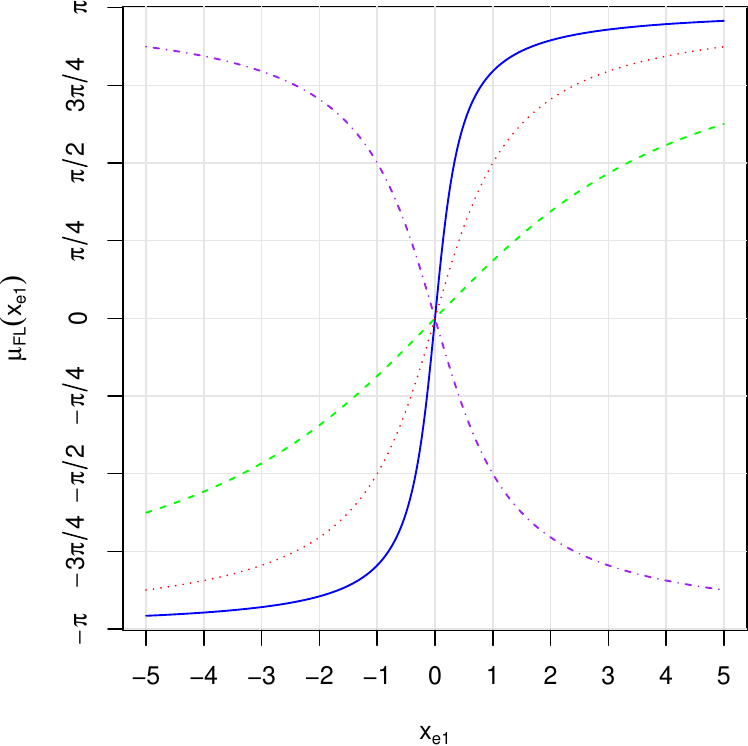}
	\end{center}
	\caption[]{
		Plots of the link functions: (a) $\mu_{DM}(\theta_x) $ \eqref{eq:mobius_2} for $\beta_0=\eta=0$ and $\delta \beta_{s2}=0.1$ (solid, blue), $\delta \beta_{s2}=0.5$ (dashed, green), $\delta \beta_{s2}=1$ (dotted, red) and $\delta \beta_{s2}=-0.5$ (dot-dashed, purple); and 
        (b) $\mu_{FL}(x_e)$ \eqref{eq:link_FL} for $\beta_0=0$, $\gamma=(0,\gamma_1)^\top $, $x_e =(1,x_{e1})^\top$, and $\gamma_1=0.3$ (dashed, green), $\gamma_1=1$ (dotted, red), $\gamma_1=3$ (solid, blue) and $\gamma_1=-1$ (dot-dashed, purple).
	} \label{fig:link_d2}
\end{figure}

Note that \citet{Kent2023} considered a transformation related to the M\"obius transformation on the circle \eqref{eq:mobius_2} through angle halving, to derive the angular central Gaussian distribution from the uniform distribution on the circle.

Figure \ref{fig:link_d2} shows the link function \eqref{eq:mobius_2} and the link function with a linear covariate, given by \eqref{eq:link_FL} below, for selected parameter values.
In Figure \ref{fig:link_d2}(a), the plots are provided for fixed values of $\beta_0$ and $\eta$ and for four selected values of $\delta \beta_{s2}$.
It can be observed that, when $\delta \beta_{s2} = 1$, the link function \eqref{eq:mobius_2} reduces to the identity mapping.
For $\beta_{s2} \leq 1$, smaller values of $\beta_{s2}$ lead to a greater concentration of points around $\mu_{DM}(\theta_x) = \beta_0 = 0$.
The cases $\delta = 1$ and $\delta = -1$ correspond to positive and negative relationships between $\mu_{DM}(\theta_x)$ and $\theta_x$, respectively.
Finally, $\eta$ serves as a location parameter for $\theta_x$.

	\begin{figure}
    \hspace{0.7cm}
		\begin{tabular}{cccc} 
			\centering 
			\hspace{-3cm} (a) & \hspace{-3cm} (b) & \hspace{-3cm} (c) & \hspace{-3cm} (d) \vspace{0cm}\\
            \includegraphics[width=3.3cm,height=3.1cm,trim=0.3cm 0.3cm 0.3cm 0cm]{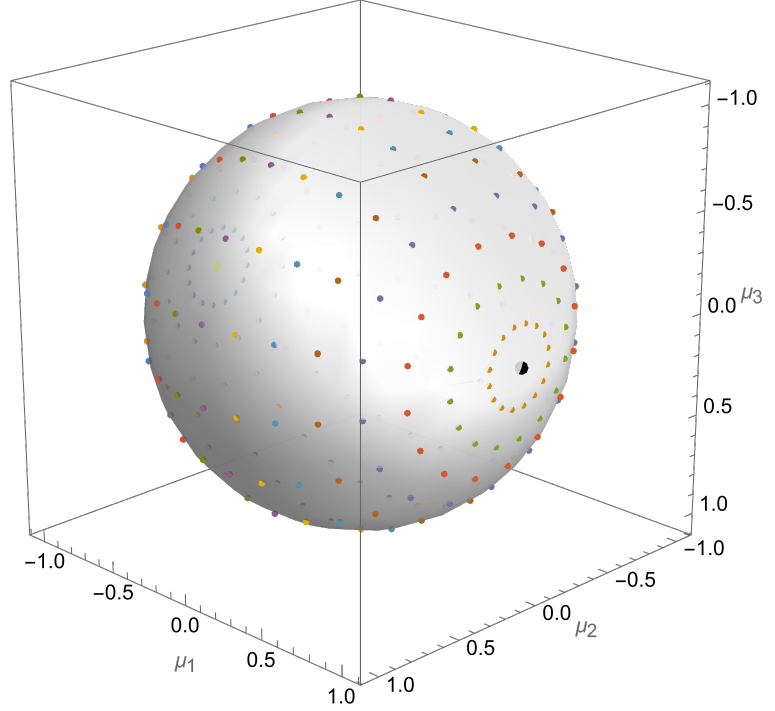} &
			\includegraphics[width=3.3cm,height=3.1cm,trim=0.3cm 0.3cm 0.3cm 0cm]{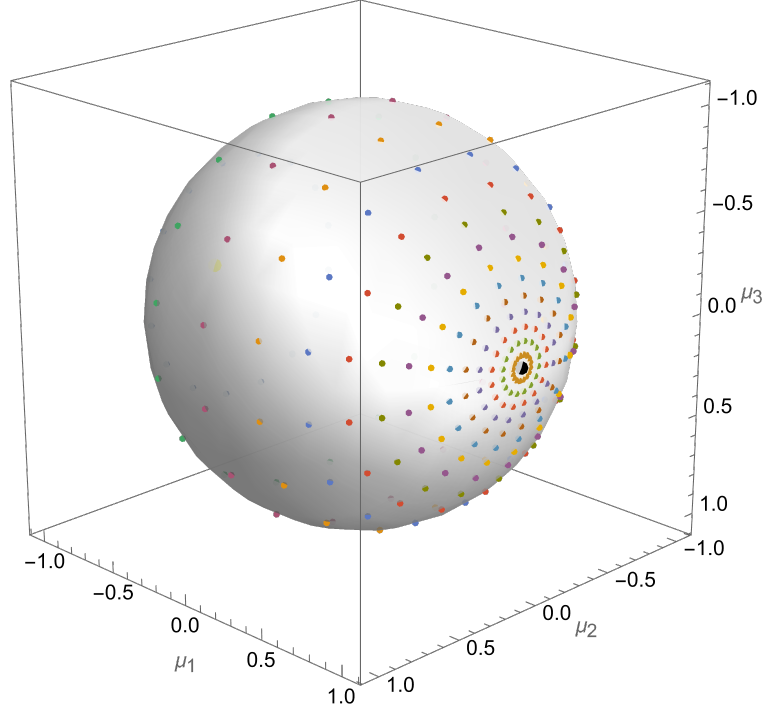} &
			\includegraphics[width=3.3cm,height=3.1cm,trim=0.3cm 0.3cm 0.3cm 0cm]{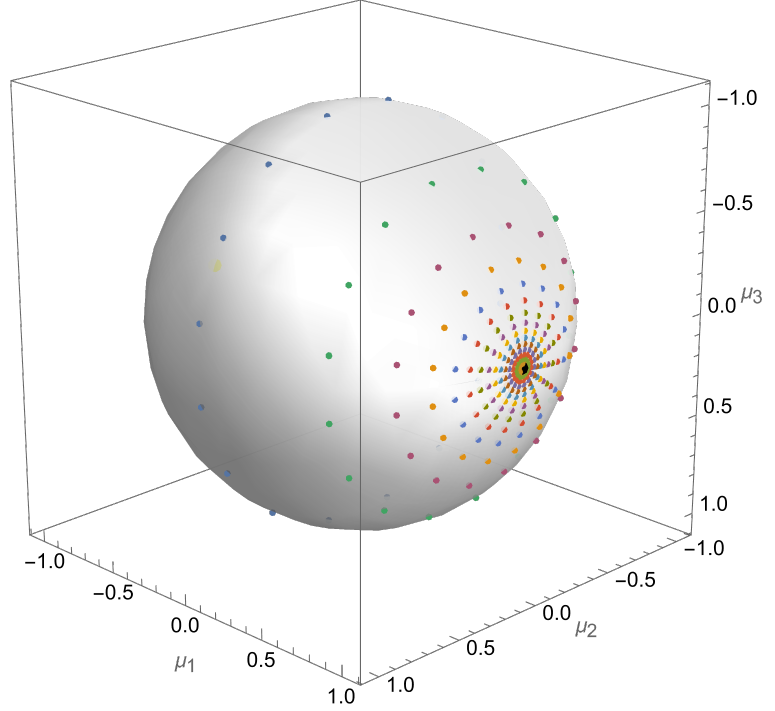} &
            \includegraphics[width=3.3cm,height=3.1cm,trim=0.3cm 0.3cm 0.3cm 0cm]{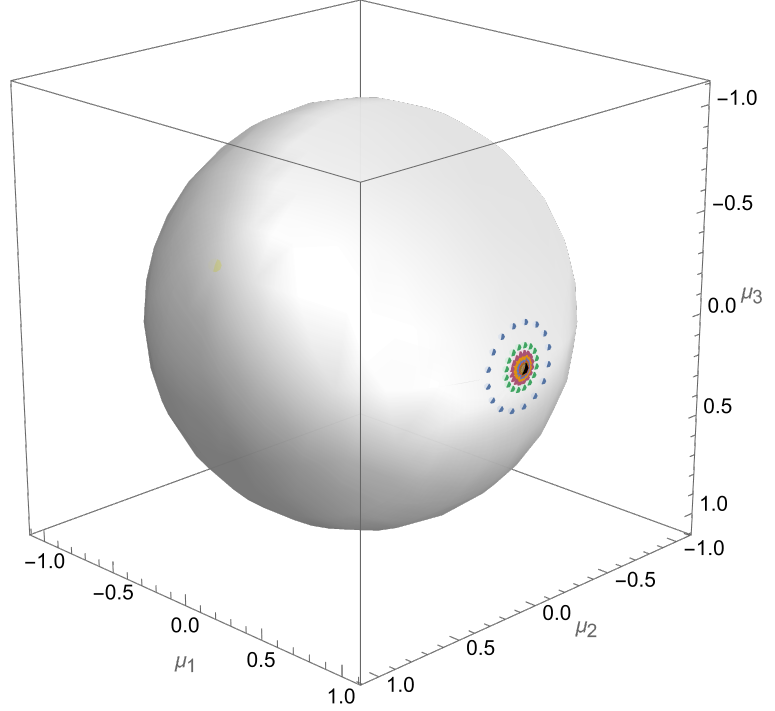} \vspace{0.2cm}\\

        \hspace{-3cm} (e) & \hspace{-3cm} (f) & \hspace{-3cm} (g) & \hspace{-3cm} (h) \vspace{0cm}\\
        \includegraphics[width=3.3cm,height=3.1cm,trim=0.3cm 0.3cm 0.3cm 0cm]{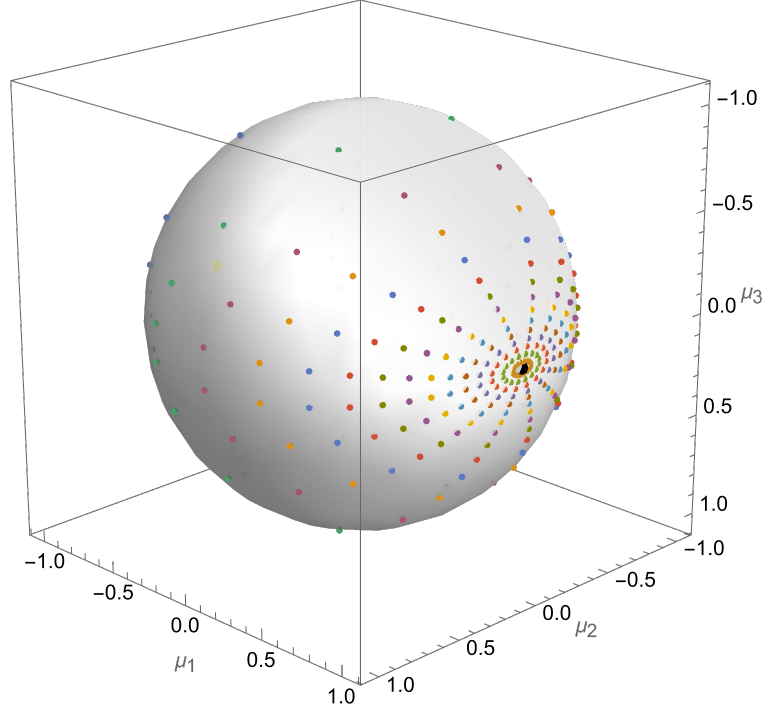} &
			\includegraphics[width=3.3cm,height=3.1cm,trim=0.3cm 0.3cm 0.3cm 0cm]{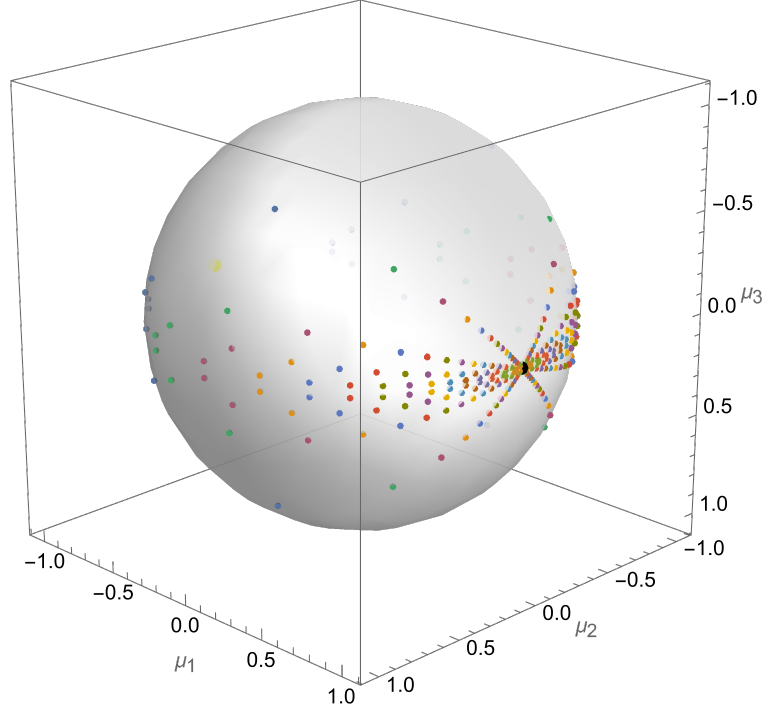} &
			\includegraphics[width=3.3cm,height=3.1cm,trim=0.3cm 0.3cm 0.3cm 0cm]{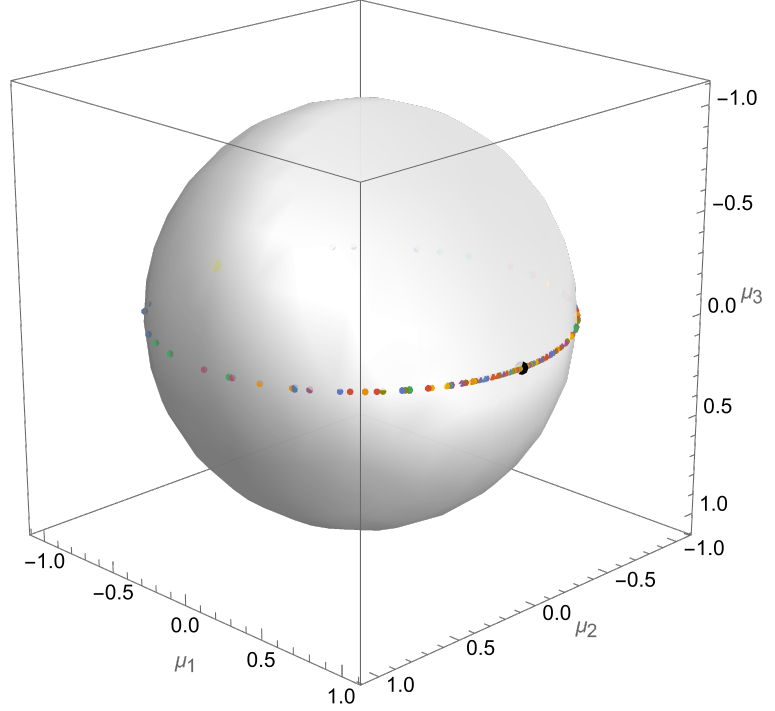} &
            \includegraphics[width=3.3cm,height=3.1cm,trim=0.3cm 0.3cm 0.3cm 0cm]{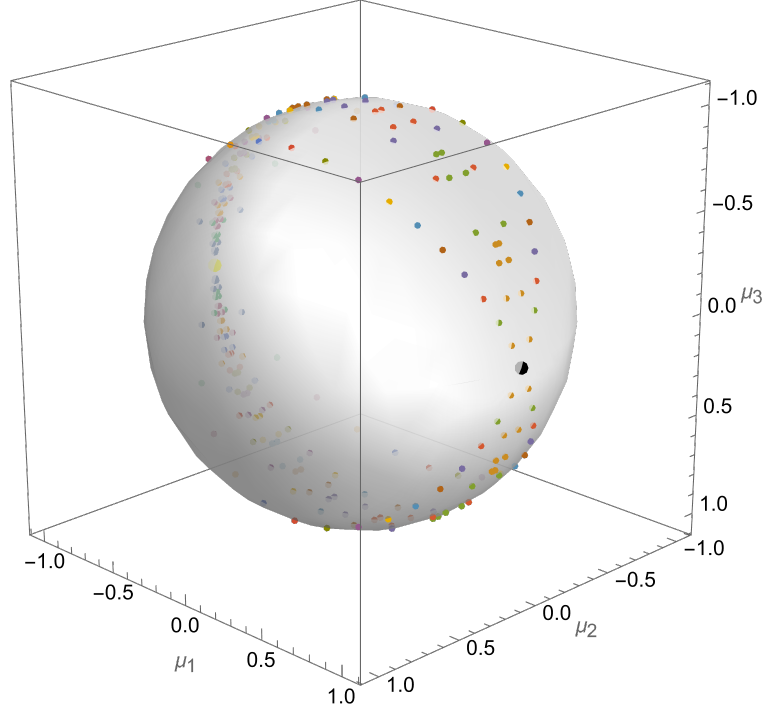} 		
			\end{tabular}
		
		\caption[]{
			Plots of the link function \eqref{eq:link_s} for $x_s=(\cos ( i \pi / 16),\sin ( i \pi / 16) \cos (2 \pi (j-0.5) /16),$ $\sin ( i \pi / 16) \sin (2 \pi (j-0.5) /16))^\top$, $i,j=1,\ldots,16$ with $p=q_s=3$, $B_0=R_s=I_3$, and: (a) $(\beta_{s2},\beta_{s3})=(1,1)$, (b) $(\beta_{s2},\beta_{s3})=(0.3,0.3)$, (c) $(\beta_2,\beta_3)=(0.1,0.1)$, 
            (d) $(\beta_2,\beta_3)=(0.01,0.01)$, 
            (e) $(\beta_{s2},\beta_{s3})=(0.3,0.3 \times 0.5)$, 
            (f) $(\beta_{s2},\beta_{s3})=(0.3,0.3 \times 0.2)$, (g) $(\beta_{s2},\beta_{s3})=(0.3,0.3 \times 0.001)$, and 
            (h) $(\beta_{s2},\beta_{s3})=(0.3,0.3 \times 10)$.
		The black and yellow dots represent the reference direction $b_{01}=e_1$ and its opposite direction $-b_{01}=-e_1$, respectively.
		}  \label{fig:link_d3s}
\end{figure}

\begin{proposition} \label{prop:mobius_p}
Suppose that the link function \eqref{eq:link_s} satisfies $p = q_s$ and $B_s = \beta_s I_{p-1}$ for $0 \leq \beta_s \leq 1$.
Then the following properties hold for the link function:

\begin{enumerate}[(i)\ ]

\item The link function can be expressed as a M\"obius transformation on the sphere:
\begin{equation}
\mu_s(x_s) = \mathcal{M}_S ( \tilde{x}_s  : B_0 , \phi e_1 ) = \mathcal{M}_S ( x_s  : \tilde{R}_0 , \phi r_{s1} ) ,  \quad x_s \in S^{p-1}, \label{eq:mobius_s2}
\end{equation}
where $\tilde{x}_s = R_s^\top x_s$, $\phi = (1 - \beta_s)/(1 + \beta_s)$, $\tilde{R}_0 = B_0 R_s^\top$, $\mathcal{M}_S$ denotes a modified version of the M\"obius transformation which maps $S^{p-1}$ onto itself \citep{Kato2020}:
\begin{equation}
\mathcal{M}_S ( x : R , \psi ) = R \left\{ \frac{1 - \|\psi \|^2}{\| x + \psi \|^2} ( x + \psi ) + \psi \right\} , \quad x \in \mathbb{R}^{p},  \label{eq:mobius_sphere}
\end{equation}
$R$ is a $p \times p$ orthogonal matrix and $\psi \in \mathbb{R}^p \setminus S^{p-1}$.

\item If $\beta_s = 1$, the link function corresponds to that of \citet{Chang1986} and takes the form:
$ \mu_s(x_s) = \tilde{R}_0 x_s. $

\item If $\beta_s = 1$ and $R_s = I_p$, the link function reduces to that of \citet{Chang1987} and \citet{Rivest1989} given by
$ \mu_s(x_s) = B_0 x_s. $

\item If $\beta_s = 1$ and $B_0 = R_s$, the link function becomes the identity mapping, i.e., $\mu_s(x_s) = x_s$.

\end{enumerate}
\end{proposition}

See \citet{Chang1986} for properties of the link function under an orthogonal transformation in Proposition \ref{prop:mobius_p}(ii) and \citet{Chang1987} and \citet{Rivest1989} for those under a rotation in Proposition \ref{prop:mobius_p}(iii).

The M\"obius transformation \eqref{eq:mobius_s2}, which appears in Proposition \ref{prop:mobius_p}(i), is a conformal mapping which projects the sphere $S^{p-1}$ onto itself.
\citet{Kato2020} discussed this transformation not as a link function but as a transformation to derive a probability distribution.
Unlike the proposed link function \eqref{eq:link_s}, the M\"obius transformation does not allow scale changes with different strengths for each axis.
The expression \eqref{eq:mobius_s2} implies that, considering the original representation of the covariate $x_s$, it would be interpretable to reparametrize $B_0$ as $B_0 = \tilde{R}_0 R_s$, where $\tilde{R}_0$ is a $p \times p$ orthogonal matrix.
With this parametrization, the parameter $\tilde{R}_0$ is a rotation and/or reflection of the transformed covariate.
See \citet{Kato2020} for more details of the M\"obius transformation on $S^{p-1}$.

Figure \ref{fig:link_d3s} displays plots of the link function \eqref{eq:link_s} with $p=q_s=3$ for $B_0=R_s=I$ and some selected values of $(\beta_{s2},\beta_{s3})$.
In Figure \ref{fig:link_d3s}(a)--(d), the link function \eqref{eq:link_s} reduces to the M\"obius transformation \eqref{eq:mobius_s2}.
If $\beta_{s2}=\beta_{s3}=1$, the link function \eqref{eq:link_s} becomes the identity mapping.
As $\beta_{s2} (= \beta_{s3})$ decreases, the points on the sphere are isotropically attracted toward the reference direction $e_1 (= b_{01})$.
If $\beta_{s2} = \beta_{s3} \simeq 0$, all the points apart from $-b_{01}(=-e_1)$ take close values to $b_{01}$. 
A comparison among the frames (b) and (e)--(h) of Figure \ref{fig:link_d3s} suggests that the scale of $x_s(=\tilde{x}_s)$ with respect to the axis $b_{0j}$ can be controlled by changing the value of $\beta_{sj}$ $(j=2,3)$.
In particular, Figure \ref{fig:link_d3s}(g) suggests that if $\beta_{s3} \simeq 0 $, then the values of $x_s$ are nearly transformed into the unit circle spanned by $\{e_1,e_2\}$; see Proposition \ref{prop:mu_dim}.
If $\beta_{s3}$ is much greater than one, the transformed points on the sphere tend to lie close to the unit circle spanned by $\{e_1, e_3\}$ and, unlike the case when $\beta_{s2} \simeq 0$, also near the opposite reference point $-b_{01}$.

Note that the M\"obius transformation \eqref{eq:mobius_sphere}, which maps the sphere onto itself, is a special case of the M\"obius transformation on the extended Euclidean space $\overline{\mathbb{R}}^{p-1}$, which is defined as
\begin{equation}
	\mathcal{M}_E (x : A , \gamma , a ,b, \varepsilon) = A \left( \gamma \, \frac{x+a}{\|x+a\|^{\varepsilon}} + b \right), \quad x \in \mathbb{R}^{p} \setminus \{-a\}, \label{eq:mobius_com}
\end{equation}
where $a,b \in \mathbb{R}^{p}$, $\gamma \in \mathbb{R} \setminus \{ 0 \}$, $\varepsilon \in \{0,2\}$, and $A$ is a $p \times p$ orthogonal matrix.
In addition, define $\mathcal{M}_E (-a) = Ab$ and $\mathcal{M}_E (\infty) = \infty$ for $\varepsilon=0$, and $\mathcal{M}_E (-a) = \infty$ and $\mathcal{M}_E (\infty) = A b$ for $\varepsilon=2$.

If $p = q_s$, the stereographic projection $\mathcal{S}$ and its inverse $\mathcal{S}^{[-1]}$ can also be viewed as special cases of \eqref{eq:mobius_com} with restricted domains.
Specifically, they can be expressed as domain-restricted versions of the following functions:
\begin{align}
\tilde{\mathcal{S}}(x) & = \left( I - 2 e_1 e_1^\top \right) \left\{ 2 \frac{x+e_1}{\|x+e_1\|^2} - e_1 \right\} , \quad x \in \mathbb{R}^p \setminus \{-e_1\},  \label{eq:s_star}\\
\tilde{\mathcal{S}}^{-1} (x) & = \left( I - 2 e_1 e_1^\top \right) \left\{ 2 \frac{x-e_1}{\|x-e_1\|^2} + e_1 \right\}, \quad x \in \mathbb{R}^p \setminus \{e_1\} . \label{eq:s_star_inv}
\end{align}
Also define $\tilde{\mathcal{S}}(-e_1) = \infty$ and $\tilde{\mathcal{S}}(\infty)=e_1$.
Then it is clear that $\mathcal{S}(x) = \tilde{\mathcal{S}}(x)$ and $\mathcal{S}^{[-1]}(x) = \tilde{\mathcal{S}}^{-1}(x)$ for $x \in S^{p-1}$.
The M\"obius transformation on $\overline{\mathbb{R}}^p$ has been studied extensively in mathematics, and its mathematical properties can be found, e.g. in Chapter 2 of \citet{Iwaniec2001} and Chapter 4 of \citet{Ratcliffe2019}.

It is known that the group of the M\"obius transformations \eqref{eq:mobius_com} is closed under composition.
Sets of the proposed link functions \eqref{eq:link_s} satisfying certain parameter conditions are also closed under composition; see Appendix \ref{sec:closure} in the online supplementary material.

\subsection{Other special cases} \label{sec:special_others}

In this subsection, we consider two special cases of the link function \eqref{eq:link} where Euclidean covariates are included.
The first case we consider is the link function with $q_s=0$, namely,
\begin{equation}
\mu_e (x_e) = B_0 \mathcal{S}^{[-1]} \left(  B_e R_e^\top x_e  \right), \quad x_e \in \mathbb{R}^{q_e}.\label{eq:link_e}
\end{equation}
As the following proposition states, this link function is related to that of \citet{Fisher1992}.
\begin{proposition} \label{prop:link_FL}
	Let $p=2$,
	$$
	\mu_e(x_e) = \begin{pmatrix} \cos (\mu_{FL} (x_e)) \\ \sin (\mu_{FL} (x_e)) \end{pmatrix}, \quad
	B_0 = \begin{pmatrix} \cos \beta_0 & -\sin \beta_0 \\ \sin \beta_0 & \cos \beta_0
	\end{pmatrix},
	$$
	where $\mu_{FL} ( x_e ),\beta_0 \in [-\pi,\pi)$.
	Then the link function \eqref{eq:link_e} can be expressed in polar-coordinate form as the link function of \citet{Fisher1992} given by
	\begin{equation}
		\mu_{FL}( x_e ) = \beta_0 + 2 \arctan \left(  \gamma^\top x_e \right), \label{eq:link_FL}
	\end{equation}
    where $\gamma = \beta_{e2} R_e \in \mathbb{R}^{q_e}$.
\end{proposition}

\citet{Fisher1992} presented a general class of link function which is essentially of the form
\begin{equation}
\tilde{\mu}_{FL} (x_e) = \beta_0 + g (\gamma^\top x_e), \label{eq:link_FL_general}
\end{equation}
where $\beta_0 \in [-\pi,\pi)$ and $\gamma$ is a $q_e$-dimensional vector of regression coefficients.
Here $g(x)$ is a function that maps the real line to the circle, $g(0)=0$, and they consider monotone functions having the property that as $x$ ranges from $-\infty$ to $\infty$, $g(x)$ ranges from $-\pi$ to $\pi$.
The special case of our link function \eqref{eq:link_FL} corresponds to the case $g(x) = 2 \arctan(x)$ in their general model \eqref{eq:link_FL_general}.
This special case was also considered by \citet{Fisher1992}; readers can refer to their paper for plots and simulations of this link function.

In Figure \ref{fig:link_d2}(b), the plots of the link function $\mu_{FL}(x_e)$ are shown for fixed values of $\beta_0$ and the first element of $\gamma$ and for four values of $\gamma_1$, the second element of $\gamma$.
As $|\gamma_1|$ decreases, the transformed point $\mu_{FL}(x_e)$ approaches $\beta_0 = 0$.
Larger values of $|\gamma_1|$ result in $\mu_{FL}(x_e)$ getting closer to $\beta_0 + \pi = \pi \, (\equiv -\pi)$.
As suggested by the case $\gamma_1 = 0.3$, the link function $\mu_{FL}(x_e)$ can be approximated by a straight line when $\gamma_1 x_e$ is small.
This follows from the fact that
$
\mu_{FL}(x_e) = 2 \arctan(\gamma_1 x_{e1}) \simeq 2 \gamma_1 x_{e1} + O((\gamma_1 x_{e1})^3)
$
for small $x_{e1}$.
The positive and negative signs of $\gamma_1$ indicate positive and negative relationships between $x_e$ and $\mu_{FL}(x_e)$, respectively.
Although not shown in the figure, it is clear that the first element of $\gamma$ serves as the location parameter for $x_{e1}$.


Another special case of the link function \eqref{eq:link} can be defined as follows.
Suppose in the link function \eqref{eq:link} that $p=q_s=2$,
$$
x_s = \begin{pmatrix} \cos \theta_{x} \\ \sin \theta_{x} \end{pmatrix}, \quad
\mu_s(x_s) = \begin{pmatrix} \cos ( \mu_{H}(\theta_x , x_e)) \\ \sin (\mu_{H} (\theta_x , x_e)) \end{pmatrix},
$$
$$
B_0 = \begin{pmatrix} \cos \beta_0 & -\sin \beta_0 \\ \sin \beta_0 & \cos \beta_0  
\end{pmatrix}, \quad
R_s = 
\begin{pmatrix} \cos \eta & -\sin \eta \\ \sin \eta & \cos \eta
\end{pmatrix}
\begin{pmatrix} 1 & 0 \\ 0 & \delta
\end{pmatrix}, \quad 
\beta_{e2} R_e = \gamma,
$$
where $\theta_{x}, \beta_0, \eta \in [-\pi,\pi)$, $\delta \in \{-1,1\}$.
Then, using similar approaches to obtaining Propositions \ref{prop:mobius_2} and \ref{prop:link_FL}, it can be seen that \eqref{eq:link} is of the form
\begin{equation}
\mu_H (\theta_x , x_e) = \beta_0 + 2 \arctan \left\{ \delta \beta_{s2} \tan \left( \frac{\theta_{x}-\eta}{2} \right) + \gamma^\top x_e \right\}, \label{eq:link_h}
\end{equation}
where $\mu_H (\theta_x , x_e) \in \in [-\pi,\pi)$.
To the best of our knowledge, this is a new link function for cylindrical data.
This link function can be considered a hybrid of the link function of \citet{Downs2002} and that of \citet{Fisher1992}.
If $\gamma=0$, the link function \eqref{eq:link_h} reduces to the link function \eqref{eq:mobius_2} of \citet{Downs2002}.
If \( \beta_{s2} = 0 \), our link function \eqref{eq:link_h} reduces to the link function \eqref{eq:link_FL} of \citet{Fisher1992}.  
For a fixed value of \( \theta_x \), equation \eqref{eq:link_h} simplifies to the link function of \citet{Fisher1992}, but with an intercept.  
For a fixed value of \( x_e \), \( \mu_H(\theta_x, x_e) \) is not symmetric with respect to \( \theta_x \), and it can be readily seen that \( \mu_H(\theta_x, x_e) = 0 \) implies  
$
\theta_x = \eta + 2 \arctan \{ -\gamma^\top x_e / (\delta \beta_{s2}) \}.
$  
Plots and interpretation of this link function are provided in Figure \ref{fig:link_h} of the online supplementary material.


\section{Maximum likelihood estimators}
\label{sec:estimators}

In this section we describe parameter estimation based on maximum likelihood for spherical regression with link function (\ref{eq:link}) and error distribution the scaled von Mises--Fisher distribution (\ref{eq:log_lik}).  With suitable modifications, a similar approach could be developed for any family of error distributions in (\ref{class}), such as the Kent family or the ESAG family. 

A preliminary transformation (Section \ref{sec:prelim_transform}) and a reparameterisation of the mean link (Section \ref{sec:reparametrization}) greatly simplify the search.
For regression with the von Mises--Fisher density, the parameters of the mean link are found by maximising $\sum_i y_i^\top \mu(x_i)$ using automatic differentiation \citep{bell2023cp} and a gradient-based numerical solver \citep{kraft1994al}.
The optimum concentration is  found by maximum likelihood using a general univariate numerical solver and accurate implementations of Bessel functions.
For regression with the scaled von Mises--Fisher density, the orthogonal vectors \( (\gamma_2(x_i), \ldots, \gamma_p(x_i)) \) are selected using parallel transport (see Section~\ref{sec:gamma}) and all parameters are optimised at once (see Section \ref{sec:SvMF_estimation}).

\subsection{Reparametrization of mean link}
\label{sec:reparametrization}
The following reparameterisation of the mean link \eqref{eq:link2} greatly simplifies estimation by avoiding the direct optimisation of $B_0$, $R_s$ and $R_e$, which are all on Stiefel manifolds.

\begin{proposition} \label{prop:reparametrization}
The parameters which determine the link function $\mu(x)$ from \eqref{eq:link2} may be written
$
(b_{01},r_{s1},\Omega).
$
Here $\Omega= B_{0,-1} ( B_s R_{s,-1}^\top , B_e R_{e}^\top )$ is a $(p,q_s+q_e)$ matrix with three constraints
$b_{01}^\top \Omega = 0_{q_s+q_e}^\top$, $\Omega\tilde{I}_s r_{s1} = 0_p$, and $\Omega\tilde{I}_s \tilde{I}_s^\top \Omega^\top$ and 
 $\Omega\tilde{I}_e \tilde{I}_e^\top \Omega^\top$ must commute,
where $\tilde{I}_s = ( I_{q_s}, O_{q_s \times q_e} )^\top$ and $\tilde{I}_e = ( O_{q_e \times q_s}, I_{q_e} )^\top$.
Using these new parameters $(b_{01},r_{s1},\Omega)$:

\noindent (i) the mean link $\mu(x)$ can be expressed as
\begin{equation}
\mu(x) = \frac{(1 - \| \tilde{t}(x) \|^2) b_{01} + 2 \tilde{t}(x)}{1 + \| \tilde{t}(x) \|^2}, \label{eq:link_repa}
\end{equation}
where 
$
\tilde{t}(x) = \Omega  \{ \tilde{I}_s x_s / (1 + r_{s1}^\top x_s ) + \tilde{I}_e x_e \}
$ and $
\tilde{I}_e = ( O_{q_e \times q_s}, I_{q_e} )^\top
$

\noindent (ii) for positive-definite $B_s^2 + B_e^2$ the singular valued decomposition (SVD) of $\Omega$ is 
$\Omega = B_{0,-1} \Sigma V^\top$ where
\[\Sigma = (B_s^2 + B_e^2)^{1/2} \quad\quad \textrm{and} \quad\quad
V^\top = (B_s^2 + B_e^2)^{-1/2} (B_s R_{s,-1}^\top , B_e R_{e}^\top)
\]
where for a diagonal matrix $A$, the meaning of $A^2$ and $A^{1/2}$ is obvious.

\noindent (iii) the original parameters can be recovered as follows:
\[
(\Sigma V^\top \tilde I_s)(\Sigma V^\top \tilde I_s)^\top = B_s R_{s,-1}^\top R_{s,-1} B_s^\top = B_s^2
\]
and $B_s^{-1}\Sigma V^\top \tilde I_s = B_s^{-1} B_s R_{s,-1}^\top = R_{s,-1}^\top$, when $B_s$ is positive-definite.
The parameters for the Euclidean covariate $B_e$ and $R_e$ can be recovered similarly.
The matrix $B_0$ can be recovered from $b_{01}$ and the left-singular vectors of $\Omega$.
When singular values of $\Omega$ are distinct, then the recovered parameters are unique up to sign of $b_{0j}$ and $(r_{sj}, r_{e(j-1)})^\top$, $j=2,...,p$.
Scales $\beta_{sj}$ or $\beta_{ej}$ that are zero are easily detected from the SVD, and the corresponding columns of $R_{s,-1}$ and $R_e$ are then ignored. 
\end{proposition}

The commutativity constraint in Proposition \ref{prop:reparametrization} ensures that right-singular vectors of $\Omega$ can be divided into the two sets of orthogonal vectors $R_{s,-1}$ and $R_e$ (i.e $(\Sigma V^\top \tilde I_s)(\Sigma V^\top \tilde I_s)^\top$ produces a diagonal matrix).
See Appendix \ref{sec:reparameterization_pf} for a proof of Proposition \ref{prop:reparametrization}.

Furthermore, given $b_{01}$ and $r_{s1}$, any $(p,q_s + q_e)$ matrix $M$ can be projected to $\mathrm{Proj}(M) = (I_p - b_{01} b_{01}^\top)M
(I_{q_s + q_e} - \tilde{I}_s r_{s1}r_{s1}^\top \tilde{I}_s^\top)$ to satisfy 
$b_{01}^\top \mathrm{Proj}(M) = 0_{q_s+q_e}^\top$ and  $\mathrm{Proj}(M)\tilde{I}_s r_{s1} = 0_p$.
Note that $\mathrm{Proj}(\Omega) = \Omega$ for any $\Omega$ satisfying the constraints.
Thus we can reduce the optimisation problem to maximising the log-likelihood given by $(b_{01}, r_{s1}, \mathrm{Proj}(M))$ where $M$ is any $(p,q_s + q_e)$ with constraints
$\|b_{01}\| = 1$, $\|r_{s1}\| = 1$ and $\mathrm{Proj}(M)\tilde{I}_s \tilde{I}_s^\top \mathrm{Proj}(M)^\top$ and 
 $\mathrm{Proj}(M)\tilde{I}_e \tilde{I}_e^\top \mathrm{Proj}(M)^\top$ must commute.

\subsection{Preliminary transformations}
\label{sec:prelim_transform}
Prior to estimation, spherical covariates and response were rotated according to the preliminary transformation by \citet[Section 4.1.3]{Scealy2019}.
Non-constant Euclidean covariates were centered and rotated by principal component analysis.
Such transformations can be incorporated into the fitted parameters for more interpretable results.
We also recommend centering and scaling Euclidean covariates by their standard deviation so that all covariates are unitless, however we do not do this automatically as such scaling cannot always be incorporated into the parameters of the mean link.

After these preliminary transformations, default initial parameters for the mean link were $B_0 = I_p$, $B_s = 0.9 I_{p-1}$, $B_e = 0.9 I_{p-1}$, with 
$R_s = (I_p, 0_{q_s -p, p})^\top$ and similarly
$R_e = (I_{p-1}, 0_{q_e - (p-1), p-1})^\top$.
However, multiple initial parameters should be tried.

\subsection{A parallel transport construction of $(\gamma_2(x), \ldots, \gamma_p(x))$ in model class (\ref{class})} \label{sec:gamma}

Given a covariate vector $x$ and mean direction $\mu(x)$ determined using (\ref{eq:link}), if a model of the form (\ref{class}), e.g. (\ref{eq:log_lik}), is to be used,  it is necessary to define orthonormal orientation vectors $\gamma_2(x), \ldots, \gamma_p(x)$, which are also orthogonal to $\mu(x)$.
For our regressions, we construct these vectors using parallel transport, which is the closest analogy to constant direction available on Riemannian manifolds.
This approach to estimating the orientation of the error distribution at each $\mu(x)$ does greatly simplify the problem, perhaps even to the point of over-simplifying.  However, the only alternative approach we are aware of is that of \citet{Paine2020} which involves the introduction of several linear predictors. In the absence of clear evidence that our parallel transport approach to estimating $\gamma_j(x)$, $j=2, \ldots , p$ is inadequate, we believe this approach is preferable on the grounds of simplicity. In our numerical examples there was no indication that the parallel transport approach is inadequate.

Denote by $R_{a,b}$ the negative of Jupp's \citeyearpar{Jupp1988} rotation matrix, 
\[R_{a,b} = I_p - \frac{(a+b)(a+b)^\top}{1+b^\top a},\]
which is simple to calculate and represents the parallel transport of a tangent vector at $a \in \mathbb{S}^{p-1}$ to a vector in the tangent space at $b \in \mathbb{S}^{p-1}$, where $\mathbb{S}^{p-1}$ has the canonical embedding in $\mathbb{R}^p$ and transport is along the minimum-distance geodesic between $a$ and $b$ (Appendix \ref{sec:parallel_tansport_form}); 
a more complicated but equivalent representation that can be useful in other settings is the rotation matrix from \citet[Lemma 2]{Amaral2007}.
When $b = -a$, there are multiple minimum-distance geodesic paths between $a$ and $b$ and $R_{a,b}$ is not well defined.


Let $\gamma_{01}, \gamma_{02}, \ldots , \gamma_{0p}$ be orthonormal vectors; we term $\gamma_{01}$ the \emph{base location} of the parallel transport.
For each covariate vector $x$ and for $\gamma_{01} \in \mathbb{S}^{p-1}$, we then define $\gamma_{j}(x;\gamma_{01})$, $j = 2, ..., p$ as
$\gamma_{j}(x;\gamma_{01}) = R_{\gamma_{01},\mu(x)}\gamma_{0j}$.
We treat the base location $\gamma_{01}$ as an unknown unit vector parameter and estimate $\gamma_{01}$ along with the other unknown parameters by maximizing the likelihood.
In some situations it may be desired to instead associate $\gamma_{01}$ with another unit vector e.g. take $\gamma_{01}=b_{01}$ or take $\gamma_{01}=\sum_{i=1}^n y_i/\vert \vert \sum_{i=1}^n y_i  \vert \vert$.
In the case where there is a spherical covariate but no Euclidean covariate vector, an alternative method of determining axes $\gamma_2(x), \ldots , \gamma_p(x)$ is to use a Gram-Schmidt orthogonalization procedure.  See Appendix B.2.
  


\subsection{Estimation of axes and shape}
\label{sec:SvMF_estimation}
First, a preliminary von-Mises Fisher regression with link \eqref{eq:link} is conducted leading to preliminary estimates of the mean link and concentration.
Let $\mu_i$ be the mean direction specified by these preliminary estimates, for each observed $y_i$, $i=1,\ldots,n$.
Given a $\gamma_{01}$ we obtain a preliminary estimate of the axes $\gamma_{02}, \ldots , \gamma_{0p}$ by applying the \citet[Section 4.1.1]{Scealy2019} method of moments estimate to rotated residuals \citep{Jupp1988}. 
For any $a \in \mathbb{S}^{p-1}$, define $P_a=I_p-a a^\top$.
For each $i$, $i=1,\ldots,n$, the rotated residual $v_i=R_{\gamma_{01}, \mu_i}^\top P_{\mu_i}y_i$ is in the tangent space of $\gamma_{01}$.
A moment estimate of the axes and their scales is provided by the spectral decomposition
$\sum_{i=1}^n v_i v_i^\top = \hat\Gamma \Lambda \hat\Gamma^\top$,
where the columns of $\hat\Gamma$ are $[\gamma_{01} \, \hat\gamma_{02} \, \ldots \hat\gamma_{0p}]$ and $\Lambda$ is a $p \times p$ diagonal matrix of eigenvalues satisfying $0=\lambda_1$, $\lambda_2 \geq \ldots \geq \lambda_p > 0$, which have a  non-linear relationship to the shape parameters $\kappa, a_1, a_2, \ldots, a_p$.
The analogous relationship between population quantities follows from \citep[Proposition 3]{Scealy2019}; see Appendix \ref{sec:rotresid_2ndmo}.
A preliminary estimate of the shape parameters $a_2, \ldots, a_p$ can then be obtained using the standard deviation of the rotated residuals in each orientation axis (echoing \citep[Proposition 2]{Scealy2019}).

Using these preliminary estimates as the initial parameters,
maximum-likelihood estimation of all mean-link parameters and distributional parameters is conducted simultaneously with the tail-parameter $a_1$ fixed apriori.
The optimisation uses automatic differentiation \citep{bell2023cp} and a gradient-based numerical solver \citep{kraft1994al}.
When $p=3$, the normalising constant $c_{g,3}(\kappa,\lambda) = c_3(\kappa) a_1$ is computed exactly as $a_1 2\pi (e^\kappa - e^{-\kappa})/\kappa$.
For automatic differentiation, when $p\neq 3$ the normalising constant $c_{g,p}(\kappa,\lambda) = c_p(\kappa) a_1$ uses approximations of the modified Bessel function of the first kind for small $\kappa$ \citep[Eq 10.25.2]{nist2025di} and large $\kappa$ \citep[p.377, Section 9.7.1]{abramowitz1964ha}. 
The positive scales $a_2, \ldots, a_p$ are  chosen such that $ \prod_{j=2}^p a_j$.
The axes $\gamma_{02}, \ldots \gamma_{0p}$ are expressed as rotation of the away from the initial (i.e. preliminary) axes, with this rotation parameterised as a $(p-1) \times (p-1)$ skew-symmetric matrix using an inverse Cayley transform.
If $p\neq 3$, then after convergence of the above optimisation, the estimated concentration $\kappa$ is refined independently of the other parameters using accurate implementations of Bessel functions provided by the R language.

\section{Applications}
\label{sec:applications}
Here we demonstrate estimation of the  link function (\ref{eq:link}) with scaled von Mises--Fisher distribution for data on $S^2$ and data on $S^4$.
In both cases we estimate the base location $\gamma_{01}$ of the parallel transport.

\subsection{Midatlantic ridge}
We applied our regression to the midatlantic ridge data ($n=70$) analysed by \citet{Rosenthal2014}.
We treated locations on the ridge as the spherical covariate and the corresponding location on the continent as the response.
A Euclidean covariate was used to indicate the western and eastern side of the ridge and we standardised this covariate to have a mean of 0 and standard deviation of 1.
The leave-one-out-cross-validated mean-squared error of our regression was $0.01$, which was noticeably better than the value of $0.074$ reported by \citet{Rosenthal2014}.

\begin{figure}
	\begin{center}
		\includegraphics[width = \textwidth-2cm]{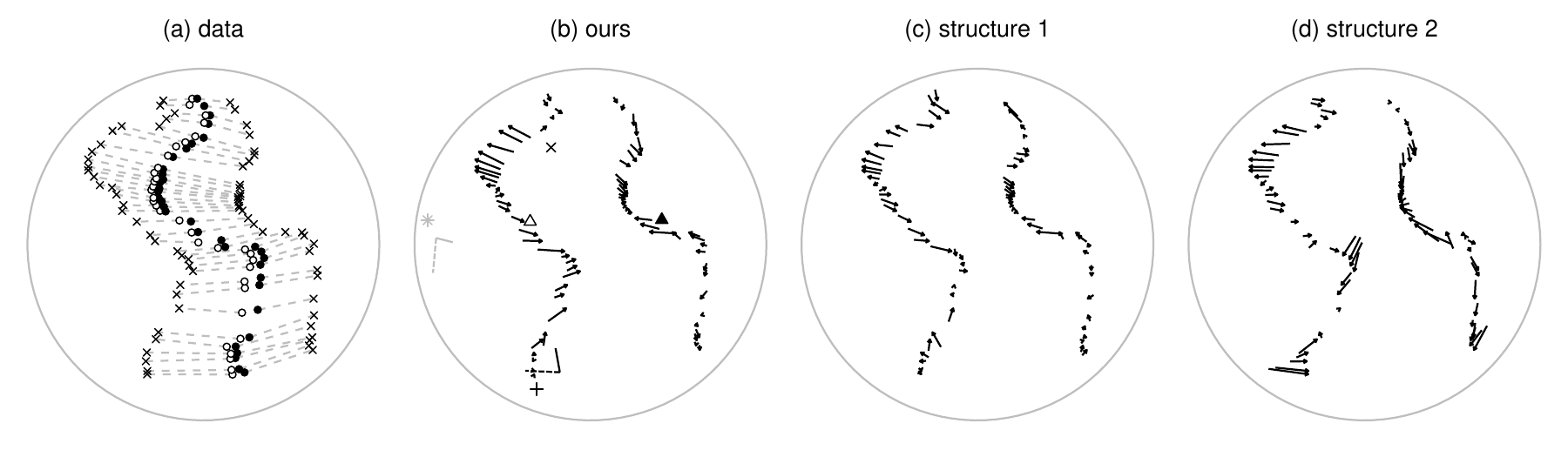}
	\end{center}
	\caption[]{
Regression for the midatlantic ridge data.
(a) midatlanic ridge (circles) and corresponding locations on the continent (crosses) from \citet{Rosenthal2014}.
(b) our regression.
\citet{Paine2020} structure 1 (c) and 2 (d) regression.
The sphere is shown orthogonally projected with north pointing up the page.
Arrows: start at the predicted mean and end at the observed continental location, and thus represent residuals.
Filled symbols: eastern side.
Unfilled symbols: western side.
Triangles: predicted continental location for the western (unfilled) and eastern (filled) side from the average ridge location.
Plus, cross and star: estimated value of $r_{s1}$, $r_{s2}$ and $r_{s3}$ respectively; $r_{s2}$ in gray is on the back side of the sphere.
Pair of black lines: direction of the estimated $b_{02}$ (solid) and $b_{03}$ (dashed) at the estimated $b_{01}$.
Pair of gray lines: estimated directions of $\gamma_{02}$ (solid) and $\gamma_{03}$ (dashed) at the estimated $\gamma_{01}$.
}
    \label{fig:midatlantic}
\end{figure}

Our regression had fewer degrees of freedom (16 compared to 22) and slightly better AIC (-463 compared to -460) than the \citet{Paine2020} Structure 1 regression, and had visually similar predicted means (Figure \ref{fig:midatlantic}).
Furthermore the estimates for our link function are more interpretable.
Suppose $\mu(x) = b_{01}$ (plus symbol in Figure \ref{fig:midatlantic}b), the pair of black lines in Figure \ref{fig:midatlantic}b then indicate the direction of change of $\mu(x)$ from infinitesimal positive changes in the $t(x)$ from Proposition \ref{prop:link2}.
For other locations of $\mu(x)$, the effect of $t(x)$ can be inferred from the gradient of the inverse stereographic projection.
From the estimated $R_e$ and $B_e$ (Table \ref{tab:midatlantic}) the western side has a relatively large positive effect on the $t_3$ in Proposition \ref{prop:link2}.
Thus, given an $x$ such that $\mu(x) = b_{01}$, switching from the eastern side to the western side moves the mean along the direction of the dashed black line in Figure \ref{fig:midatlantic}b.
The influence of the spherical covariate (midatlantic ridge location) can be interpreted from the estimated scales in $B_s$ (Table \ref{tab:midatlantic}) and the columns of $R_s$ (plus, cross and star in Figure \ref{fig:midatlantic}b).
As ridge locations get closer to $r_{s2}$ (cross symbol) the value of $t_2(x)$ increases, moving the mean further along the directions given by $b_{02}$ (black line).
Moving ridge locations closer to $r_{s3}$ (star symbol), the value of $t_3(x)$ increases, moving the mean further in the $\hat{b}_{03}$ (dashed black line) directions.
For ridge locations further from $r_{s1}$ (plus symbol), $1+\hat{r}_{s1}^\top x_s$ is smaller, which magnifies the effect of other changes in the ridge location.
The estimated scales of the scaled von-Mises Fisher, $\hat{a}_2 = 2.38$ and $\hat{a}_3 = 0.42$, indicate that the error of the predicted mean was larger in directions parallel to the gray line in Figure \ref{fig:midatlantic}b.

\begin{table}
\centering
\begin{tabular}[t]{lrrrr}
\toprule
\multicolumn{3}{c}{ } & \multicolumn{2}{c}{$R_e^\top$} \\
\cmidrule(l{3pt}r{3pt}){4-5}
  & diag($B_s$) & diag($B_e$) & western side & $1$\\
\midrule
$t_2$ & 0.90 & 0.07 & 0.13 & -0.99\\
$t_3$ & 0.77 & 0.25 & 0.99 & 0.13\\
\bottomrule
\end{tabular}
\caption{Selected estimated mean link parameters. Parameters in the top and bottom row effect $t_2$ and $t_3$ from Proposition \ref{prop:link2}, respectively.}
\label{tab:midatlantic}
\end{table}

\subsection{Earthquake moment tensors on $S^4$}
\begin{figure}
	\begin{center}
		\includegraphics[width = \textwidth-1cm]{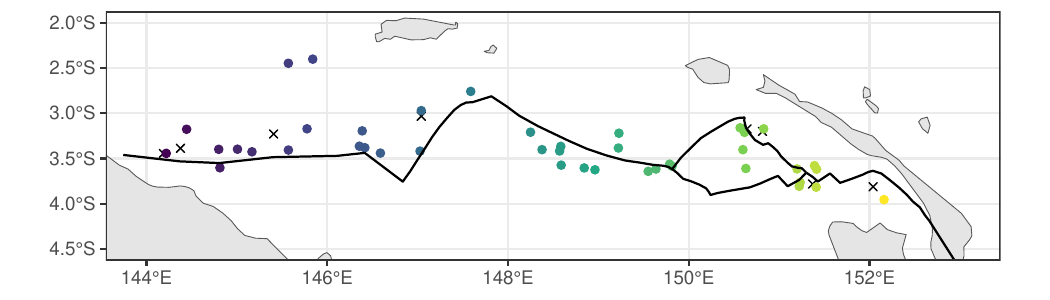}
	\end{center}
	\caption[]{Locations of earthquakes along the Bismarck Sea Seismic Line (solid line) near Papua New Guinea. Eight earthquakes (marked x) were outliers and 42 (bullet) were used for regression.
    Colours match Figure \ref{fig:eathquake_res} and are given by longitude.
	}
    \label{fig:earthquake_locations}
\end{figure}

In seismology, the moment tensor of an earthquake is given as a $3\times 3$ symmetric matrix representing the force-couples that describe the physics of deformation at the source of the earthquake.
Moment tensors are routinely calculated with the moment tensor's trace set to zero \citep[e.g.][]{lay1995mo} and studied with the magnitude (Frobenius norm) rescaled to be 1 \citep[e.g.][]{tape2019ei}.
The resulting symmetric matrices can be expressed as unit vectors in $\mathbb{R}^5$ by applying contrasts from a Helmert submatrix \citep[p305]{mardia2000di} to the diagonal elements of the moment tensors, Mtt, Mrr, and Mff, and scaling the off-diagonal elements, Mrf, Mrt, and Mtf, by $\sqrt{2}$, where \emph{r}, \emph{t} and \emph{f} represent the directions up, south and east, respectively.
Here we study 42 shallow earthquakes (depth $<$ 20km) along the Bismarck Sea Seismic Line (BSSL) \citep{hejrani2017ce}.
See Figure \ref{fig:earthquake_locations}.
For Euclidean covariates we use the direction (strike) of the BSSL at the closest point to the earthquake, earthquake location (latitude, longitude), and a covariate that is equal to earthquake longitude when it is above 148 and zero otherwise.
We do not use any spherical covariates.

\begin{figure}
	\begin{center}
		\includegraphics[width = \textwidth-2cm]{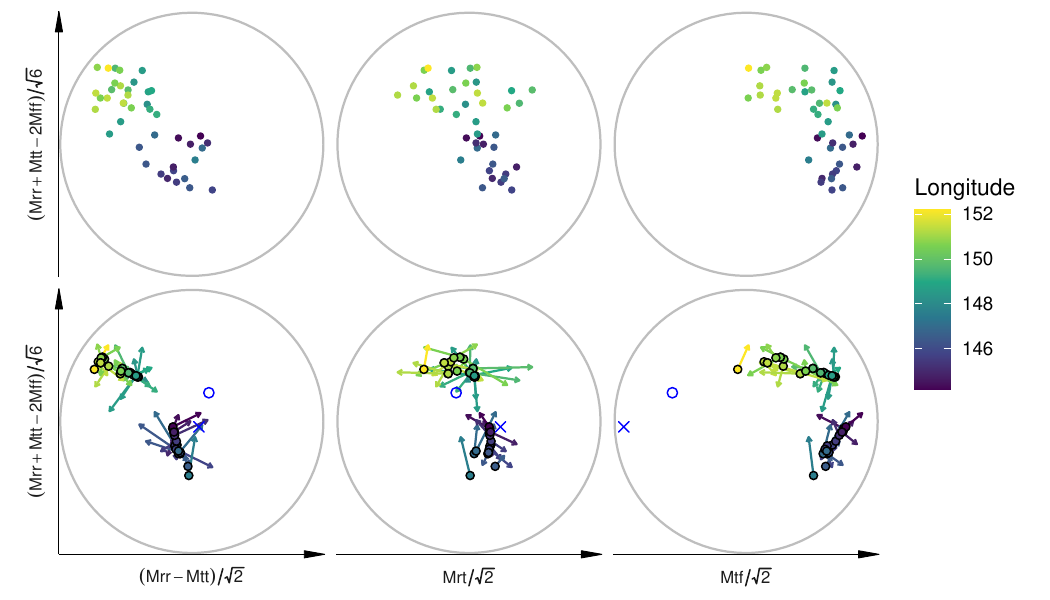}
	\end{center}
	\caption[]{42 normalised earthquake moment tensors on $S^4$.
Moment tensors are shown orthogonally projected onto three pairs of basis directions given by expressing the moment tensors as unit vectors in $\mathbb{R}^5$.
Top: observed earthquake moment tensors.
Bottom: predicted earthquake moment tensor according to our regression.
Colour: longitude of earthquakes.
Arrows: represent residuals and point from mean earthquake moment tensor to the corresponding observed earthquake moment tensor.
Blue x: $\hat{b}_{01}$.
Blue circle: $\hat{\gamma}_{01}$.
Grey circle boundary: intersection of $S^4$ with the plane given by the pair of axes.
	} 
    \label{fig:eathquake_res}
\end{figure}

All covariates were standardised to have mean of zero and standard deviation of one.
The regression captured the bipartite behaviour of earthquake moment tensors west and east of $148^\circ$E longitude (see Figure \ref{fig:eathquake_res}), including a decrease of the horizontal force couple (Mtf) with longitude east of $148^\circ$E. This bipartite behaviour is not surprising given that the earthquakes in both these regions have predominately strike-slip mechanisms where the dipping plates on average have a different angle west and east of $148^\circ$E longitude (see Figure 16 in Hejrani et al. 2017, specifically the insert of nodal planes comparing region 2 versus 3 and 4).  The estimated scales were $\hat{a}_2 = 2.08, \hat{a}_3=1.4, \hat{a}_4=1.0, \hat{a}_5 = 0.34$. A parametric bootstrap likelihood ratio test rejected the von Mises--Fisher regression ($p = 0.002$), suggesting that an anisotropic distribution was valuable here.
Our regression had 38 degrees of freedom, much fewer than a \citet{Paine2020} structure 1 regression, which would have required 84 parameters. 
Further results, including  confidence intervals for $\hat{a}_2, \ldots, \hat{a}_5$ estimated by parametric bootstrap, are given in Appendix \ref{sec:earthquake_supp} of the online supplementary material.

\section{Conclusion}
\label{sec:conclusion}
Using a scaled form of the M\"obius transformation, we proposed a new link function for regression with spherical responses.
Our link function allows both Euclidean and spherical covariates, has interpretable parameters and generalises many existing link functions.
In our modelling we used the scaled von Mises--Fisher distribution as the error distribution but this family can be replaced by any family of elliptically symmetric distributions satisfying \eqref{class}, which includes the Kent and ESAG families.
We avoided the introduction of many additional parameters by using parallel transport to determine the axes of symmetry of elliptically-symmetric error distributions as a function of the covariates; and a reparameterisation of the mean link greatly simplified maximum likelihood estimation.  
In applications to data with responses on $S^2$ and $S^4$, our regression models were, respectively, more interpretable and more than halved the degrees of freedom compared to the alternative.




\section*{Competing interests}
No competing interest is declared.

\section*{Author contributions statement}

SK was the principal contributer to the construction of the link function, supported by KLH and to a lesser extent ATAW.  SK also provided nearly all of the proofs of the proporties of the link function. The unified approach to elliptically symmetric families was due to ATAW.  KLH was responsible for the numerical work, while the parallel transport idea for estimating axes was due to KLH and ATAW.  JLS provided expertise in the geophysical data example and all four authors contributed to the preparation and/or checking of the manuscript. 

\section*{Acknowledgments}
SK is grateful to JSPS KAKENHI Grant Number 25K15037 for financial support.
KLH, JLS and ATAW are grateful to the Australian Research Council for supporting this research through grant DP220102232.

\vfill \eject

\vskip 0.2truein

\begin{center}

\textbf{\Large Supplementary Material}

\vskip 0.2truein

\textbf{\Large Regression for spherical responses with linear and spherical covariates using a scaled link function}

\vskip 0.2truein

{\large {\scshape  Shogo\ Kato}\,$^{a}$, \quad {\scshape Kassel\ L.\ Hingee}\,$^{b}$, \vspace{0.1cm}\\
{\scshape Janice\ L.\ Scealy}\,$^{b}$, \quad {\scshape and} \quad {\scshape Andrew\ T.A.\ Wood}\,$^{b}$ \vspace{0.5cm}\\
\textit{\(^{a}\) Institute of Statistical Mathematics, Tokyo, Japan} \vspace{0.1cm}\\
\textit{\(^{b}\) Australian National University, Canberra, Australia}} \vspace{0.5cm}\\ 
\end{center}

\appendix 
\setcounter{figure}{0}
\setcounter{table}{0}
\setcounter{theorem}{0}
\setcounter{page}{1}
\setcounter{equation}{0}

\renewcommand{\thefigure}{\Alph{section}.\arabic{figure}}
\renewcommand{\thetable}{\Alph{section}.\arabic{table}}
\renewcommand{\thetheorem}{\Alph{section}.\arabic{theorem}}
\renewcommand{\theequation}{\Alph{section}.\arabic{equation}}
\numberwithin{equation}{section}


\section{Further results and proofs for the new link function}

\subsection{Further results on the link function}

\subsubsection{Remarks on the derivation of the link function}

Here we provide two remarks on how the proposed link function given in Definition \ref{def:link} is derived.

The first remark is the use of the extended stereographic projection \eqref{eq:stereo} rather than the ordinary stereographic projection which maps the sphere onto the Euclidean space.
Using the stereographic projection defined on the sphere instead of the one on the disc, the link function \eqref{eq:link} can be expressed as
\begin{equation}
\mu(x) = B_0 \mathcal{S}^{[-1]} ( \check{B}_s \check{\mathcal{S}} (\check{R}_s^\top x_s) + B_e R_e^\top x_e  ), \label{eq:mu_alternative}
\end{equation}
where $\check{B}_s = (B_s,O_{(p-1)\times (q_s-p)})$, $\check{R}_s = (R_s,r_{s,p+1},\ldots,r_{sq})$, and $r_{s,p+1}, \ldots, r_{sq_s}$ are orthonormal vectors satisfying $R_s^\top r_{si} = 0_p$ for $i=p+1,\ldots,q_s$.
Here $\check{\mathcal{S}}$ denotes the stereographic projection \eqref{eq:stereo} with $p$ replaced by $q_s$.
This expression implies that although the stereographic projection 
$\mathcal{S}$ on the disc can be replaced by the ordinary stereographic projection 
$\check{\mathcal{S}}$ on the sphere, the alternative expression \eqref{eq:mu_alternative} 
includes the redundant parameters $r_{s,p+1},\ldots,r_{s,q}$ and the last $p-q_s$ components of $\check{\mathcal{S}}(\check{R}_s^\top x_s)$ turn out to be unnecessary. 
Therefore the original expression \eqref{eq:link} is preferable for its simplicity.

The second remark concerns the regression coefficient of the Euclidean variable $x_e$.
For $p=q_e+1$, instead of using the transformation $B_e R_e^\top x_e$ in \eqref{eq:link}, it is also possible to adopt a more common transformation, $\Sigma^{1/2} x_e$, for the Euclidean covariate $x_e$, where $\Sigma$ is a positive definite matrix.
Note that $\Sigma$ can be decomposed as $\Sigma = Q \Lambda Q^\top$, where $\Lambda$ is a diagonal matrix of eigenvalues and $Q$ is an orthogonal matrix. It is straightforward to observe that $\Lambda$ and the subsequent $Q^\top$ play roles similar to $B_e$ and $R_e^\top$, respectively.
In addition, the pre-multiplication by $Q$ can be essentially represented as $B_0$ apart from the sign change.
This suggests that the transformation $B_e R_e^\top x_e$ in \eqref{eq:link} can be interpreted similarly to multiplication by $\Sigma^{1/2}$ and offers a somewhat consistent form with the transformation of the spherical covariate.
The restriction that $B_0$ is a rotation matrix follows the conventions in the models of \citet{Fisher1992,Downs2002}, which will be shown to be special cases of ours, and helps avoid redundancy in the parameter space.

\subsubsection{On the roles of the scales of \( B_s \) and \( B_e \)}

In Section \ref{sec:general} of the main article, the roles of the scales of \( B_s \) and \( B_e \) are briefly discussed. 
The theoretical justification for this discussion is provided in the following proposition.

\begin{proposition} \label{prop:beta}
Let 
$f(\beta) = b_{01}^\top \mu(x ; \beta B_s , \beta B_e)$, $\beta \geq 0$,
where \( \mu(x; B_s , B_e ) \) denotes the link function given by equation \eqref{eq:link}.  
Then \( f(\beta) \) is monotonically decreasing with respect to \( \beta \). 
In particular, \( f(0) = 1 \).
If \( t(x) \neq 0 \), then \( \lim_{\beta \rightarrow \infty} f(\beta) = -1 \), whereas if \( t(x) = 0 \), \( f(\beta) = 1 \) for all \( \beta \geq 0 \).
If \( x_s = -r_{s1} \) and $B_s \neq O_{(p-1) \times (p-1)}$, then \( f(\beta) = -1 \) for any \( \beta > 0 \). 
\end{proposition}

See Section \ref{sec:proof_beta} for the proof.

\subsubsection{Relationship with the link function of \citet{Downs2003}} \label{sec:downs}

As briefly discussed in Section \ref{sec:special_sphere} of the main article, the link function \eqref{eq:link_s} with a single spherical covariate with $d=3$ includes the link function of \citet{Downs2003} as a special case.
The link function of \citet{Downs2003} is defined by
\begin{equation}
\mu_D(x_s) = P_D \left( \mathcal{M}_D \left( P_D^{-1} (x_s) \right) \right), \quad x_s \in S^2, \label{eq:downs}
\end{equation}
where $ \mathcal{M}_D(z) = (\alpha_{11} z + \alpha_{12})/(\alpha_{21}z + \alpha_{22}) $ is the M\"obius transformation on the complex plane, $z \in \overline{\mathbb{C}}$, $\alpha_{11},\alpha_{12},\alpha_{21},\alpha_{22}\in \mathbb{C}$, $\alpha_{11}\alpha_{22}-\alpha_{12}\alpha_{21}$ is nonzero, and $P_D$ and $P_D^{-1}$ are given by
$$
P_D(z) = \left( \frac{2x_c}{x_c^2+y_c^2+1}, \frac{2y_c}{x_c^2+y_c^2+1}  , \frac{x_c^2+y_c^2-1}{x_c^2+y_c^2+1} \right)^\top ,  \quad z = x_c + i y_c \in \mathbb{C},
$$
$$
P_D^{-1}(x_s) = \frac{x_1}{1-x_3} + i \frac{x_2}{1-x_3}, \quad x_s=(x_1,x_2,x_3)^\top \in S^2 \setminus \{ e_3 \},
$$
respectively.
Also, define $P_D(\infty) = e_3$ and $P_D^{-1}(e_3) = \infty$.
Note that $P_D^{-1}$ is a stereographic projection which maps the sphere $S^2$ onto the extended complex plane and $P_D$ is its inverse mapping.
This transformation is also considered in \citet{Downs2009} to derive a probability distribution on $S^2$.
As the following proposition shows, the link function of \citet{Downs2003} belongs to an isotropic special case of our link function for $p=q_s=3$.

\begin{proposition} \label{prop:mobius_3}
The link function \eqref{eq:downs} of \citet{Downs2003} can be expressed using the M\"obius transformation on the sphere $S^2$ as
\begin{equation}
\mu_D(x_s) = \mathcal{M}_S (x_s : R_{0D} , \psi_D), \quad x_s \in S^2, \label{eq:mobius_s3}  
\end{equation}
where $\mathcal{M}_S$ is defined as in \eqref{eq:mobius_sphere}.
Here, for $\alpha_{21} \neq 0$, $R_{0D}$ and $\phi_D$ are 
\begin{equation}
R_{0D} = - B_0(\omega_{neq}) H (e_2) H (u_1) H(v_{neq}), \label{eq:R0D_neq}
\end{equation}
\begin{equation}
\phi_D =  H(v_{neq}) \left(  \frac{2 u_1}{\| u_1 \|^2 } - \frac{2 v_{neq}}{ \|v_{neq}\|^2} - e_3  \right),  \label{eq:phiD_neq}
\end{equation}
respectively, where
$$
B_0(\omega) = 
\begin{pmatrix}
\cos \omega & - \sin \omega & 0 \\
\sin \omega & \cos \omega & 0 \\
0 & 0 & 1
\end{pmatrix} , \quad  
H(v) = I - 2 \frac{v v^\top}{\|v \|^2},
$$
$$
v_{neq}= H (u_1) \left\{ \gamma_{neq} \frac{u_1}{ \| u_1 \|^2 } - H(e_1) B_0(\omega_{neq} )^\top u_2 \right\},
$$
$ u_1 = (\mbox{Re}(\alpha_{22}/\alpha_{21}) , \mbox{Im}(\alpha_{22}/\alpha_{21}),1)^\top $,
$ u_2 = (\mbox{Re}(\alpha_{11}/\alpha_{21}) , \mbox{Im}(\alpha_{11}/\alpha_{21}), 1)^\top $,
$ \gamma_{neq} e^{i \omega_{neq} } = (\alpha_{12}\alpha_{21} -\alpha_{11}\alpha_{22})/\alpha_{21}^2,$ $\gamma_{neq} \geq 0$, $-\pi \leq \omega_{neq} < \pi$.
For $\alpha_{21} = 0$,
\begin{equation}
R_{0D} = - B_0(\omega_{eq}) H (v_{eq}),  \quad
\phi_D =  \left| \frac{\alpha_{11}}{\alpha_{22}} \right| \frac{2 v_{eq}}{\|v_{eq}\|^2} -e_3,  \label{eq:R0D_phiD_eq}
\end{equation}
where
$ \omega_{eq}=\arg (\alpha_{11}/\alpha_{22})$ and
\[
v_{eq} = B_0 (\omega_{eq})^\top (\mbox{Re}(\alpha_{12}/\alpha_{22}), \mbox{Im}(\alpha_{12}/\alpha_{22}) , |\alpha_{11}/\alpha_{22}|-1)^\top.
\]
\end{proposition}
The stereographic projection and affine transformation are also used in an early version of the preprint of \citet{Kato2015c} to derive a probability distribution on the sphere.
However their mapping has a different form from the spherical case of our link function \eqref{eq:link_s} and is used for neither a link function nor a regression model.

\subsubsection{Closure under composition} \label{sec:closure}

As discussed in the last paragraph of Section \ref{sec:special_sphere}, subsets of the proposed link functions \eqref{eq:link_s} possess closure properties under composition.
Details of these properties are given below.

\begin{proposition} \label{prop:closure}
Let $\mu(x_s : B_0 , R_{s1} , B ) $ denote the proposed link function \eqref{eq:link_s} with $p = q_s$.
Then the following closure properties hold for $\mu(x_s : B_0 , R_{s1} , B ) $:
\begin{enumerate}[(i)\ ]
\item 
\begin{equation} \mu \left( \mu ( x_s : B_{01}, R_1 , B_1) : B_{02} , B_{01} , B_2 \right) = \mu (x_s : B_{02} , R_1 , B_1 B_2 ), \label{eq:closure1}
\end{equation}

\item 
\begin{align}
\begin{split} \label{eq:closure2}
\lefteqn{ \mu \left( \mu ( x_s : B_{01}, R_1 ,  \beta_1 I_{p-1}) : B_{02} , R_2 , \beta_2 I_{p-1} \right) } \hspace{2cm}\\
 & = \mu (x_s : B_0' , R' , \beta' I_{p-1} ) = \mathcal{M}_S ( x_s : \tilde{R} , \tilde{\psi}  ), 
\end{split}
\end{align}
where 
$ B_0' = \tilde{R} R' $, $ R' $ is a $ p \times p $ matrix with determinant $ \det( \tilde{R} ) $, whose first column is $ \tilde{\psi} / \| \tilde{\psi} \| $,
$$
\beta' = \frac{1-\| \tilde{\psi} \|}{1+\| \tilde{\psi} \|}, \quad
\tilde{\psi} = \frac{\mathcal{M}_S ( R_1 B_{01}^\top \psi_2 : I_p , \psi_1 )}{ \| \mathcal{M}_S ( R_1 B_{01}^\top \psi_2 : I_p , \psi_1 ) \|^2 } ,
\quad
\psi_j = \frac{1-\beta_j}{1+\beta_j} r_j, \quad j=1,2,
$$
$$
\tilde{R} = B_{02} R_2^\top H(\psi_2) H \left( B_{01} R_1^\top \frac{\psi_1 }{\|\psi_1\|^2} + \frac{ \psi_2 }{ \|\psi_2\|^2 } \right) B_{01} R_1^\top H(\psi_1) H( \tilde{\psi} ).
$$
\end{enumerate}
\end{proposition}
For the isotropic subclasses of the proposed link functions \eqref{eq:link_s} discussed in Proposition \ref{prop:closure}(ii), there is variability in the choice of $R'$.  
Possible ways to construct $R'$ include using a method similar to that given in equation (8) of \citep{Scealy2019}, and applying Gram--Schmidt orthogonalization.

\subsection{Proofs of results for the link function}

\subsubsection{Proof of Proposition \ref{prop:link2}}
\begin{proof}[Proof.]
Part of the link function \eqref{eq:link} can be calculated as
\begin{align*}
B_{s} \mathcal{S} (R_{s}^\top x_{s}) + B_e ( R_e^\top x_e + c_e ) & =  \frac{B_s R_{s,-1}^\top x_s}{1+r_{s1}^\top x_s} + B_e (R_e^\top x_e + c_e) \\
 & = t(x).
\end{align*}
Then, for $x_s \neq -r_{s1}$, it follows that
\begin{align*}
\mu(x) & = B_0 \mathcal{S}^{[-1]}(t(x)) = B_0 \left[ \frac{1}{1+\|t(x)\|^2} \left( 1 - \|t(x)\|^2 , 2 t_2(x) , \ldots , 2 t_p(x)  \right)^\top \right] \\
  & =  \frac{1}{1+\|t(x)\|^2} (b_{01} , \ldots , b_{0p}) \left( 1 - \|t(x)\|^2 , 2 t_2(x) , \ldots , 2 t_p(x)  \right)^\top \\
   & = \frac{(1-\|t(x)\|^2) b_{01} + 2 \sum_{j=2}^p t_j(x) b_{0j} }{1+\|t(x)\|^2}.
\end{align*}
If $x_s=-r_{s1}$, then if follows from Definition \ref{def:link} that $\mu(x)=-b_{01}$ for $B_s \neq O_{(p-1) \times (p-1)}$ and $\mu(x) = b_{01}$ for $B_s =O_{(p-1) \times (p-1)}$.
\end{proof}

\subsubsection{Proof of Proposition \ref{prop:mu_dim}}

\begin{proof}[Proof.]
	
	Before proving Proposition \ref{prop:mu_dim}, we first show the following lemma.
	
\noindent \textbf{Lemma 1}. 
		\textit{For $p<q_s$, it holds that $R_s^\top S^{q-1} = D^p$}.
	\begin{proof}[Proof.]		
		
		We will show the claim by proving (a) $R_s^\top S^{q_s-1} \subset D^p$ and (b) $R_s^\top S^{q_s-1} \supset D^p$.
		\begin{enumerate}[(a)\ ]
			\item 
			For any $x \in S^{q_s-1}$,
			\begin{align*}
				\| R_s^\top x \|^2 & = x^\top R_s R_s^\top x =  (r_{s1}^\top x)^2 + \cdots + (r_{sp}^\top x)^2 \\
				& \leq (r_{s1}^\top x)^2 + \cdots + (r_{sp}^\top x)^2 + (r_{s,p+1}^\top x)^2 + \cdots + (r_{sq_s}^\top x)^2 \\
                & = x^\top \check{R}_s^\top \check{R}_s x = x^\top x = 1,
			\end{align*}
			where $\check{R}_s$ and $r_{s,p+1},\ldots,r_{s,q_s}$ are defined as in \eqref{eq:mu_alternative}.
			Thus we have $R_s^\top S^{q_s-1} \subset D^p$.
			
			\item Let $y \in D^p$.
			Define a $q_s$-dimensional vector $\check{y} $ by $\check{y}=(y^\top , (1-\|y\|^2)^{1/2} , 0_{q_s-p-1}^\top )^\top$.
			Let $ x = \check{R}_s \check{y}.$
			Then, noting that $\check{R}_s^\top \check{R}_s = I_p $, it is easy to see that
			$$
			\| x \|^2 = \| \check{R}_s \check{y} \|^2  = \check{y}^\top \check{R}_s^\top \check{R}_s \check{y} = \check{y}^\top \check{y} = \|y\|^2 + \sqrt{1-\|y\|^2}^2 = 1. 
			$$
			Thus $x \in S^{q_s-1}$.
			For such $x$, it holds that
			$$
			R_s^\top x = R_s^\top \check{R}_s \check{y} = R_s^\top (R_s y + \sqrt{1-\|y\|^2} r_{s,p+1} ) = R_s^\top R_s y = y.
			$$
			Therefore, for any $y \in D_p$, there exists $x \in S^{q_s-1}$ such that $y = R_s^\top x $.
			This implies $R_s^\top S^{q_s-1} \supset D^p$.
		\end{enumerate}
		
		Thus, since both (a) and (b) hold, it follows that $R_s^\top S^{q_s-1} = D^p$ for any $p<q_s$.
	\end{proof}
	Now we are ready to prove Proposition \ref{prop:mu_dim}.
	
	If $p=q_s$, it is straightforward to see that $R_s$ is an orthogonal matrix.
	Thus we have $R_s^\top S^{p-1} = S^{p-1}$.
	As discussed in the second last paragraph of Section \ref{sec:general}, the stereographic projection $\mathcal{S}$ with $p=q_s$ is a bijective mapping which maps the unit sphere $S^{p-1}$ onto $\overline{\mathbb{R}}^{p-1}$.
    Clearly, $R_e^\top \overline{\mathbb{R}}^{q_e} = \overline{\mathbb{R}}^{p-1}$.    
	Using these results, we have
	\begin{align}
    \begin{split} \label{eq:mu_transform}
		\mu ( \overline{\mathbb{R}}^{q_e} \times S^{q_s-1} ) & = B_0 \mathcal{S}^{[-1]} ( B_s \mathcal{S} (R_s^\top S^{p-1}) + B_e (R_e^\top \overline{\mathbb{R}}^{q_e} + c_e) ) \\
        & = B_0 \mathcal{S}^{[-1]} ( B_s \mathcal{S} ( S^{q_s-1} ) + B_e (\overline{\mathbb{R}}^{p-1} + c_e)  ) \\
        & = B_0 \mathcal{S}^{[-1]} ( B_s \overline{\mathbb{R}}^{p-1} + B_e \overline{\mathbb{R}}^{p-1}  ). 
        \end{split}
	\end{align}
	
	For the case $p<q_s$, it follows from Lemma 1 that $R_s^\top S^{q_s-1} = D^p$.
	The facts $\mathcal{S} (S^{p-1}) = \overline{\mathbb{R}}^{p-1} $ and $S^{p-1} \subset D^p$ imply $\mathcal{S} (D^p) = \overline{\mathbb{R}}^{p-1}$.
	Then, following the same calculations as in equation \eqref{eq:mu_transform}, we have
    $
    \mu ( \overline{\mathbb{R}}^{q_e} \times S^{q_s-1} ) =  B_0 \mathcal{S}^{[-1]} ( B_s \overline{\mathbb{R}}^{p-1} + B_e \overline{\mathbb{R}}^{p-1}  ).
    $
    	
	Thus, for $ p \geq q $, it holds that $ \mu ( \overline{\mathbb{R}}^{q_e} \times S^{q_s-1} ) =  B_0 \mathcal{S}^{[-1]} ( B_s \overline{\mathbb{R}}^{p-1} + B_e \overline{\mathbb{R}}^{p-1} )$.
    We consider the three cases of the elements of $ (B_s,B_e) $ to carry out further calculations.

	\begin{enumerate}[(i)\ ]
		\item  If $ \beta_{ep}^2+\beta_{sp}^2 > 0$, then it follows from the definition that $\beta_{ej}^2+\beta_{sj}^2 > 0 $ for any $2 \leq j \leq p$.
        In this case, $ B_s \overline{\mathbb{R}}^{p-1} + B_e \overline{\mathbb{R}}^{p-1} = \overline{\mathbb{R}}^{p-1} $.
		This implies that
		$$
		\mu ( \overline{\mathbb{R}}^{q_e} \times S^{q_s-1} ) = B_0 \mathcal{S}^{[-1]} ( \overline{\mathbb{R}}^{p-1} ) = B_0 S^{p-1} = S^{p-1}.
		$$

        \item 
        Let \( \beta_{ek}^2+\beta_{sk}^2 > 0 \) and \( \beta_{e,k+1}=  \beta_{s,k+1} = 0 \) for \( 2 \leq k \leq p-1 \).
		Then $ B_s \overline{\mathbb{R}}^{p-1} + B_e \overline{\mathbb{R}}^{p-1} = (\mathbb{R}^{k-1} \times \{ 0_{p-k} \}) \cup \{\infty\} $.
		For $t(x) = (t_2(x),\ldots,t_k(x), 0_{p-k}^\top )^\top \in (\mathbb{R}^{k-1} \times \{ 0_{p-k} \}) \cup \{\infty\} $, its inverse stereographic projection can be expressed as
		\begin{align*}
			\mathcal{S}^{[-1]}(t(x)) & = \frac{1}{1+\|t(x)\|^2} (1-\| t(x)\|^2, 2 t_2(x), \ldots, 2 t_k(x), 0 ,\ldots , 0)^\top \\
			& = \begin{pmatrix}
			    \breve{S}^{[-1]} (\breve{t}(x)) \\
                0_{p-s}
			\end{pmatrix},
		\end{align*}
		where $\breve{t}(x) = (t_2(x),\ldots, t_k(x))^\top$ and $\breve{S}^{[-1]} (\breve{t}(x))$ is the inverse stereographic projection \eqref{eq:inverse} with $p$ replaced by $k$. 
		Since $\breve{t}(x) \in \mathbb{R}^{k-1}$, $\breve{\mathcal{S}}$ maps $\mathbb{R}^{k-1}$ onto $S^{k-1} \setminus \{ -e_1 \}$.
		Also $\mathcal{S}^{[-1]}(\infty) = -e_1$.
		It follows that $\mathcal{S}^{[-1]} \left( (\mathbb{R}^{k-1} \times \{ 0_{p-k} \}) \cup \{\infty\} \right) = S^{k-1}  \times \{0_{p-k}\}$, implying that $\mathcal{S}^{[-1]}$ maps $(\mathbb{R}^{k-1} \times \{ 0_{p-k} \}) \cup \{\infty\}$ onto $S^{k-1} \times \{ 0_{p-k} \} $.
        For $ (u_1,\ldots,u_k , 0_{p-k}^\top)^\top \in S^{k-1} \times \{ 0_{p-k} \} $, it follows that $B_0 (u_1,\ldots,u_k , 0_{p-k}^\top)^\top = u_1 b_{01} + \cdots + u_k b_{0k} $.
        Therefore we have
        $$
        \mu ( \overline{\mathbb{R}}^{q_e} \times S^{q_s-1} ) =  B_0 (S^{k-1} \times \{ 0_{p-k} \}) = \tilde{S}^{k-1}. 
        $$
		
		\item If $\beta_{ep} = \beta_{sp} = 0$, it is straightforward from the expression \eqref{eq:link2} that $\mu(x)=b_{01}$.
		
		
		
		
	\end{enumerate}
\end{proof}

\subsubsection{Proof of Proposition \ref{prop:mobius_p}}

\begin{proof}[Proof.]
	First we show (i).
    It follows from the expression \eqref{eq:link_s} that $\mu_s(x_s)$ can be calculated as
	\begin{align}
		\mu_s (x_s ) = & \ \frac{B_0 }{1 + \| B_s R_{s,-1}^\top x_s / (1+r_{s1}^\top x_s) \|^2} \\
        & \times \left[ \left\{ 1 - \left\| \frac{B_s R_{s,-1}^\top x_s}{1+r_{s1}^\top x_s} \right\|^2  \right\} e_1 + 2 B_s  \frac{R_s^\top x_s - (r_{s1}^\top x_s) e_1 }{1+r_{s1}^\top x_s}  \right]  \\
        = & \  \frac{B_0}{ (1+r_{s1}^\top x_s)^2 +\| B_s R_{s,-1}^\top x_s \|^2 } \biggl[ \left\{ (1+r_{s1}^\top x_s)^2 - \| B_s R_{s,-1}^\top x_s \|^2 \right\} e_1  \nonumber \\
        & + 2 (1+r_{s1}^\top x_s) B_s \left( R_s^\top x_s - (r_{s1}^\top x_s) e_1 \right)  \biggr] \nonumber \\
		= & \ \frac{B_0}{ (1+ \tilde{x}_{s1} )^2 + \beta_s^2 \| R_{s,-1}^\top x_s \|^2 } \nonumber \\
		& \times \left[ \left\{ (1+\tilde{x}_{s1})^2 - \beta_s^2 \| R_{s,-1}^\top x_s \|^2 \right\} e_1 + 2 (1+ \tilde{x}_{s1}) \beta_s \left( \tilde{x}_s - \tilde{x}_{s1} e_1 \right)  \right], \label{eq:mu_mobius_proof}
	\end{align}
	where $\tilde{x}_{sj}$ denotes the $j$th element of $\tilde{x}_s$.
	Here $\| R_{s,-1}^\top x_s\|^2$ can be calculated as
	$$
	\| R_{s,-1}^\top x_s \|^2 =  \sum_{j=s}^p (r_j x_{sj})^2 = \sum_{j=s}^p \tilde{x}_{sj}^2 = 1 - \tilde{x}_{s1}^2, 
	$$ 
	where the last equality holds because $\|\tilde{x}_s\|^2=1$ based on the assumption $p=q_s$.
	Substituting this result into \eqref{eq:mu_mobius_proof} and collect the coefficients of $\tilde{x}_s$ and $e_1$, we have
	$$
	\mu_s(x_s) = B_0 \, \frac{1}{ 1+\beta_s^2 + (1-\beta_s^2) \tilde{x}_{s1}} \left[ 2 \beta_s \tilde{x}_s + \{ 1-\beta_s^2 + (1-\beta_s)^2 \tilde{x}_{s1} \} e_1 \right]. \label{eq:mu_mobius_proof2}
	$$
	Then it is a straightforward exercise to see that this expression is the same as the intermediate expression of \eqref{eq:mobius_s2} by substituting $\phi=(1-\beta_s)/(1+\beta_s)$ into the intermediate expression of \eqref{eq:mobius_s2} and collecting the coefficients of $\tilde{x}$ and $e_1$.

	Finally, substituting $\tilde{x}_s=R_s^\top x_s$, it follows that
	\begin{align*}
		\frac{1-\|\psi e_1 \|^2}{\|\tilde{x}_s+\psi e_1 \|^2} (\tilde{x}_s + \psi e_1) + \psi e_1  &= 
		\frac{1-\|\psi e_1 \|^2}{\|R_s^\top x_s +\psi e_1 \|^2} (R_s^\top x_s + \psi e_1) + \psi e_1  \\ 
		& = R_s^\top \left\{ \frac{1-\|\psi R_s e_1 \|^2}{\|x_s +\psi R_s e_1 \|^2}   ( x_s + \psi R_s e_1) + \psi R_s e_1 \right\} \\
		& = R_s^\top \left\{ \frac{1-\|\psi r_{s1}\|^2}{\|x_s+\psi r_{s1}\|^2} (x_s+ \psi r_{s1}) + \psi r_{s1} \right\}. 
	\end{align*}
	Thus we obtain the final expression of \eqref{eq:mobius_s2}, and this concludes the proof of (i).

    Proposition \ref{prop:mobius_p}(ii), (iii) and (iv) can be proved by substituting $\beta_s=1$, $(\beta_s,R_s)=(1,I_p)$, and $(\beta_s,B_0)=(1,R_s)$, into \eqref{eq:mobius_s2} respectively.
\end{proof}

\subsubsection{Proof of Proposition \ref{prop:mobius_2}}
\begin{proof}[Proof.]
	It follows from equation \eqref{eq:link_linear} that the link function \eqref{eq:link_s} satisfies
    \begin{equation}
    \mathcal{S} \left( B_0^\top \mu_{DM} (x_s) \right) = B_s \mathcal{S} \left( R_s^\top x_s \right). \label{eq:link_linear_s}
    \end{equation}
    Under the assumptions of the proposition, the right-hand side of \eqref{eq:link_linear_s} can be calculated as
	\begin{align*}
		B_s \mathcal{S} ( R_s^\top x_s ) & = \beta_{s2} \mathcal{S} 
		\begin{pmatrix}
			\cos (\theta_x - \eta) \\
			\delta \sin (\theta_x - \eta )
		\end{pmatrix} 
		= \frac{ \delta \beta_{s2} \sin (\theta_x- \eta ) }{1+\cos (\theta_x - \eta )}
		= \delta \beta_{s2} \tan \left( \frac{\theta_x-\eta}{2} \right),
	\end{align*}
	where the last equality follows because 
	$$
	\tan \frac{\theta}{2} = \frac{\sin (\theta/2)}{\cos (\theta/2)} = \frac{\cos (\theta/2) \sin (\theta/2)}{\cos^2 (\theta /2)} = \frac{ \sin \theta}{1+\cos \theta}.
	$$
    Similarly, the left-hand side of (\ref{eq:link_linear_s}) can be expressed as
	\begin{equation}
	\mathcal{S} \left( B_0^\top \mu_{DM} (x_s) \right) = \tan \left( \frac{ \mu_{DM}(\theta_x)-\beta_0}{2} \right). \label{eq:s_tangent}
	\end{equation}
    Summarizing these results, we have
	$$
	\tan \left( \frac{ \mu_{DM}(\theta_x)-\beta_0 }{2} \right) = \delta \beta_{s2} \tan \left( \frac{\theta_x-\eta}{2} \right). 
	$$
	The claimed expression (\ref{eq:mobius_2}) is available through straightforward calculations of this equation.
\end{proof}

\subsubsection{Proof of Proposition \ref{prop:link_FL}}
\begin{proof}[Proof.]

It follows from equation \eqref{eq:link_linear} and the assumptions of the proposition that the link function \eqref{eq:link_e} satisfies
\begin{equation}
\mathcal{S} \left( B_0^\top \mu_{FL} (x_s) \right) = \gamma^\top x_e . \label{eq:link_linear_e}
\end{equation}
Using an approach similar to the derivation of equation \eqref{eq:s_tangent}, the left-hand side of \eqref{eq:link_linear_e} can be expressed as
$$
\mathcal{S} \left( B_0^\top \mu_{FL} (x_s) \right) = \tan \left( \frac{ \mu_{FL}(\theta_x)-\beta_0}{2} \right).
$$
Substituting this result into \eqref{eq:link_linear_e} and rearranging the resulting expression in terms of $\mu_{FL}$, we obtain equation \eqref{eq:mobius_2}, as required.


\end{proof}

\subsubsection{Proof of Proposition \ref{prop:reparametrization}}
\label{sec:reparameterization_pf}
\begin{proof}[Proof of part (i).]
We start from the form of the mean link in Proposition \ref{prop:link2}.
First note that
\begin{align*}
B_{0,-1} \, t(x) & = B_{0,-1} \left\{ \frac{B_s R_{s,-1}^\top x_s}{1+r_{s1}^\top x_s} + B_e R_e^\top x_e \right\} = \frac{ \Omega \tilde{I}_s x_s}{1+r_{s1}^\top x_s} + \Omega \tilde{I}_e x_e  \\
& = \Omega \left\{ \frac{ \tilde{I}_s x_s}{1+r_{s1}^\top x_s} + \tilde{I}_e x_e \right\} = \tilde{t}(x).
\end{align*}
Then it follows easily that
\begin{align*}
\| t(x) \|^2 & = t(x)^\top t(x) = t(x)^\top I_{p-1} t(x) = t(x)^\top B_{0,-1}^\top B_{0,-1} t(x) = \left\{ B_{0,-1} t(x) \right\}^\top  B_{0,-1} t(x) \\
 &  = \tilde{t}(x)^\top \tilde{t}(x) \\
 & = \| \tilde{t}(x)\|^2.
\end{align*}
The remaining term in the numerator of \eqref{eq:link2} is 
$$
2\sum_{j=2}^p t_{j}(x) b_{0j} = 2B_{0,-1} t(x) = 2\tilde{t}(x).
$$
It follows that \eqref{eq:link2} can be expressed as \eqref{eq:link_repa}.
\end{proof}

\begin{proof}[Proof of part (ii).]
The columns of $R_{s,-1}$ and $R_{e}$ each have a norm of $1$ and thus the rows of $(B_s R_{s,-1}^\top , B_e R_{e}^\top)$ have norm given by diagonal elements of $(B_s^2 + B_e^2)^{1/2}$.
Furthermore 
\begin{align*}
& (B_s^2 + B_e^2)^{-1/2} (B_s R_{s,-1}^\top , B_e R_{e}^\top) \left\{(B_s^2 + B_e^2)^{-1/2} (B_s R_{s,-1}^\top , B_e R_{e}^\top)\right\}^\top\\
=& (B_s^2 + B_e^2)^{-1/2} (B_s R_{s,-1}^\top , B_e R_{e}^\top) (B_s R_{s,-1}^\top , B_e R_{e}^\top)^\top (B_s^2 + B_e^2)^{-1/2}\\
=& (B_s^2 + B_e^2)^{-1/2} \left(B_s R_{s,-1}^\top  R_{s,-1} B_s + B_e R_{e}^\top R_e B_e \right) (B_s^2 + B_e^2)^{-1/2}\\
=& (B_s^2 + B_e^2)^{-1/2} \left(B_s^2 + B_e^2 \right) (B_s^2 + B_e^2)^{-1/2}\\
=& I_{p-1}
\end{align*}
and thus $V = \left\{(B_s^2 + B_e^2)^{-1/2} (B_s R_{s,-1}^\top , B_e R_{e}^\top)\right\}^\top$ satisfies $V^\top V = I_{p-1}$.
Rewriting $\Omega= B_{0,-1} ( B_s R_{s,-1}^\top , B_e R_{e}^\top )$ in terms of $V$ yields 
\[
\Omega = B_{0,-1} (B_s^2 + B_e^2)^{1/2} V^\top,
\]
which is a singular value decomposition of $\Omega$ because $B_{0,-1}^\top B_{0,-1} = I_{p-1}$ and $\Sigma = (B_s^2 + B_e^2)^{1/2}$ is diagonal.
\end{proof}

\begin{proof}[Proof of part (iii).]
Suppose we have a unit $p$-vector $b_{01}$, a unit $q_s$-vector $r_{s1}$ and a $p \times (q_s + q_e)$ matrix $\Omega$, where $\Omega$ satisfies
$b_{01}^\top \Omega = 0_{q_s+q_e}^\top$, $\Omega\tilde{I}_s r_{s1} = 0_p$, and $\Omega\tilde{I}_s \tilde{I}_s^\top \Omega^\top$ and 
 $\Omega\tilde{I}_e \tilde{I}_e^\top \Omega^\top$ commute.

Because $\Omega\tilde{I}_s \tilde{I}_s^\top \Omega^\top$ and 
 $\Omega\tilde{I}_e \tilde{I}_e^\top \Omega^\top$ commute, they are simultaneously diagonalisable by some orthogonal $p \times p$ matrix $U$ \citep[Thm 1.3.12, p50]{horn1985ma}. 
Thus the columns of $U$ are eigenvectors of both $\Omega\tilde{I}_s \tilde{I}_s^\top \Omega^\top$ and $\Omega\tilde{I}_e \tilde{I}_e^\top \Omega^\top$, and furthermore the singular value decompositions of $\Omega\tilde{I}_s$ and $\Omega\tilde{I}_e$ can be written as \citep[Proof of 7.3.1, p411-412]{horn1985ma}
\[
\Omega\tilde{I}_s = U \Sigma_s V_s^\top \quad\quad \textrm{and} \quad\quad \Omega\tilde{I}_e = U \Sigma_e V_e^\top,
\]
where $\Sigma_s$ and $\Sigma_e$ are positive-semi-definite diagonal $p \times p$ matrices, $V_s^\top V_s = I_{q_s}$, and $V_e^\top V_e = I_{q_e}$.
We use $\Omega = [\Omega \tilde I_s \,\, \Omega \tilde I_e]$ to obtain the SVD of $\Omega$
\begin{align*}
\Omega 
&= \left[\Omega \tilde I_s \,\, \Omega \tilde I_e\right]\\
&= U \left[\Sigma_s V_s^\top \,\, \Sigma_e V_e^\top\right]\\
&= U (\Sigma_s^2 + \Sigma_e^2)^{1/2} (\Sigma_s^2 + \Sigma_e^2)^{1/2 \dagger} \left[\Sigma_s V_s^\top \,\, \Sigma_e V_e^\top\right],
\end{align*}
where $(\Sigma_s^2 + \Sigma_e^2)^{1/2 \dagger}$ has diagonal elements that are the reciprocal of the corresponding non-zero diagonal elements of $(\Sigma_s^2 + \Sigma_e^2)^{1/2}$ and are zero elsewhere.
Here $U$ is the matrix of left-singular vectors of $\Omega$, the diagonal elements of $(\Sigma_s^2 + \Sigma_e^2)^{1/2}$ are the singular values of $\Omega$ and the right-singular vectors of $\Omega$ are rows of $(\Sigma_s^2 + \Sigma_e^2)^{1/2\dagger} \left[\Sigma_s V_s^\top \,\, \Sigma_e V_e^\top\right]$, which are unit vectors and orthogonal to each other because $V_s^\top V_s = I_{q_s}$, and $V_e^\top V_e = I_{q_e}$.

Because $b_{01}^\top \Omega = 0_{q_s+q_e}^\top$, at least one of the singular values of $\Omega$ is zero.
Without loss of generality suppose that the corresponding column of $U$ is the $p$th column and it is parallel to $b_{01}$.
We can drop the $p$th column from $U$, $V_s$ and $V_e$, and both the $p$th row and $p$th column from $\Sigma_s$ and $\Sigma_e$.
Denote these reduced matrices by an overhead tilde.
Then 
\begin{equation}
\Omega = \tilde U (\tilde\Sigma_s^2 + \tilde\Sigma_e^2)^{1/2} (\tilde\Sigma_s^2 + \tilde\Sigma_e^2)^{1/2 \dagger} \left[\tilde\Sigma_s \tilde V_s^\top \,\, \tilde\Sigma_e \tilde V_e^\top\right].
\end{equation}

Because $\Omega\tilde{I}_s r_{s1} = 0_p$, we have that 
\[0_p = \Omega\tilde{I}_s r_{s1} = \tilde U \tilde \Sigma_s \tilde V_s^\top r_{s1}
\]
so $\tilde V_s^\top r_{s1} = 0_p$ when $\tilde \Sigma_s$ is positive definite
and thus $\left[r_{s1} \,\, \tilde V_s\right]^\top \left[r_{s1} \,\, \tilde V_s\right] = I_p$.

Consequently $B_0 = \left[b_{01} \,\, \tilde U\right]$, $B_s = \tilde\Sigma_s$, $B_e = \tilde\Sigma_e$, $R_s = \left[r_{s1} \,\, \tilde V_s\right]$ and $R_e = \tilde V_e$ satisfy all the requirements of parameters of the mean link.
If the singular values $\Omega$ are distinct, then the parameters of the mean link are unique up to the sign of the left and right singular vectors.
\end{proof}

\noindent Computational aside:
Because $\Omega = [\Omega \tilde I_s \,\, \Omega \tilde I_e]$, it follows that 
\[
\Omega \Omega^\top = \Omega \tilde I_s \tilde I_s^\top \Omega ^\top +  \Omega \tilde I_e \tilde I_e^\top \Omega ^\top 
\]
So that an orthogonal matrix $U$ diagonalises $\Omega \tilde I_s \tilde I_s^\top \Omega ^\top$ and $\Omega \tilde I_e \tilde I_e^\top \Omega ^\top$ if and only if $U$ also diagonalises $\Omega \Omega^\top$.
Thus the commutativity constraint in Proposition \ref{prop:reparametrization} is equivalent to requiring that $\Omega \tilde I_s \tilde I_s^\top \Omega ^\top$ and $\Omega \Omega^\top$ commute.
Because $b_{01}^\top \Omega = 0_{q_s+q_e}^\top$, the difference 
\[\Omega \tilde I_s \tilde I_s^\top \Omega ^\top \Omega \Omega^\top - \Omega \Omega^\top \Omega \tilde I_s \tilde I_s^\top \Omega ^\top\]
has $(p-1)(p-2)/2$ degrees for freedom.

\subsubsection{Proof of Proposition \ref{prop:beta}} \label{sec:proof_beta}

\begin{proof}[Proof.]
It immediately follows from Proposition \ref{prop:link2} that \( f(\beta) \) can be expressed as
\[
f(\beta) = \frac{1 - \beta^2 \|t(x)\|^2}{1 + \beta^2 \|t(x)\|^2}.
\]
Clearly, this function is monotonically decreasing with respect to \( \beta \).
If $\beta=0$, then $t(x)=0$ for any $x$ and therefore $f(0)=1$.
If $t(x) \neq 0$, it is clear that \( \lim_{\beta \rightarrow \infty} f(\beta) = -1 \).
If $t(x)=0$, then $f(\beta) =1$ for any $\beta \geq 0$.
Finally, if $x_s = -r_{s1}$ and $B_s  \neq O_{(p-1) \times (p-1)}$, then it follows from the last sentence of Proposition \ref{prop:link2} that $f(\beta)=-1$ for any $\beta >0$. 
\end{proof}

\subsubsection{Proof of Proposition \ref{prop:mobius_3}}
\begin{proof}[Proof.]
First, we consider an alternative expression for the link function \eqref{eq:downs} using only Euclidean variables.  
For this purpose, we identify the extended complex plane $\overline{\mathbb{C}}$ with the embedded Euclidean space $(\mathbb{R}^2 \times \{0\}) \cup \{\infty\}$ in computing the link function \eqref{eq:downs}.
Specifically, we define $\tilde{P}_D(x) = P_D ( x_c + i y_c ) $, where $x =(x_c , y_c , 0)^\top$ with $x_c, y_c \in \mathbb{R}$, and set $\tilde{P}_D (\infty) = e_3 $.
The function $\tilde{P}_D^{-1}$ is then defined as the inverse of $\tilde{P}_D$.  
Notably, within their restricted domains, these functions can be expressed as  
\[
\tilde{P}_D (x) = H(e_3) \left\{ 2 \, \frac{x+e_{3}}{\|x+e_{3}\|^2} - e_{3} \right\}, \quad x =(x_1,x_2,0)^\top , \quad x_1,x_2 \in \mathbb{R},
\]
\[
\tilde{P}_D^{-1}(x) = H(e_3) \left\{ 2 \, \frac{x-e_{3}}{\|x-e_{3}\|^2} + e_{3} \right\}, \quad x \in S^2 \setminus \{e_3\}.
\]

Similarly, let $\tilde{M}_D(x) = (\mbox{Re}[\mathcal{M}_D (x_c + i y_c)] , \mbox{Im}[\mathcal{M}_D (x_c + i y_c) ] , 0 )^\top$, where $x =(x_c,y_c,0)^\top$, $x_c ,y_c \in \mathbb{R}$.
Note that $\mathcal{M}_D(z)$ can be expressed as 
\begin{equation}
\mathcal{M}_D (z) = 
\left\{
\begin{array}{ll}
\frac{\alpha_{12}\alpha_{22}-\alpha_{11}\alpha_{22}}{\alpha_{21}^2} \frac{\overline{z+\alpha_{22}/\alpha_{21}}}{|z+\alpha_{22}/\alpha_{21} |^2} + \frac{\alpha_{11}}{\alpha_{21}}, & \alpha_{21} \neq 0, \\
\frac{\alpha_{11}}{\alpha_{22}} z + \frac{\alpha_{12}}{\alpha_{22}}, & \alpha_{21} =0,
\end{array}
\right.  \label{eq:mobius_z_alt}
\end{equation}
where $\overline{z}$ denotes the complex conjugate of $z$.  
This expression implies that $\tilde{\mathcal{M}}_D (x)$ is of the form  
\[
\tilde{\mathcal{M}}_D (x) = \left\{
    \begin{array}{ll}
    \tilde{A}_2 \left( \tilde{\gamma}_2 \, \frac{x+ \tilde{a}_2}{\|x+ \tilde{a}_2\|^2} + \tilde{b}_2 \right), & \alpha_{21} \neq 0,  \\
    \tilde{A}_2' \left( \tilde{\gamma}'_2 x + \tilde{b}'_2 \right), & \alpha_{21} = 0,
    \end{array}
    \right.
\]
where $x = (x_c, y_c , 0)^\top$ with $x_c , y_c \in \mathbb{R}$, and for $\alpha_{21} \neq 0$, $x \neq -\tilde{a}_2$.  
The parameters are given by  
$\tilde{A}_2 = B_0 (\omega_{neq}) H(e_2)$,
$\tilde{\gamma}_2 = \gamma_{neq}$,
$\tilde{a}_2=(\mbox{Re}(\alpha_{22}/\alpha_{21}),\mbox{Im}(\alpha_{22}/\alpha_{21}),0)^\top$,  
$\tilde{b}_2=\tilde{A}_2^\top  (\mbox{Re}(\alpha_{11}/\alpha_{21}),\mbox{Im}(\alpha_{11}/\alpha_{21}),0)^\top$,  
$\tilde{A}_2'=B_0(\omega_{eq})$,  
$\tilde{\gamma}_2'=|\alpha_{11}/\alpha_{22}|$,  
$\tilde{b}'_2 = \tilde{A}_2'^\top (\mbox{Re}(\alpha_{12}/\alpha_{22}),$ $\mbox{Im}(\alpha_{12}/\alpha_{22}),0)^\top$.  
In addition, assume
$\tilde{\mathcal{M}}_D (-\tilde{a}_2) = \infty$ and  
$\tilde{\mathcal{M}}_D (\infty) = \tilde{A}_2 \tilde{b}_2$ for $\alpha_{21} \neq 0$,  
while $\tilde{\mathcal{M}}_D (\infty) = \infty$ for $\alpha_{21} = 0$.  
Thus, we obtain an alternative expression for $\mu_D(x_s)$ that avoids calculations of complex numbers:
\begin{equation}  
\mu_D(x_s) = \tilde{P}_D \left( \tilde{\mathcal{M}}_D \left( \tilde{P}_D^{-1} (x_s) \right) \right), \quad x_s \in S^2.  
\end{equation}

Note that the functions $\tilde{P}_D$, $\tilde{\mathcal{M}}_D$, and $\tilde{P}^{-1}_D$ belong to the set of M\"obius transformations $\{ \mathcal{M}_E \}$ defined in \eqref{eq:mobius_com}.
The set of M\"obius transformations $\{ \mathcal{M}_E \}$ is closed under composition.
Specifically, for any parameters $(A_j,\gamma_j,a_j,b_j , \varepsilon_j)$ $(j=1,2)$ of the M\"obius transformation \eqref{eq:mobius_com}, the following holds: 
\begin{equation}
\mathcal{M}_E \left( \mathcal{M}_E (x : A_1 , \gamma_1 , a_1 , b_1 , \varepsilon_1) \, : \, A_2, \gamma_2 , a_2 , b_2 , \varepsilon_2 \right) = \mathcal{M}_E (x : A', \gamma', a' ,b' , \varepsilon') \label{eq:mobius_closure}
\end{equation}
for some $(A',\gamma',a',b',\varepsilon')$.
This implies the existence of parameters $(\breve{A} , \breve{\gamma}, \breve{a} , \breve{b},\breve{\varepsilon})$ and $(\bar{A} , \bar{\gamma} , \bar{a} ,\bar{b} , \bar{\varepsilon})$ for the M\"obius transformation \eqref{eq:mobius_com} such that  
\begin{equation}  
\tilde{\mathcal{M}}_D \left( \tilde{P}_D^{-1} (x_s) \right) = \mathcal{M}_E (x_s : \breve{A} , \breve{\gamma}, \breve{a} , \breve{b} , \breve{\varepsilon}  ) \quad \mbox{and} \quad  
\mu_D(x_s) = \mathcal{M}_E \left( x_s : \bar{A} , \bar{\gamma} , \bar{a} ,\bar{b} , \bar{\varepsilon} \right).  \label{eq:mobius_expression_downs}  
\end{equation}  
Thus, $\mu_D(x_s)$ can be expressed in the form of the M\"obius transformation $\mathcal{M}_E$.   
We now compute the specific forms of $\mu_D(x_s)$ for the two cases: $\alpha_{21} \neq 0$ and $\alpha_{21}=0$.  \vspace{0.2cm}

\noindent \textbf{The case $\alpha_{21} \neq 0:$}
In this case, the parameter $\varepsilon$ in $\mathcal{M}_E$ from \eqref{eq:mobius_com} is equal to 2 for all of $\tilde{P}_D$, $\tilde{\mathcal{M}}_D$, and $\tilde{P}_D^{-1}$.  
Then, the parameters of the M\"obius transformation on the right-hand side of \eqref{eq:mobius_closure} for $\varepsilon_1=\varepsilon_2=2$ can be computed as follows:
\begin{equation}
\begin{gathered}
A' = A_2 A_1 H( v), \quad \gamma' = \frac{\gamma_1 \gamma_2}{\|v\|^2}, \quad a'= a_1 + \frac{\gamma_1 v}{\|v\|^2}, \\
b' = H(v) \left( \frac{\gamma_2 v}{\|v\|^2} + A_1^\top b_2 \right), \quad \varepsilon'=2, \quad v= A_1^\top a_2 + b_1. \label{eq:epsilon_2}
\end{gathered}
\end{equation}
Using this result, it is a tedious but straightforward exercise to see that the parameters of $\mathcal{M}_E$ in \eqref{eq:mobius_expression_downs} for $\alpha_{21} \neq 0$ can be calculated as 
\[
\breve{A}= B_0(\omega_{neq}) H(e_2) H(e_3) H(u_1) =  - B_0(\omega_{neq}) H(e_1) H(u_1),
\]
\[
\breve{\gamma} = \frac{2\gamma_{neq}}{1+|\alpha_{22}/\alpha_{21}|^2}, \quad 
\breve{a} = -e_3 + \frac{2a}{1+|\alpha_{22}/\alpha_{21}|^2}, \quad \breve{\varepsilon} = \bar{\varepsilon} =2,
\]
\begin{align}
\breve{b} & = H(u_1) \left( 2 \gamma_{neq} \frac{u_1}{\|u_1\|^2} + H(e_3) H (e_2) B_0 (\omega_{neq}) \breve{u}_2 \right) \\
& = H(u_1) \left( 2 \gamma_{neq} \frac{u_1}{\|u_1\|^2} - H(e_1) B_0 (\omega_{neq} ) \breve{u}_2   \right),
\end{align}
\[
\bar{A} = H(e_3) \breve{A} H (v_{neq}) = B_0(\omega_{neq}) H (e_2) H (u_1) H(v_{neq}),
\]
\[
\bar{\gamma} = \frac{4 \gamma_{neq}}{\|v_{neq}\|^2 \| u_1\|^2 }, \quad 
\bar{a} = -e_3 + 2 \frac{u_1}{\|u_1\|^2} + \frac{2 \gamma_{neq}}{ \| u_1\|^2} + \frac{v_{neq}}{\|v_{neq}\|^2},
\]
\[
\bar{b} = H(v_{neq}) \left\{ \frac{2 v_{neq}}{ \| v_{neq} \|^2} - \breve{A}^\top e_3  \right\}, 
\]
where $
\breve{u}_2= (\mbox{Re}(\alpha_{11}/\alpha_{21}),\mbox{Im}(\alpha_{11}/\alpha_{21}),0)^\top$.
Also, it can be shown that 
\begin{equation}
\bar{A}=-R_{0D}, \quad \bar{\gamma}=-(1-\| \phi_D\|^2 ), \quad \bar{a}=-\bar{b} = \phi_D, \label{eq:R0D_phiD}
\end{equation}
where $R_{0D}$ and $\phi_D$ are defined as in \eqref{eq:R0D_neq} and \eqref{eq:phiD_neq}, respectively.
Substituting these values into the second equation of \eqref{eq:mobius_expression_downs}, we obtain equation \eqref{eq:downs} for $\alpha_{21} \neq 0$. \vspace{0.2cm}

\noindent \textbf{The case $\alpha_{21} = 0:$}
If $\alpha_{21} = 0$, then $\tilde{\mathcal{M}}_D$ corresponds to the M\"obius transformation $\mathcal{M}_E$ with $\varepsilon = 0$.  
In this case, it is straightforward to express the parameters in \eqref{eq:mobius_expression_downs} as  
$$  
\breve{A} = B_0(\omega_{eq}) , \quad \breve{\gamma} = 2 \left| \frac{\alpha_{11}}{\alpha_{22}} \right|, \quad  \breve{a} = -e_3, \quad \breve{\varepsilon} = \bar{\varepsilon} = 2,  
$$  
$$  
\breve{b} = \left| \frac{\alpha_{11}}{\alpha_{22}} \right| e_3 + H(e_3) B_0 (\omega_{eq})^\top \begin{pmatrix} \mbox{Re}(\alpha_{12}/\alpha_{22}) \\ \mbox{Im} (\alpha_{12}/\alpha_{22}) \\ 0 \end{pmatrix}, \quad  
\bar{A} = B_0(\omega_{eq}) H (v_{eq}) ,  
$$  
$$  
\bar{\gamma} = \frac{4 |\alpha_{11}/\alpha_{22}|}{ \|v_{eq}\|^2 } , \quad  
\bar{a} = -e_3 + 2 \left| \frac{\alpha_{11}}{\alpha_{22}} \right| \frac{v_{eq}}{\|v_{eq}\|^2}, \quad  
\bar{b} = H(v_{eq}) \left( \frac{2 v_{eq}}{\|v_{eq}\|^2} + e_3 \right).  
$$  
In addition, the parameters $(\bar{A}, \bar{\gamma}, \bar{a}, \bar{b})$ satisfy equation \eqref{eq:R0D_phiD} for $R_{0D}$ and $\phi_D$ as defined in \eqref{eq:R0D_phiD_eq}.  
Thus, $\mu_D(x_s)$ can be expressed as the M\"obius transformation on the sphere \eqref{eq:mobius_s3} for $\alpha_{21} = 0$.
\end{proof}

\subsubsection{Proof of Proposition \ref{prop:closure}}
\begin{proof}[Proof.]

\begin{enumerate}[(i)\ ]
\item 
Recalling the definition of $\mu$ given in the first line of \eqref{eq:link_s}, the left-hand side of \eqref{eq:closure1} can be computed as
\begin{align*}
	\mu \left( \mu ( x_s : B_{01}, R_1 , B_1) : B_{02} , B_{01} , B_2 \right) & =  B_{02} \mathcal{S}^{[-1]} \big( B_2 \mathcal{S} \big[ B_{01}^\top  B_{01} \mathcal{S}^{[-1]} \{ B_1 \mathcal{S} (R_1^\top x_s ) \}  \big] \big) \\
	&   =  B_{02} \mathcal{S}^{[-1]} \big( B_1 B_2 \mathcal{S} (R_1^\top x_s ) \big) \\
	& = \mu (x_s : B_{02} , B_1 B_2 , R_1),
\end{align*}
where the second and third equalities hold because $\mathcal{S} (\mathcal{S}^{[-1]} (x_s)) = \mathcal{S}^{[-1]} (\mathcal{S}(x_s)) =  x_s$.

\item Proposition \ref{prop:mobius_p}(i) implies that  
\begin{equation}
\mu(x_s : B_{0j} , R_j , \beta_j I_{p-1} ) = \mathcal{M}_S (x_s : B_{0j} R_j^\top , \psi_j ) , \quad j=1,2. \label{eq:closure2_proof}
\end{equation}
Using this result, we obtain  
\begin{align*}
\lefteqn{ \mu \left( \mu ( x_s : B_{01}, R_1 ,  \beta_1 I_{p-1}) : B_{02} , R_2 , \beta_2 I_{p-1} \right) } \hspace{1cm} \\
 & = \mathcal{M}_S \left( \mathcal{M}_S (x_s : B_{01} R_1^\top , \psi_1 ) : B_{02} R_2^\top , \psi_2 \right).
\end{align*}
Then, by the closure property of the M\"obius transformation stated in Lemma 2.1 of \citet{Kato2020}, we have  
$$
\mathcal{M}_S \left( \mathcal{M}_S (x_s : B_{01} R_1^\top , \psi_1 ) : B_{02} R_2^\top , \psi_2 \right) = \mathcal{M}_S ( x_s : \tilde{R} , \tilde{\psi}  ).
$$
Note that although $\mathcal{M}_S$ differs in form from the function given in equation (2.3) of \citet{Kato2020} on a general domain, they have exactly the same form when restricted to the sphere.

Finally, it follows from Proposition \ref{prop:mobius_p}(i) that  
$$
\mathcal{M}_S ( x_s : \tilde{R} , \tilde{\psi}  ) = \mu (x_s : B_0' , R' , \beta' I_{p-1} ),
$$
since $\tilde{R} = B_0' R'^\top$, $|\tilde{\psi}| = (1-\beta')/(1+\beta')$, and $\tilde{\psi}/|\tilde{\psi}|$ corresponds to the first column of $R'$.  
The assumption $\det (R')=\det(\tilde{R})$ ensures that $B_0'$ is a rotation matrix as required.
\end{enumerate}

\end{proof}

\section{Additional plots of the link functions} \label{sec:plot}

Additional plots of the proposed link function \eqref{eq:link} are given to illustrate the shapes of the link functions and to discuss the interpretation of their parameters.
	\begin{figure}
		\hspace{0.8cm} \begin{tabular}{cccc} 
			\centering 
			
            \hspace{-3cm} (a) & \hspace{-3cm} (b) & \hspace{-3cm} (c) & \hspace{-3cm} (d) \vspace{0cm}\\
            \includegraphics[width=3.3cm,height=3.1cm,trim=0.3cm 0.3cm 0.3cm 0cm]{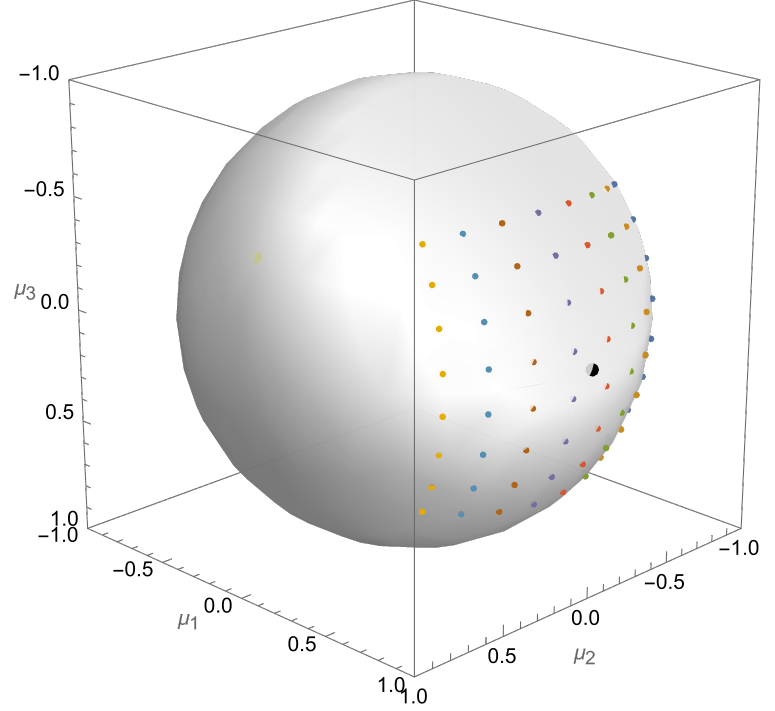} &
			\includegraphics[width=3.3cm,height=3.1cm,trim=0.3cm 0.3cm 0.3cm 0cm]{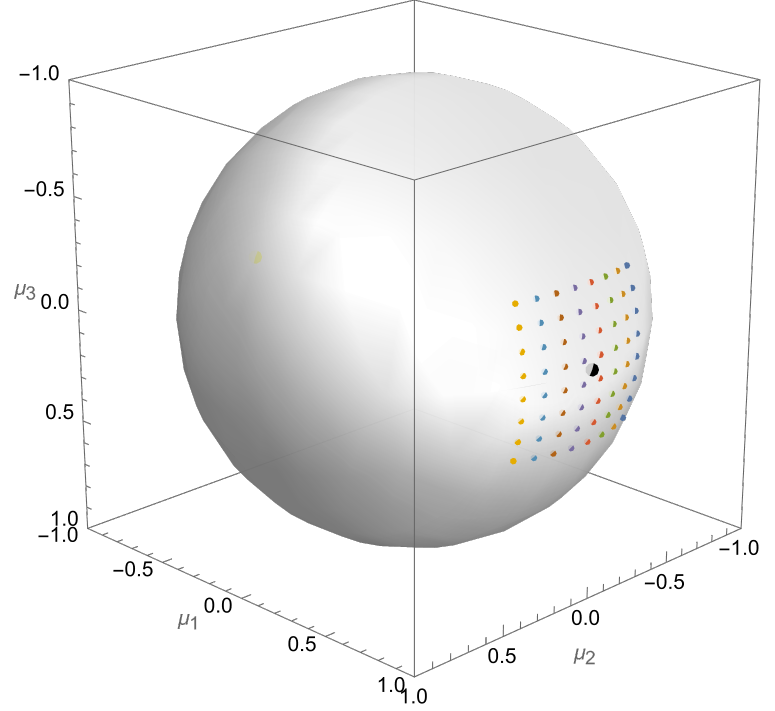} &
			\includegraphics[width=3.3cm,height=3.1cm,trim=0.3cm 0.3cm 0.3cm 0cm]{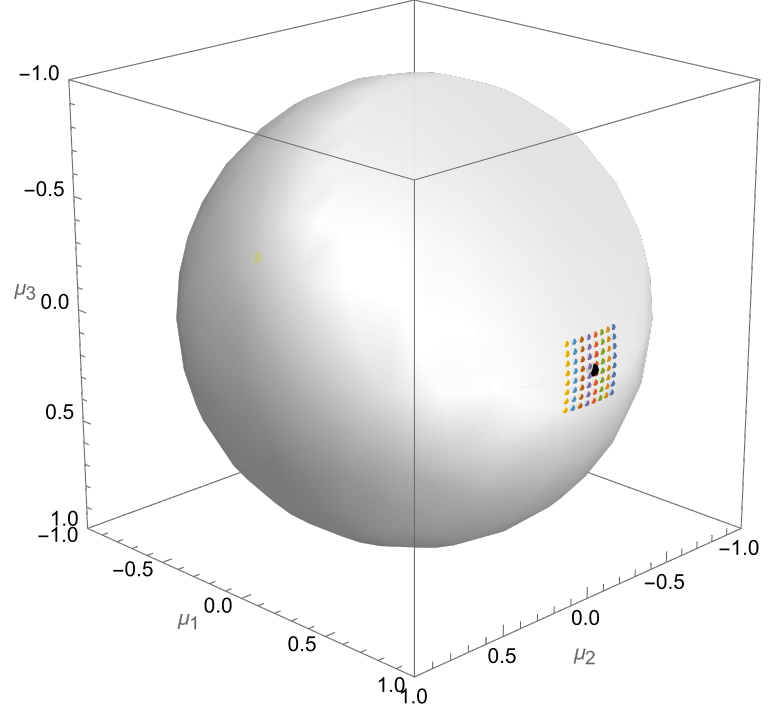} &
            \includegraphics[width=3.3cm,height=3.1cm,trim=0.3cm 0.3cm 0.3cm 0cm]{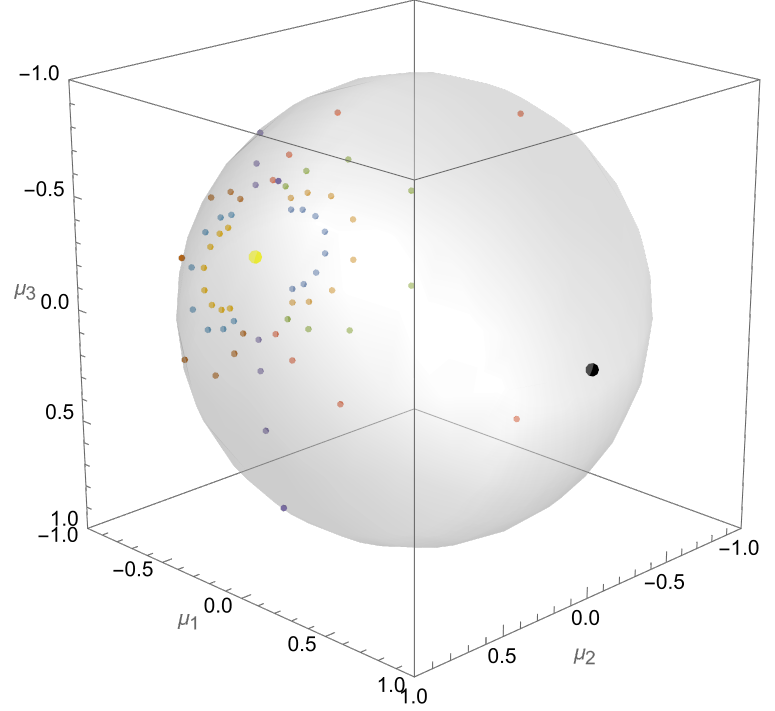} \\

            \hspace{-3cm} (e) & \hspace{-3cm} (f) & \hspace{-3cm} (g) & \hspace{-3cm} (h) \vspace{0cm}\\
            	\includegraphics[width=3.3cm,height=3.1cm,trim=0.3cm 0.3cm 0.3cm 0cm]{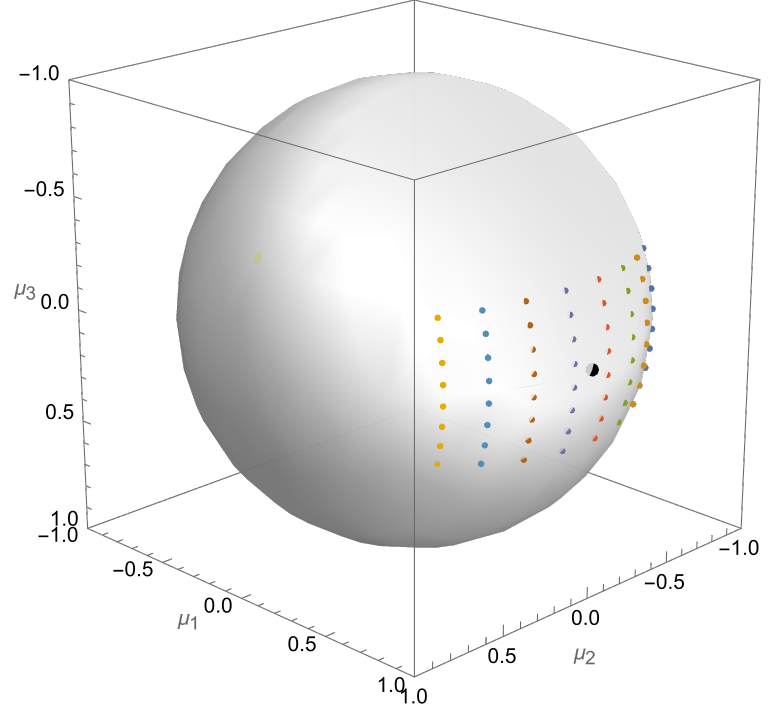} &
			\includegraphics[width=3.3cm,height=3.1cm,trim=0.3cm 0.3cm 0.3cm 0cm]{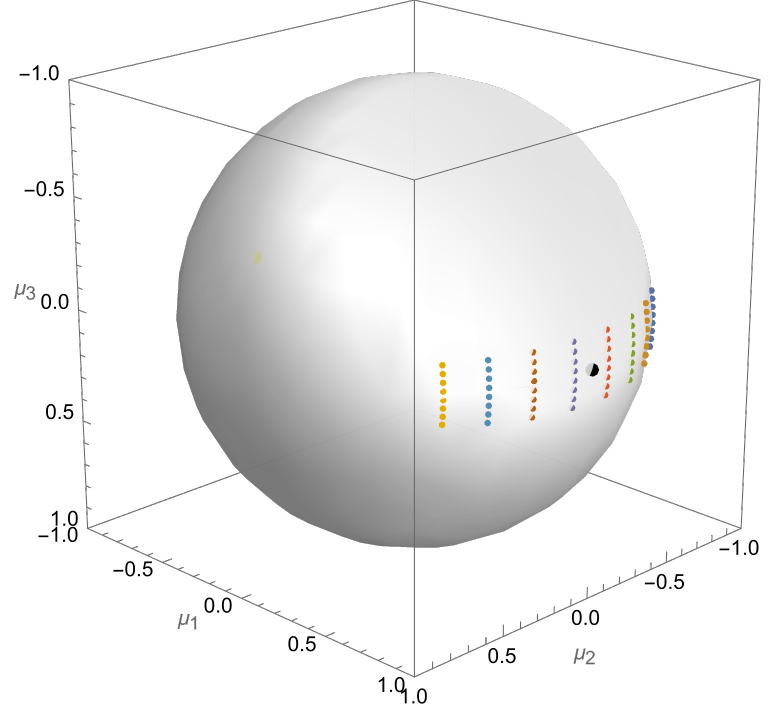} &
			\includegraphics[width=3.3cm,height=3.1cm,trim=0.3cm 0.3cm 0.3cm 0cm]{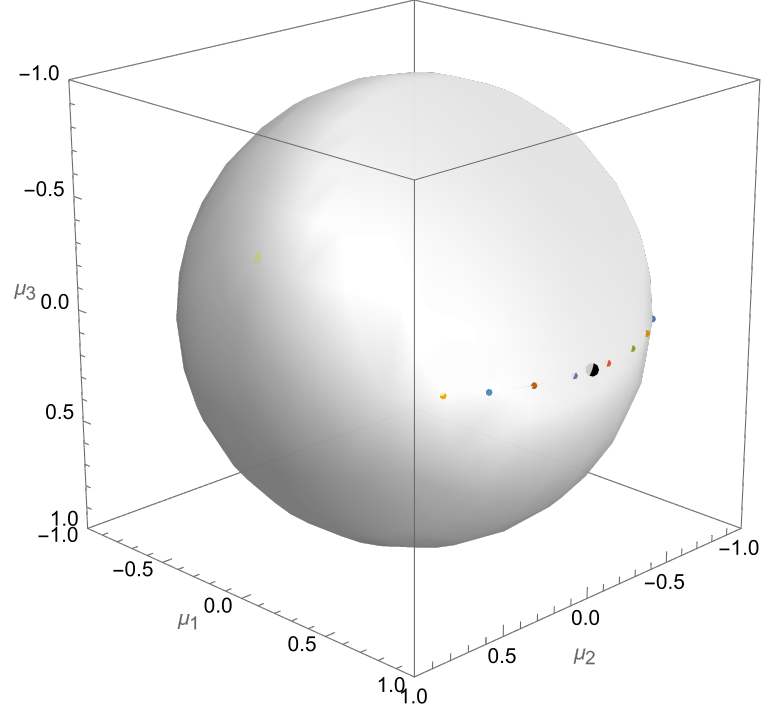} &
            \includegraphics[width=3.3cm,height=3.1cm,trim=0.3cm 0.3cm 0.3cm 0cm]{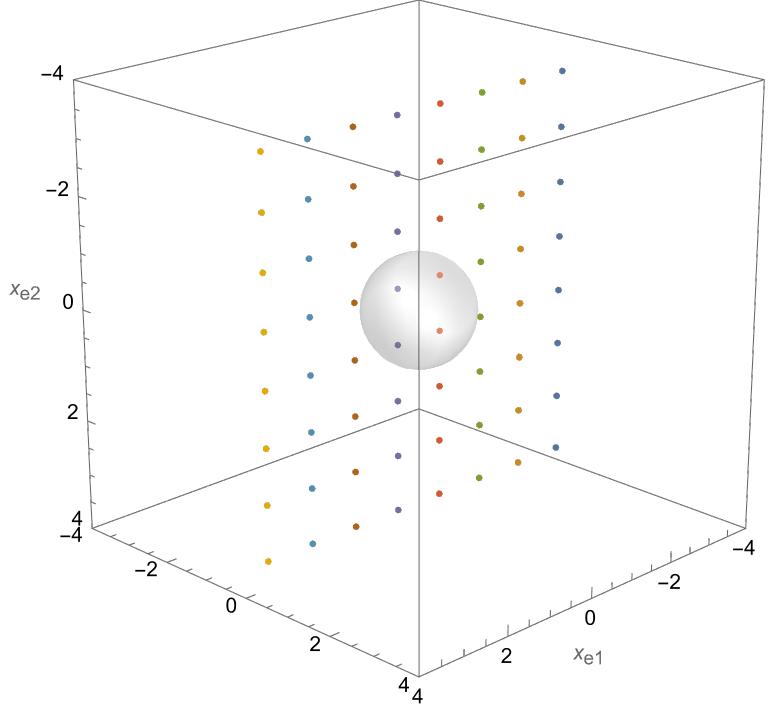} \\
			
			\end{tabular}
		
		\caption[]{
			Plots of the link function \eqref{eq:link_e} for $x_e=(i-4.5,j-4.5)^\top$, $i,j=1,\ldots,8$ with $(p,q_e)=(3,2)$, $B_0=I_3$, $R_e=I_2$ and: (a) $(\beta_{e2},\beta_{e3})=(0.1,0.1)$, (b) $(\beta_{e2},\beta_{e3})=(0.05,0.05)$, (c) $(\beta_2,\beta_3)=(0.02,0.02)$, (d) $(\beta_{e2},\beta_{e3})=(1.5,1.5)$, (e) $(\beta_{e2},\beta_{e3})=(0.1,0.1 \times 0.5)$, (f) $(\beta_{e2},\beta_{e3})=(0.1,0.1 \times 0.02)$, and (g) $(\beta_{e2},\beta_{e3})=(0.1,0.1 \times 0.001)$.
		The black and yellow dots represent the reference direction $b_{01}=e_1$ and its opposite $-b_{01}=-e_1$, respectively.
        The frame (h) exhibits the plot of $\{x_e \}$ in the two-dimensional place embedded in $\mathbb{R}^3$.
		} \label{fig:link_d3e}
\end{figure}
Figure \ref{fig:link_d3e} provides plots of the link function \eqref{eq:link_e} with $p-1=q_e=2$.  
The original linear covariates $\{x_e\}$ are displayed in Figure \ref{fig:link_d3e}(h), while the transformed points are shown in the other frames of the figure.  
Frames \ref{fig:link_d3e}(a)--(d) suggest that as $\beta_{e2}(=\beta_{e3})$ tends to $0$ or $\infty$, the $x_e$'s are isotropically attracted toward the reference direction $b_{01}(=e_1)$ and its opposite direction $-b_{01}(=-e_1)$, respectively.  
Figures \ref{fig:link_d3e}(e)--(g) indicate that the smaller the value of $\beta_{e3}$, the smaller the dispersion of $\mu_e(x_e)$ with respect to $b_{03}$.


	\begin{figure}
    \hspace{0.8cm}
		\begin{tabular}{cccc} 
			\centering 
			
            \hspace{-3cm} (a) & \hspace{-3cm} (b) & \hspace{-3cm} (c) &
            \hspace{-3cm} (d)
            \vspace{0cm}\\
            \includegraphics[width=3.3cm,height=3.3cm,trim=0cm 0cm 2.1cm 0cm,clip]{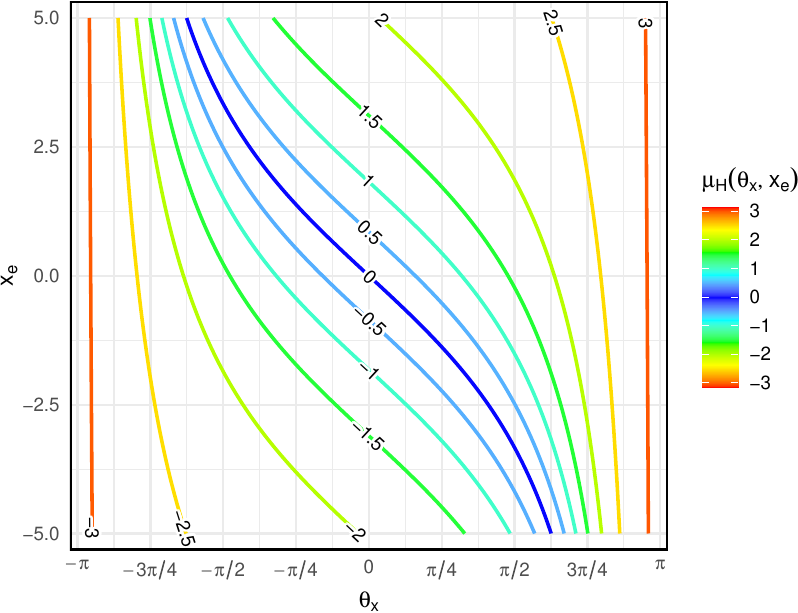} &
			\includegraphics[width=3.3cm,height=3.3cm,trim=0cm 0cm 2.1cm 0cm,clip]{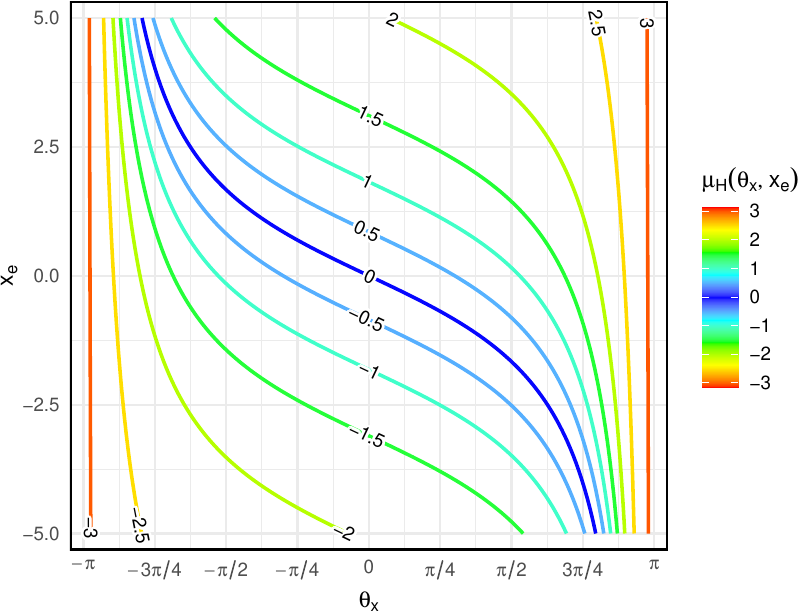} &
			\includegraphics[width=3.3cm,height=3.3cm,trim=0cm 0cm 2.1cm 0cm,clip]{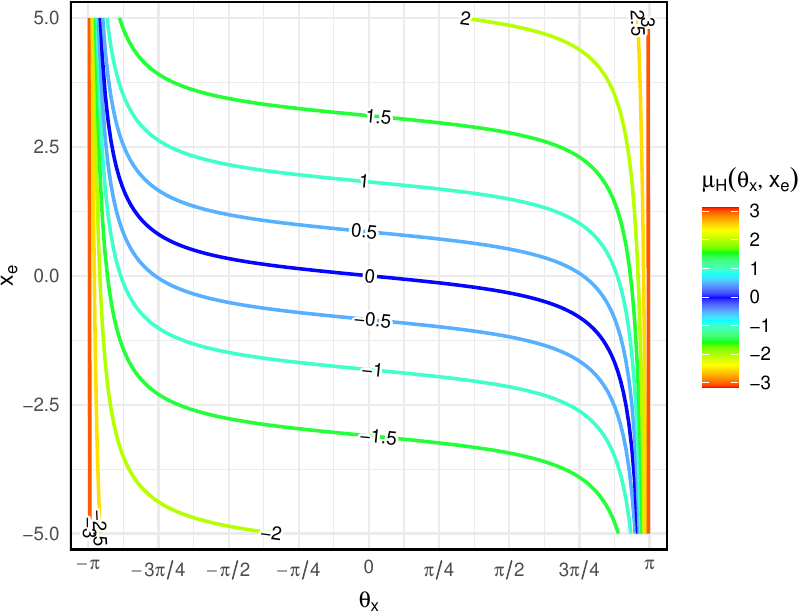} &
            \includegraphics[width=3.3cm,height=3.3cm,trim=0cm 0cm 2.1cm 0cm,clip]{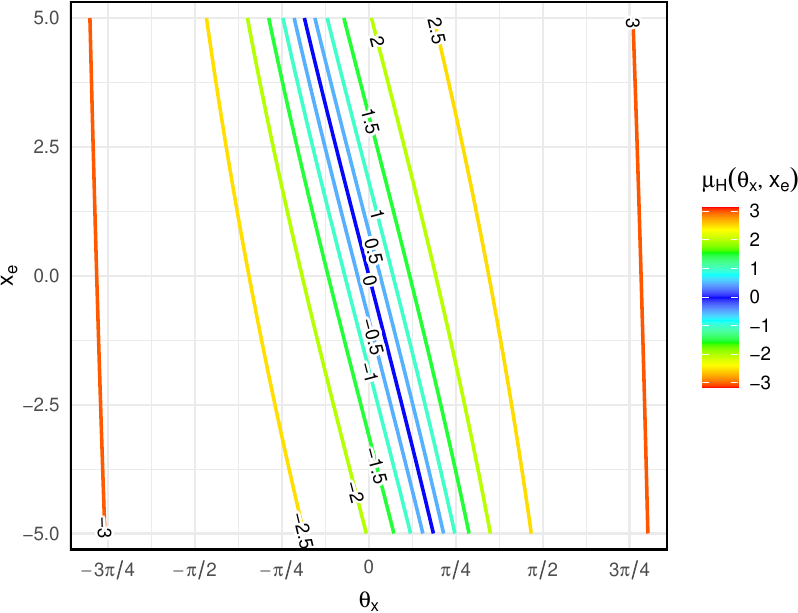}\\

            \hspace{-3cm} (e) & \hspace{-3cm} (f) & \hspace{-3cm} (g) & \vspace{0cm}\\
			\includegraphics[width=3.3cm,height=3.3cm,trim=0cm 0cm 2.1cm 0cm,clip]{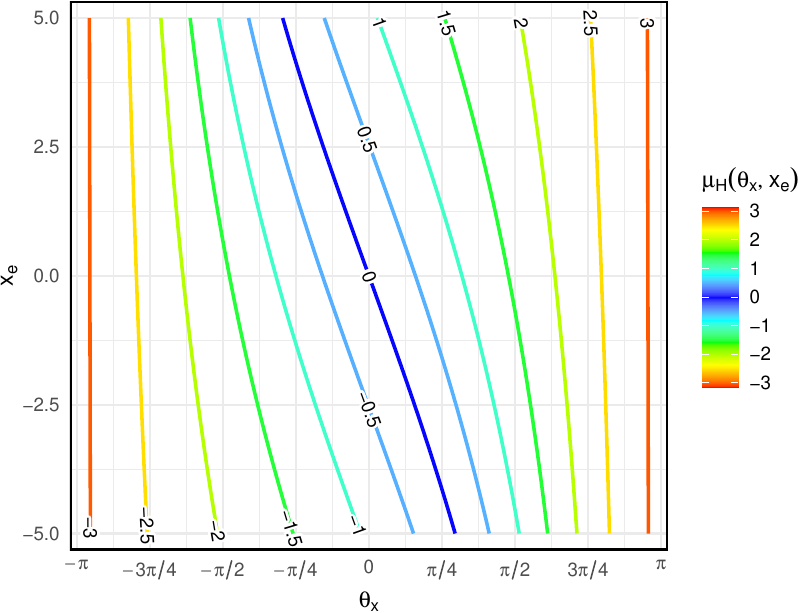} &
			\includegraphics[width=3.3cm,height=3.3cm,trim=0cm 0cm 2.1cm 0cm,clip]{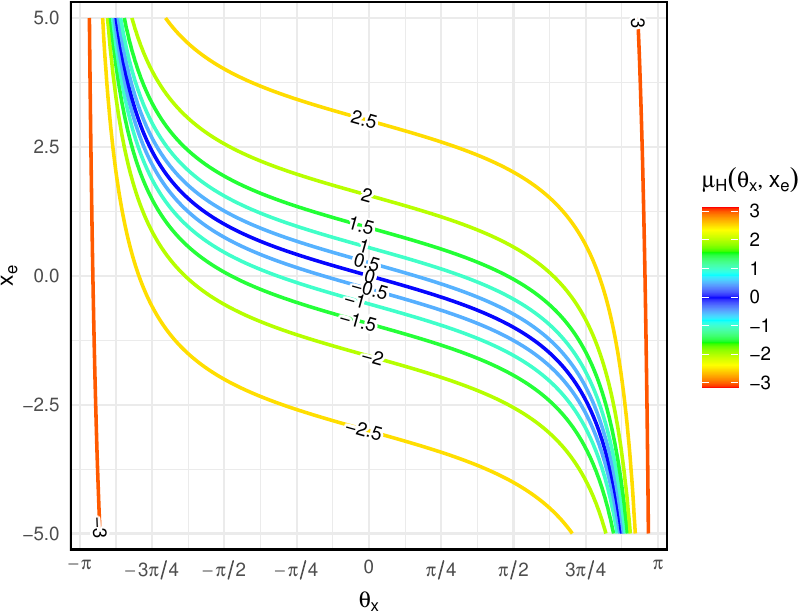} &
			\includegraphics[width=3.3cm,height=3.3cm,trim=0cm 0cm 2.1cm 0cm,clip]{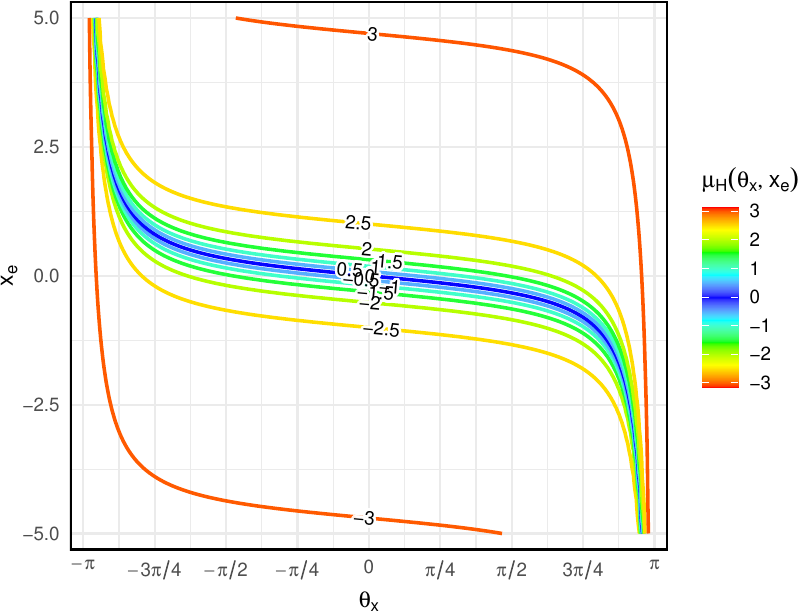} &
			
			\end{tabular}
		
		\caption[]{
			Plots of the link function \eqref{eq:link_h} for $q_e=1$, $\beta_0=\eta=0$, $\delta=1$,  and:
            (a) $(\beta_{s2},\gamma) = (1,0.3)$,
            (b) $(\beta_{s2},\gamma) = (0.5,0.3)$,
            (c) $(\beta_{s2},\gamma) = (0.1,0.3)$,
            (d) $(\beta_{s2},\gamma) = (5,0.3)$,
            (e) $(\beta_{s2},\gamma) = (1,0.1)$,
            (f) $(\beta_{s2},\gamma) = (1,1)$,
            (g) $(\beta_{s2},\gamma) = (1,3)$.
		} \label{fig:link_h}
\end{figure}

Figure \ref{fig:link_h} illustrates the link function \eqref{eq:link_h} with seven selected combinations of $(\beta_{s2}, \gamma)$.  
Frame (a) suggests that for $\beta_{s2}=1$ and $\gamma=3$, $\mu_{H}(\theta_x, x_e)$ is widely dispersed with respect to both $\theta_x$ and $x_e$.  
Frames (b) and (c) imply that as the value of $\beta_{s2}$ decreases, $\mu_{H}(\theta_x, x_e)$ becomes more concentrated around $0$ $(=\beta_0)$.  
In contrast, if $\beta_{s2}$ exceeds one, the points $\mu_{H}(\theta_x, x_e)$ are attracted toward $\pi$ $(=\beta_0+\pi)$; see frame (d).  
Frames (e)--(g) suggest that as $\gamma$ decreases (increases), the concentration of $\mu_{H}(\theta_x, x_e)$ around $0$ $(\pi)$ increases.  
In all frames, it can be observed that $(\theta_x,x_e)=(0,0)$ implies $\mu_H(\theta_x,x_e) = 0$.  
Moreover, when $\theta_x \simeq \pi$, it follows that $\mu_H(\theta_x,x_e) \simeq \pi$ regardless of the parameter values.

\section{Miscellaneous further results}

\subsection{Fisher information}

Finally, we note that that the Fisher information is block diagonal with two blocks.  One block corresponds to orientation parameters and the other block corresponds to concentration and shape parameters.  More specifically, suppose that we have a local parametrization $\theta$ so that $\mu=\mu(x;\theta), \,\gamma_2=\gamma_2(x;\theta), \ldots , \gamma_p=\gamma_p(x;\theta)$ are functions of the covariate vector $x$ and unconstrained parameter vector $\theta$.  Suppose also that $\kappa=\kappa(x;\phi), \, \lambda_1=\lambda_1(x;\phi), \ldots , \lambda_p=\lambda_p(x;\phi)$, where $\phi$ is unconstrained.  Then, if the population distribution  is of the form \eqref{class}, under very mild conditions,  the joint Fisher information $I(\theta, \phi)$ for $\theta$ and $\phi$ is block diagonal of the form
\[
I(\theta, \phi)=\begin{pmatrix}
    I(\theta) & 0\\
    0^\top & I(\phi)
\end{pmatrix},
\]
where $O$ is the matrix of zeros of dimension $\textrm{dim}(\theta) \times \textrm{dim}(\phi)$.
This follows directly from results in \citet{Rivest1984a} because Rivest's (1984) symmetry condition is satisfied by the distributions in any family of the form \eqref{class}.

\subsection{Matrix representation of parallel transport on the sphere}
\label{sec:parallel_tansport_form}

\citet[Example 8.1.1]{absil2008op} give the parallel transport along a geodesic of the sphere from $x(0)$ to $x(t)$ as
\begin{equation}
\xi(t) = -x(0) \sin(\|\dot{x}(0)\|t) u^\top \xi(0) + u \cos(\|\dot{x}(0)\| t) u^\top \xi(0) + (I_p - u u^\top) \xi(0),
\label{eq:par_transport_AMS}
\end{equation}
where $t$ parameterises the geodesic curve, $\xi(0)$ is a tangent vector to $x(0)$ represented in $\mathbb{R}^p$ space (i.e.~$\xi(0)^\top x(0) = 0$), $u = \frac{\dot{x}(0)}{\|\dot{x}(0)\|}$ is the direction of the geodesic at $x(0)$ as a unit vector, and $\xi(t)$ is the tangent vector after transport along the curve.
Note that the original formula by \citet{absil2008op} incorrectly had the cosine term as $u \cos(\|\dot{x}(0)\| t) x(0)^\top \xi(0)$.

Given the start $a = x(0)$ and end $b = x(t) \neq -a$ of the geodesic and unit speed $\|\dot{x}(0)\|=1$, the expression for geodesics on the sphere \citet[Example 5.4.1]{absil2008op} yields
\[
u = \frac{b - \cos(t) a}{\sin(t)},
\]
where $\cos(t) = b^\top a$.

We substitute this into \eqref{eq:par_transport_AMS} to obtain:
\begin{align*}
\xi(t) &= -x(0) \sin(\|\dot{x}(0)\|t) u^\top \xi(0) + u \cos(\|\dot{x}(0)\| t) u^\top \xi(0) + (I_p - u u^\top) \xi(0)\\
&= \left\{I_p + \sin(\|\dot{x}(0)\|t) (-x(0) u^\top) + \cos(\|\dot{x}(0)\| t) uu^\top - u u^\top \right\} \xi(0)\\
&= \left\{I_p + \sin(t) (-a u^\top) + (\cos(t) - 1) u u^\top \right\} \xi(0)\\
&= \left\{I_p + \sin(t) \left(-a \frac{(b - \cos(t) a)^\top}{\sin(t)}\right) + (\cos(t) - 1) \frac{(b - \cos(t) a)(b - \cos(t) a)^\top}{\sin^2(t)}\right\} \xi(0)\\
&= \left\{I_p - a (b - \cos(t) a)^\top + (\cos(t) - 1) \frac{(b - \cos(t) a)(b - \cos(t) a)^\top}{(1-\cos(t))(1+\cos(t))}\right\} \xi(0)\\
&= \left\{I_p - a (b - \cos(t) a)^\top - \frac{(b - \cos(t) a)(b - \cos(t) a)^\top}{1+\cos(t)}\right\} \xi(0)\\
&= \left\{I_p - (a b^\top - \cos(t) aa^\top) - \frac{bb^\top + \cos^2(t)aa^\top - \cos(t) ab^\top - \cos(t) b a^\top}{1+\cos(t)}\right\} \xi(0)\\
%
%
&= \left\{I_p - a b^\top - \frac{bb^\top - \cos(t) ab^\top}{1+\cos(t)} + \cos(t) aa^\top - \frac{\cos^2(t)aa^\top - \cos(t) b a^\top}{1+\cos(t)}\right\} \xi(0)\\
&= \left\{I_p - a b^\top - \frac{bb^\top - \cos(t) ab^\top}{1+\cos(t)}\right\} \xi(0) + 0\\
&= \left\{I_p - \frac{(1+\cos(t)) ab^\top + bb^\top - \cos(t) ab^\top}{1+\cos(t)}\right\} \xi(0)\\
&= \left\{I_p - \frac{ab^\top + bb^\top}{1+\cos(t)}\right\} \xi(0)\\
&= \left\{I_p - \frac{ab^\top + bb^\top + ba^\top + aa^\top}{1+\cos(t)}\right\} \xi(0)\\
&= \left\{I_p - \frac{(a+b)(a+b)^\top}{1+\cos(t)}\right\} \xi(0)\\
&= \left\{I_p - \frac{(a+b)(a+b)^\top}{1+b^\top a}\right\} \xi(0)
\end{align*}
where we have used that $a\top \xi(0) = 0$.
Thus parallel transport of vectors from the tangent space of $a$ to tangent space of $b$ can be written as the matrix 
\[I_p - \frac{(a+b)(a+b)^\top}{1+b^\top a},\]
which is the negative of the rotation matrix used for rotated residuals by \citet{Jupp1988}.

Using $a^\top \xi(0) = 0$ we can also write the third line of the above derivation as
\begin{align*}
&\left\{I_p + \sin(t) (-a u^\top) + (\cos(t) - 1) u u^\top \right\} \xi(0)\\
=&\left\{I_p + \sin(t) (u a^\top -a u^\top) + (\cos(t) - 1)(a a^\top + u u^\top) \right\} \xi(0)\\
= & Q^\top \xi(0)
\end{align*}
where $Q$ is the rotation matrix from  \citet[Lemma 2]{Amaral2007}, which has the additional property that $Q^\top a = b$.

\subsection{Second moment of rotated residuals}
\label{sec:rotresid_2ndmo}

Let $\Gamma_0$ be the matrix with columns $\gamma_{01}, \gamma_{02}, \ldots, \gamma_{0p}$ and let $\Gamma_{0,-1}$ omit the first column of $\Gamma_0$.
Given a mean direction $\mu$,
the parallel transport of the orientation axes 
 to $\mu$ is given by the columns of $R_{\gamma_{01},\mu}\Gamma_{0,-1}$ (i.e.~$R_{\gamma_{01},\mu}\gamma_{02}, \ldots, R_{\gamma_{01},\mu}\gamma_{0p}$).
Let $y_\mu$ be a random vector with mean $\mu$ distributed according to the scaled von-Mises Fisher distribution with orientation axes given by the columns of  $R_{\gamma_{01},\mu}\Gamma_{0,-1}$.
It follows from \citep[Proposition 3]{Scealy2019} that 
\[
\mathbb{E}\left[(P_{\mu} y_\mu) (P_{\mu} y_\mu)^\top \right] = R_{\gamma_{01},\mu}\Gamma_{0,-1} D \Gamma_{0,-1}^\top R_{\gamma_{01},\mu}^\top
\]
where $P_\mu = I_p - \mu \mu^\top$ projects $y$ to the tangent space at $\mu$ and $D$ is a $p \times p$ diagonal matrix with eigenvalues satisfying $0=\lambda_1$, $\lambda_2 \geq \ldots \geq \lambda_p > 0$.
Thus
\[
\mathbb{E}\left[(R_{\gamma_{01},\mu}^\top P_{\mu} y_\mu) (R_{\gamma_{01},\mu}^\top P_{\mu} y_\mu)^\top \right] = \Gamma_{0,-1} D \Gamma_{0,-1}^\top = \Gamma_{0} D \Gamma_{0}^\top.
\]
Here the right hand side does not depend on $\mu$ and thus $\Gamma_{0} D \Gamma_{0}^\top$ is the second moment of the rotated residuals $R_{\gamma_{01},\mu}^\top P_{\mu} y_\mu$ across different means $\mu$.

Furthermore, we can standardise the rotated residuals to have isotropic second moment using the psuodo-inverse $(D^{\frac{1}{2}})^\dagger$ of $D^{\frac{1}{2}}$
\[
\mathbb{E}\left[\left((D^{\frac{1}{2}})^\dagger\Gamma_{0} R_{\gamma_{01},\mu}^\top P_{\mu} y_\mu\right) \left((D^{\frac{1}{2}})^\dagger\Gamma_{0}^\top R_{\gamma_{01},\mu}^\top P_{\mu} y_\mu\right)^\top \right] = \tilde I_p,
\]
where $\tilde I_p$ is a diagonal matrix with $0$ in the first diagonal element and $1$ elsewhere.

\subsection{Variations on the proposed link function}
The Euclidean part of the proposed link function \eqref{eq:link} was originally derived so that the special case \eqref{eq:link_e}, involving only linear covariates, becomes an extension of the model proposed by \citet{Fisher1992}.  
Apart from this, other approaches exist for defining link functions involving a spherical response and linear covariates.  
In this section, we present a couple of such link functions.

The first link function is defined by 
\begin{equation}
 \mu_W (x) = \frac{\tilde{\mu}(x)}{\| \tilde{\mu}(x) \|}, \quad x \in \mathbb{R}^{q_e}.
 \label{eq:mu_wood}
\end{equation}
Here
\begin{equation}
\tilde{\mu}(x) = B_0 \tilde{\mathcal{S}}^{-1} \left( \tilde{B}_e \tilde{\mathcal{S}} ( \tilde{R}_e^\top x) \right), \label{eq:mu_tilde}
\end{equation}
where $B_0$ is defined as in Definition~\ref{def:link}, 
$\tilde{R}_e$ is a $q_e \times p$ matrix satisfying $\tilde{R}_e^\top \tilde{R}_e = I_p$, 
$\tilde{B}_e = \mathrm{diag}(\beta_{e1}, \ldots, \beta_{ep})$ with $ \beta_{ei} \geq 0$ $(i=1,\ldots,p)$, 
and $\tilde{\mathcal{S}}(x)$ and $\tilde{\mathcal{S}}^{-1}(x)$ are defined in \eqref{eq:s_star} and \eqref{eq:s_star_inv}, respectively.
Note that $\tilde{\mathcal{S}}$ and $\tilde{\mathcal{S}}^{-1}$ are the extended stereographic projection and its inverse, respectively.
If $\tilde{B}_e = \beta I_p$, then the transformation $\tilde{\mu}(x)$ consists solely of elements from the M\"obius group defined in \eqref{eq:mobius_com}.  
Since the M\"obius group \eqref{eq:mobius_com} is closed under composition, $\tilde{\mu}(x)$ can also be expressed in the form of \eqref{eq:mobius_com} by appropriately substituting its parameters.  
This implies that $\tilde{\mu}(x)$ is a conformal mapping, which preserves the angles between curves before and after the transformation.  
The resulting link function is then obtained by normalizing $\tilde{\mu}(x)$, in a manner somewhat analogous to the approach of \citet{Rosenthal2014}.

The second link function we consider is
\begin{equation}
\mu_H(x) = B_0 \mathcal{S}^{[-1]}\left( \frac{ B_e \tilde{R}_{e,-1}^\top x }{\tilde{r}_{e1}^\top x + c_e}  \right), \quad x \in  \mathbb{R}^{q_e}, \label{eq:mu_hingee}
\end{equation}
where $(\tilde{r}_{e1}, \tilde{R}_{e,-1}) = \tilde{R}_e$, $c_e \in \mathbb{R}$, and $\mathcal{S}^{[-1]}$, $B_0$, and $B_e$ are defined as in Definition~\ref{def:link}, and $\tilde{R}_e$ is defined in \eqref{eq:mu_tilde}. 
This link function has a form somewhat similar to the link function \eqref{eq:link_s} with spherical covariance only.  
The constant $c_e$ determines the origin of $x_e$, which is mapped to the reference direction $b_{01}$, and it also works as a scale parameter for $x_e$.
Note that this link function can also be expressed as
\begin{equation}
\mu_H(x) = B_0 \mathcal{S}^{[-1]} \left( B_e \mathcal{S} \left(  \tilde{R}_e^\top \frac{x}{c_1} \right)  \right), \label{eq:mu_hingee2}
\end{equation}
implying that $\mu_H(x)$ is similar in form to the link function \eqref{eq:link_s} with only a spherical covariate.
A difference is that $\mu_H(x)$ includes the scale parameter $c_e$, which is commonly introduced for linear variables.
The expression \eqref{eq:mu_hingee2} implies that $\mu_H(x)$ has the alternative representation \eqref{eq:link2} in which $t(x)$ is replaced by 
$$
\tilde{t}(x) = \frac{ B_e \tilde{R}_{e,-1}^\top x}{\tilde{r}_{e1}^\top x + c_e}.
$$

It is straightforward to incorporate linear variables into the two link functions $\mu_W(x)$ and $\mu_H(x)$ by adding $B_s \mathcal{S} (R_s^\top x_s)$ inside the arguments of $\tilde{\mathcal{S}}^{-1}$ and $\mathcal{S}^{-1}$, respectively.

Although the three link functions $\mu_e(x)$ defined in \eqref{eq:link_e}, $\mu_W(x)$, and $\mu_H(x)$ all reduce the dimension of $x$ from $q_e$ to $p-1$, they achieve this reduction in different ways.  
The function $\mu_e(x)$ reduces the dimension from $q_e$ to $p-1$ solely through the multiplication $R_e^\top x$.  
In contrast, both $\mu_W(x)$ and $\mu_H(x)$ first reduce the dimension from $q_e$ to $p$ via the multiplication $\tilde{R}_e^\top x$.  
Then, $\mu_W(x)$ performs an additional reduction to $p-1$ dimensions using the projection $\tilde{\mu}(x)/\|\tilde{\mu}(x)\|$, while $\mu_H(x)$ achieves the final reduction through the stereographic projection $\mathcal{S}$.

In this paper, we mainly focus on the link function \eqref{eq:link_e} due to its relationship with the link function of \citet{Fisher1992}.  
However, it would be of interest to investigate the other two link functions, $\mu_W(x)$ and $\mu_H(x)$, which have different forms and interpretations.

\subsection{The Rosenthal et al. (2014) model}

As mentioned in the Introduction, \citet{Rosenthal2014} defined a link function for spherical covariates as
\begin{equation}
\mu_R (x_s) = \frac{A x_s}{ \| A x_s \|}, \quad x_s \in S^{q_s-1}, \label{eq:rosenthal}
\end{equation}
where $A$ is a $q_s \times q_s$ non-singular matrix with determinant one.
Unlike our link function, theirs does not involve linear covariates.
Moreover, even in the special case \eqref{eq:link_s} of our link function with only a spherical covariate, their formulation requires that the dimensions of the response and covariates be the same, i.e., $p = q_s$.
Their link function has tractable features in terms of group actions.
Its interpretation is quite different from that of our special case \eqref{eq:link_s}.
The link function \eqref{eq:rosenthal} has the symmetric property $\mu(x_s) = - \mu(-x_s)$, implying that the two points $x_s$ and $-x_s$ are attracted equally in opposite directions.
In contrast, our link function does not exhibit this symmetry; instead, the points are attracted toward a single direction (see Figure~\ref{fig:link_d3s}), providing an analogy somewhat similar to an affine transformation in linear regression model.
The error distribution in the model of \citet{Rosenthal2014} is the von Mises--Fisher distribution, whereas our model allows for either this distribution or the more flexible scaled von Mises--Fisher distribution.

\subsection{Comparison with other regression models}

We compare the proposed regression model defined in Definition \ref{def:link} with other regression models with circular or spherical responses.
The general form of the proposed link function \eqref{eq:link} accommodates both a $(q_s - 1)$-dimensional spherical variable and $q_e$ linear variables.  
To the best of our knowledge, no existing parametric link function shares the same configuration of response and covariates.  

Link functions for general covariates have been proposed by \citet{Paine2020}.  
However, their link functions are defined only for responses in $S^2$, and their extension to higher dimensions is not discussed in detail.  
Moreover, their approach assumes a linear relationship between the mean direction of the response and the covariates, and therefore both the form and interpretation of their link functions are different from ours.
Their error distributions are Kent distribution and elliptically symmetric angular Gaussian distributions.
Similar to the distribution of \citet{Scealy2019} adopted in one of our models, both distributions have contours of constant density which
exhibit ellipse-like symmetry.
But the distribution of \citet{Scealy2019} has a simpler form of the density in terms of both functional part and normalizing constant.


\citet{Scealy2011} considered several link functions involving multiple linear variables.
However, their link functions take values in the positive orthant of the sphere and differ from ours, which take values on the entire sphere.
The same holds for the link functions adopted in \citet{Scealy2017,Scealy2019}, which are identical to one of the link functions in \citet{Scealy2011}.
These link functions are specifically designed for either square-root transformed compositional data or suitably transformed data where the responses are contained well within the positive orthant of the sphere.
The error distribution for the regression model of \citet{Scealy2019} is the same as ours, while \citet{Scealy2017,Scealy2019} adopted Kent distribution.

As discussed in Sections~\ref{sec:special_sphere} and \ref{sec:special_others}, the proposed link function includes the link functions of \citet{Chang1986,Chang1987,Fisher1992,Rivest1989,Downs2002,Downs2003} as special cases.
All of these regression models adopt either the von Mises--Fisher distribution or the von Mises distribution, which is the circular case of the von Mises--Fisher distribution.
Therefore our model with the von Mises--Fisher error distribution encompasses these existing regression models as special cases.
In addition, one variant of our model assumes the scaled von Mises--Fisher distribution, thereby offering greater flexibility than the existing models.
While the use of the scaled von Mises--Fisher distribution may not generally be recommended in the circular case ($p = 2$), it becomes important when $p \geq 3$, particularly in situations where the error structure is not isotropic.

Note that the link function of \citet{Downs2002} serves as the foundation for many other regression and related statistical models, including those of \citet{Kato2008bb,Kato2010a,Rueda2016,Rueda2019,Jha2021}.
Therefore, the special case of our link function with $p = q_s = 2$ and $q_e = 0$ also corresponds to the link function or related functions used in those models.



\subsection{Further results for the earthquake moment tensor regression}
\label{sec:earthquake_supp}
\begin{figure}
    \centering
    \includegraphics[width=\linewidth-2cm]{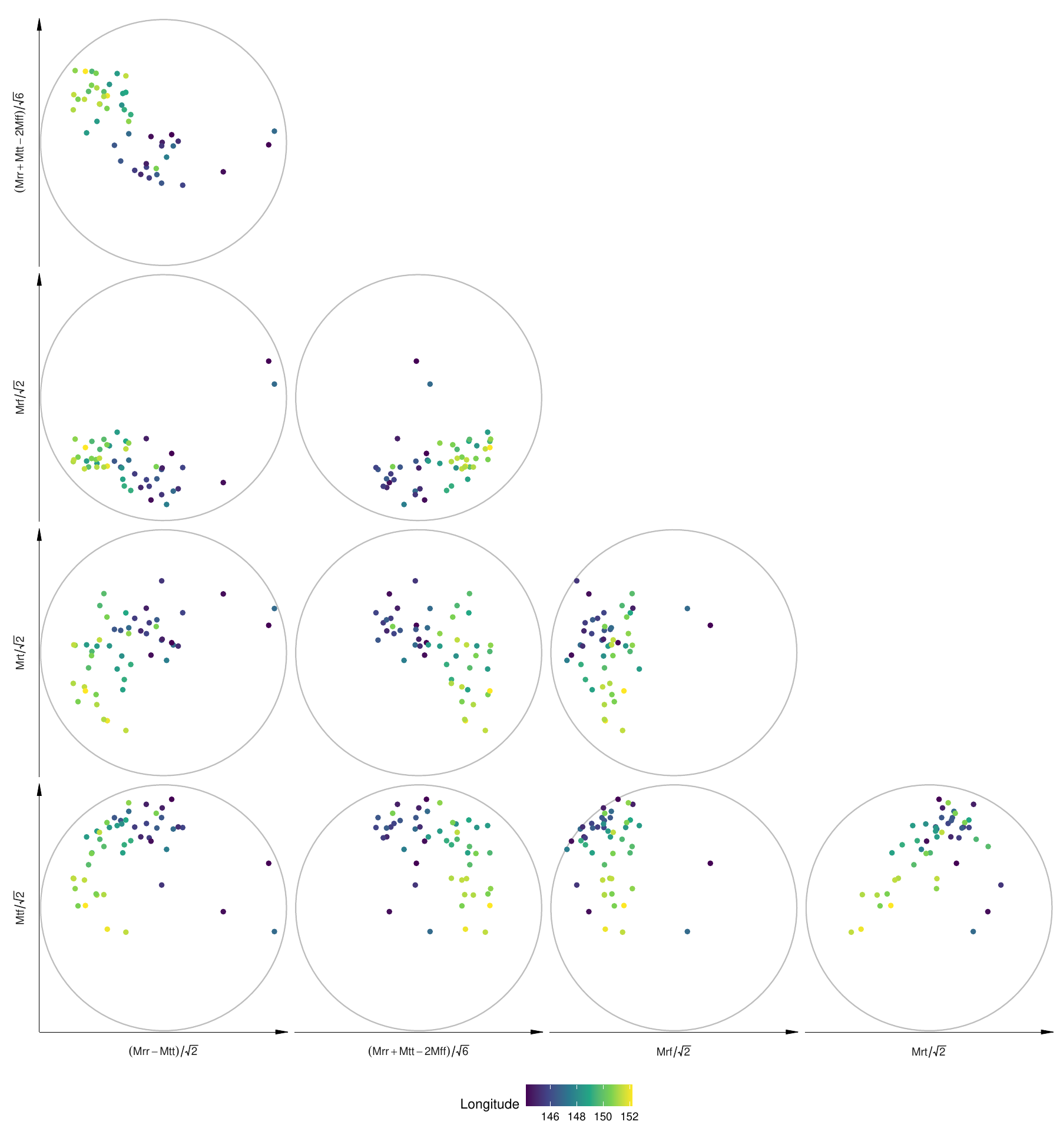}
    \caption{50 normalised earthquake moment tensors on $S^4$.
Moment tensors are shown orthoganally projected onto pairs of their natural basis directions.
Grey circle boundary: intersection of $S^4$ with the plane given by the pair of axes.}
    \label{fig:earthq_mtensors}
\end{figure}

\begin{figure}
    \centering
    \includegraphics[width=\linewidth-2cm]{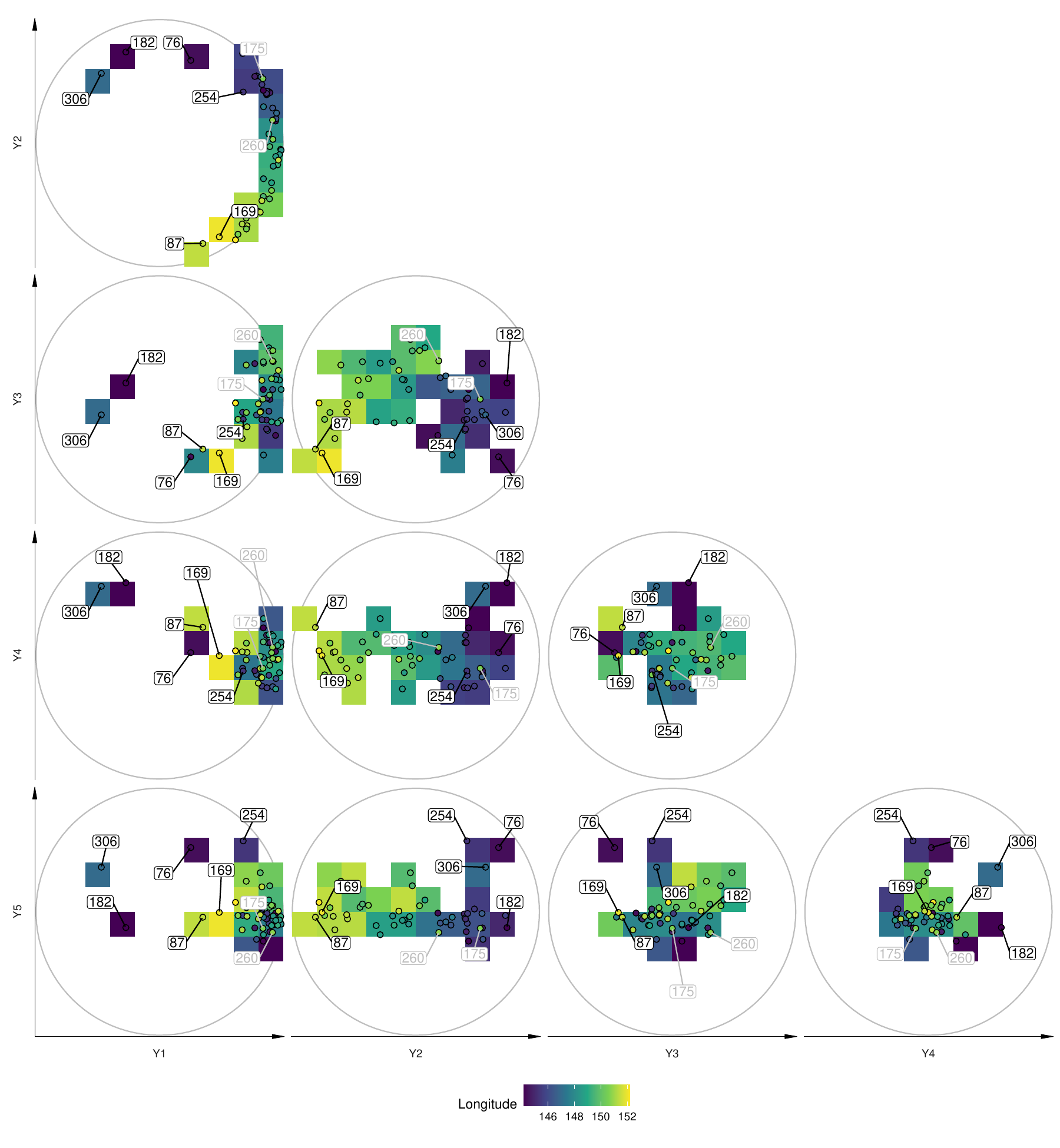}
    \caption{50 normalised earthquake moment tensors on $S^4$ shown orthogonally projected onto pairs of standardised coordinates.
Point colour: longitude of earthquake.
Background colour: average longitude of earthquakes.
Numbers: earthquake number in \citet{hejrani2017ce}.
Grey numbers: earthquakes with unusual moment tensors given their high longitude.
Grey circle boundary: intersection of $S^4$ with the plane given by the pair of axes.}
    \label{fig:earthq_mtensors_outlier}
\end{figure}

The location of the BSSL was obtained from a global database of tectonic plate boundaries \citep{bird2003up,ahlenius2014te}.
The 50 normalised earthquake moment tensors are shown on $S^4$ in Figure \ref{fig:earthq_mtensors} in their natural coordinates, which are obtained by applying the contrasts from a Helmert submatrix to the diagonal elements of the moment tensors (Mtt, Mrr, Mff) and scaling the off-diagonal elements (Mrf, Mrt, Mtf) by $\sqrt{2}$.
A similar plot (Figure \ref{fig:earthq_mtensors_outlier}) using coordinates that standardise mean and covariance \citep[See Section 4.1.3]{Scealy2019} revealed five outliers in standard coordinate Y1, one outlier in standard coordinate Y5, and two high-longitude earthquakes in a region with low-longitude earthquakes (see the Y2-Y4 plot).
These eight outliers were omitted from the regression.

\begin{figure}
    \centering
    \includegraphics[width=0.7\linewidth]{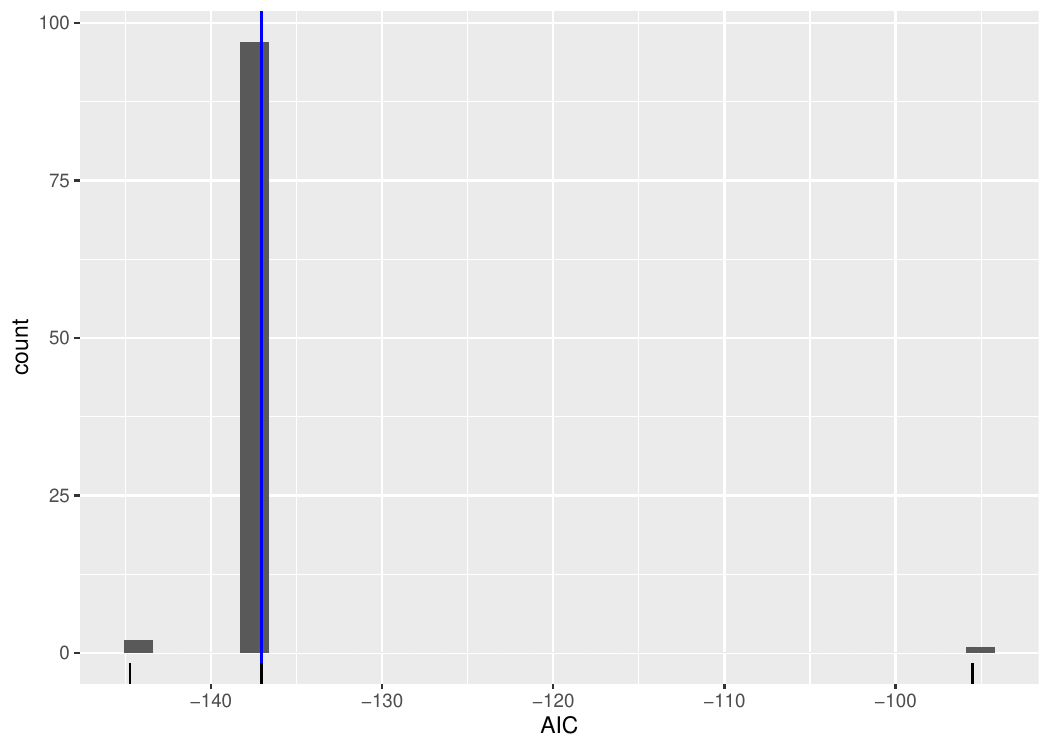}
    \caption{Histogram of AIC obtained by optimisation method from 100 random starts.
    The default start obtained the same AIC (blue line) as most other starts.
    One start obtained a better AIC.
    }
    \label{fig:earthq_restarts}
\end{figure}

Optimisation of the regression from 100 different starting locations (Figure \ref{fig:earthq_restarts}) found three local optima, with the default starting parameters arriving at the second best local optimum.
The regression in the best local optimum had an AIC of -144.7 and an estimated concentration of $59$.
The estimated mean parameters are given in Table \ref{tab:earthq_mean}.
The estimated scales and orientation parameters of the scaled von Mises--Fisher are given in Table \ref{tab:earthq_SvMF}.
Parametric bootstrap with 1000 resamples (i.e.~simulations) was used to obtain 95\% confidence intervals for the scales.
The regression had 38 degrees of freedom.

\begin{table}[]
    \centering
\hspace{0.5cm} \begin{minipage}[t]{0.6\textwidth}
\begin{tabular}[t]{lllll}
\toprule
\multicolumn{1}{c}{ } & \multicolumn{4}{c}{$R_e$} \\
\cmidrule(l{3pt}r{3pt}){2-5}
Strike & 0.07 & 0.05 & 0.21 & -0.97\\
Latitude & 0.02 & 0.04 & -0.14 & 0.09\\
Longitude & 0.49 & 0.08 & -0.84 & -0.16\\
Longitude.L148 & -0.57 & 0.78 & -0.24 & -0.06\\
ones & -0.65 & -0.62 & -0.41 & -0.17\\
\bottomrule
\end{tabular}
\end{minipage}
\hspace{-0.5cm}
\begin{minipage}[t]{0.3\textwidth}
\begin{tabular}[t]{llll}
\toprule
\multicolumn{4}{c}{$B_e$} \\
\cmidrule(l{3pt}r{3pt}){1-4}
2.76 & 0 & 0 & 0\\
0 & 1.39 & 0 & 0\\
0 & 0 & 0.59 & 0\\
0 & 0 & 0 & 0.06\\
\bottomrule
\end{tabular}
\end{minipage}

\begin{tabular}[t]{llllll}
\toprule
\multicolumn{1}{c}{ } & \multicolumn{5}{c}{$B_0$} \\
\cmidrule(l{3pt}r{3pt}){2-6}
(Mrr - Mtt)/$\sqrt{2}$ & 0.05 & 0.55 & -0.13 & 0.41 & 0.72\\
(Mrr + Mtt - 2Mff)/$\sqrt{6}$ & -0.04 & -0.5 & 0.7 & 0.02 & 0.5\\
Mrt /$\sqrt{2}$ & 0.24 & -0.46 & -0.22 & 0.81 & -0.17\\
Mrf /$\sqrt{2}$ & 0.27 & 0.48 & 0.65 & 0.29 & -0.43\\
Mtf /$\sqrt{2}$ & -0.93 & 0.07 & 0.09 & 0.31 & -0.15\\
\bottomrule
\end{tabular}

    \caption{Estimated parameter matrices of the mean link. The columns of $R_e$ are in a Euclidean space where each basis direction corresponds to a standardised covariate: \emph{Longitude.L148} corresponds to the covariate that equals longitude when above 148 and zero otherwise; \emph{ones} is the covariate that is always a value of $1$.
    Similarly the predicted mean direction, which is obtained using a left-multiplication of $B_0$, has basis directions given by the representation of moment tensors as unit vectors in $\mathbb{R}^5$.
    }
    \label{tab:earthq_mean}
\end{table}

\begin{table}[]
    \centering
\begin{tabular}[t]{lllll}
\toprule
\multicolumn{1}{c}{ } & \multicolumn{4}{c}{Scales} \\
\cmidrule(l{3pt}r{3pt}){2-5}
  & a2 & a3 & a4 & a5\\
\midrule
Estimate & 2.08 & 1.40 & 1.00 & 0.34\\
Lower & 1.95 & 1.17 & 0.75 & 0.19\\
Upper & 3.04 & 1.94 & 1.31 & 0.40\\
\bottomrule
\end{tabular}

\begin{tabular}[t]{llllll}
\toprule
  & $\gamma_{01}$ & $\gamma_{02}$ & $\gamma_{03}$ & $\gamma_{04}$ & $\gamma_{05}$\\
\midrule
(Mrr - Mtt)/$\sqrt{2}$ & 0.13 & -0.51 & 0.21 & 0.81 & -0.13\\
(Mrr + Mtt - 2Mff)/$\sqrt{6}$ & 0.22 & 0.51 & -0.59 & 0.49 & 0.31\\
Mrt /$\sqrt{2}$ & -0.10 & 0.49 & 0.75 & 0.19 & 0.38\\
Mrf /$\sqrt{2}$ & 0.78 & 0.28 & 0.20 & -0.08 & -0.51\\
Mtf /$\sqrt{2}$ & -0.56 & 0.39 & -0.04 & 0.24 & -0.69\\
\bottomrule
\end{tabular}

    \caption{Estimated scales (top) and orientation directions (bottom).
    Top: lower and upper ends of 95\% confidence intervals were estimated by parametric bootstrap.
    Bottom: The directions $\gamma_{0j}$ are in the same coordinate system as the response, which has basis directions related to the elements of the moment tensors.
    }
    \label{tab:earthq_SvMF}
\end{table}

\begin{figure}
    \centering
    \includegraphics[width=\linewidth-2cm]{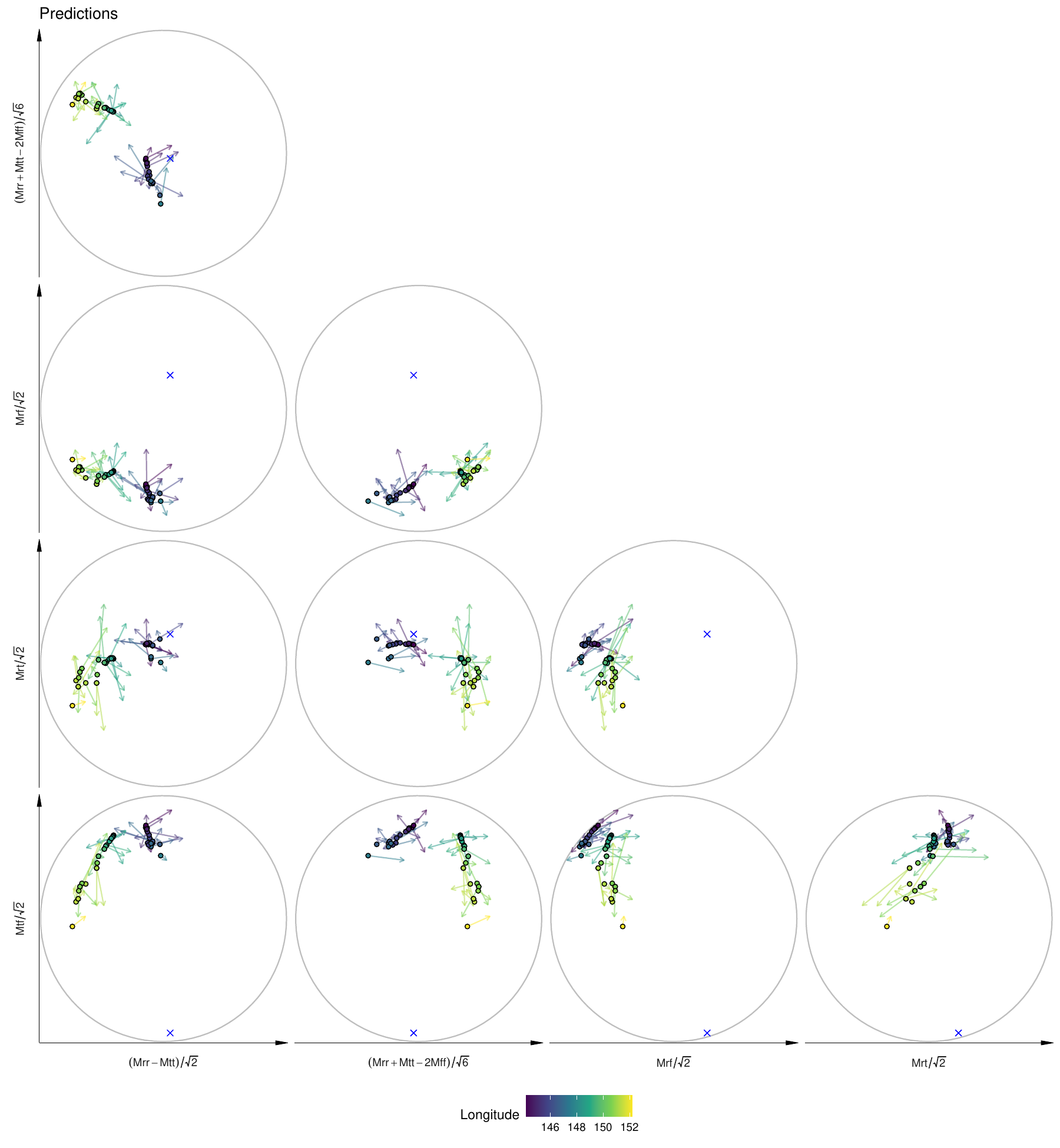}
    \caption{42 predicted normalised earthquake moment tensors.
Colour: longitude of earthquakes.
Arrows: represent residuals and point from mean earthquake moment tensor to the corresponding observed earthquake moment tensor.
Blue x: $\hat{b}_{01}$.
Grey circle boundary: intersection of $S^4$ with the plane given by the pair of axes.}
    \label{fig:earthq_predictions}
\end{figure}

\begin{figure}
    \centering
    \includegraphics[width=\linewidth]{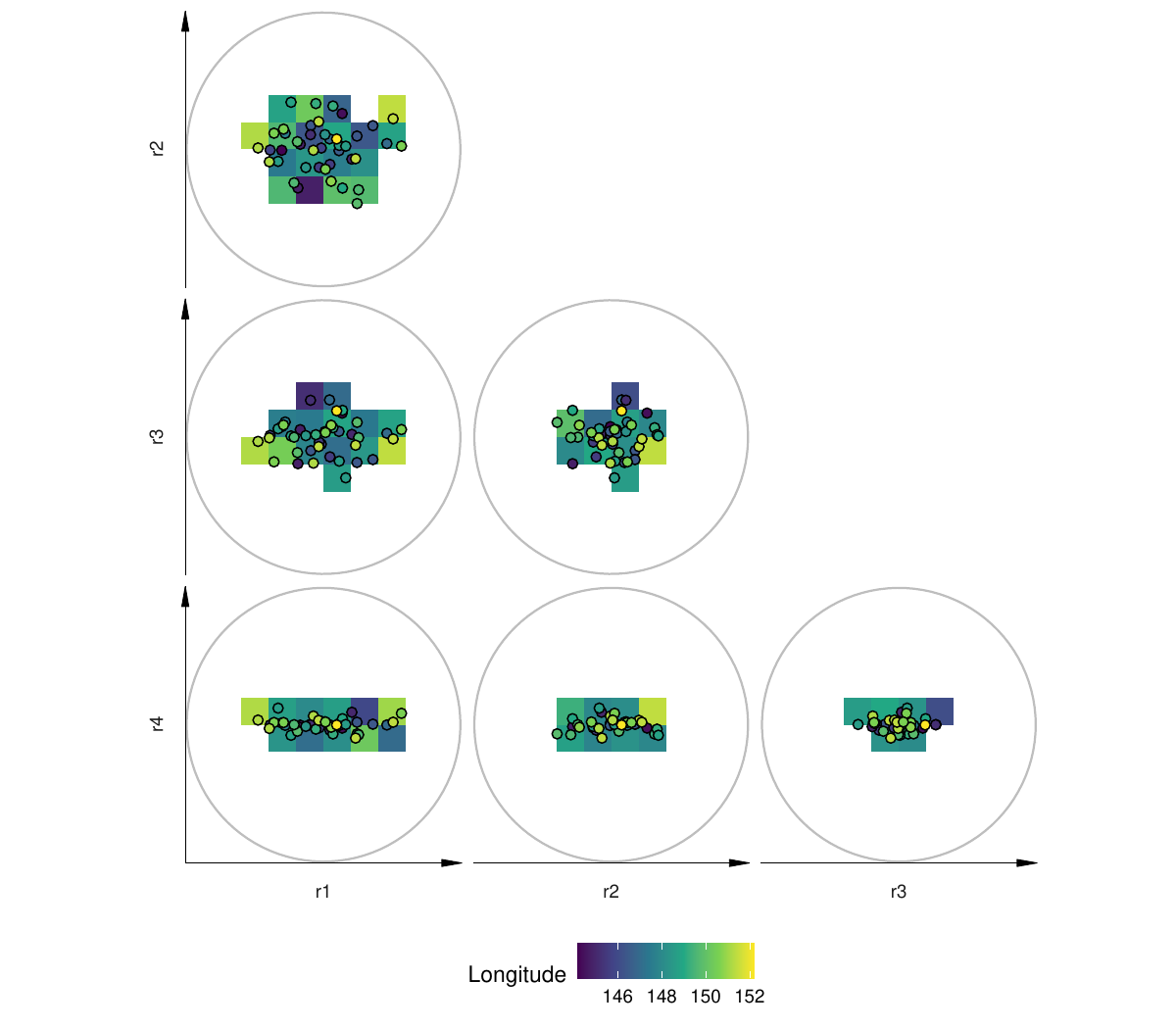}
    \caption{Rotated residuals in the tangent space at $\hat{\gamma}_{01}$. The directions r1, r2, r3, r4, correspond to the directions $\hat{\gamma}_{02}$, $\hat{\gamma}_{03}$, $\hat{\gamma}_{04}$, $\hat{\gamma}_{05}$, respectively.
Point colour: longitude of earthquake.
Background colour: average longitude of earthquakes.
Grey circle boundary: projection of the intersection of $S^4$ with the plane given by the pair of axes.
    }
    \label{fig:earthq_rresids}
\end{figure}

\begin{figure}
    \centering
    \includegraphics[width=\linewidth-2cm]{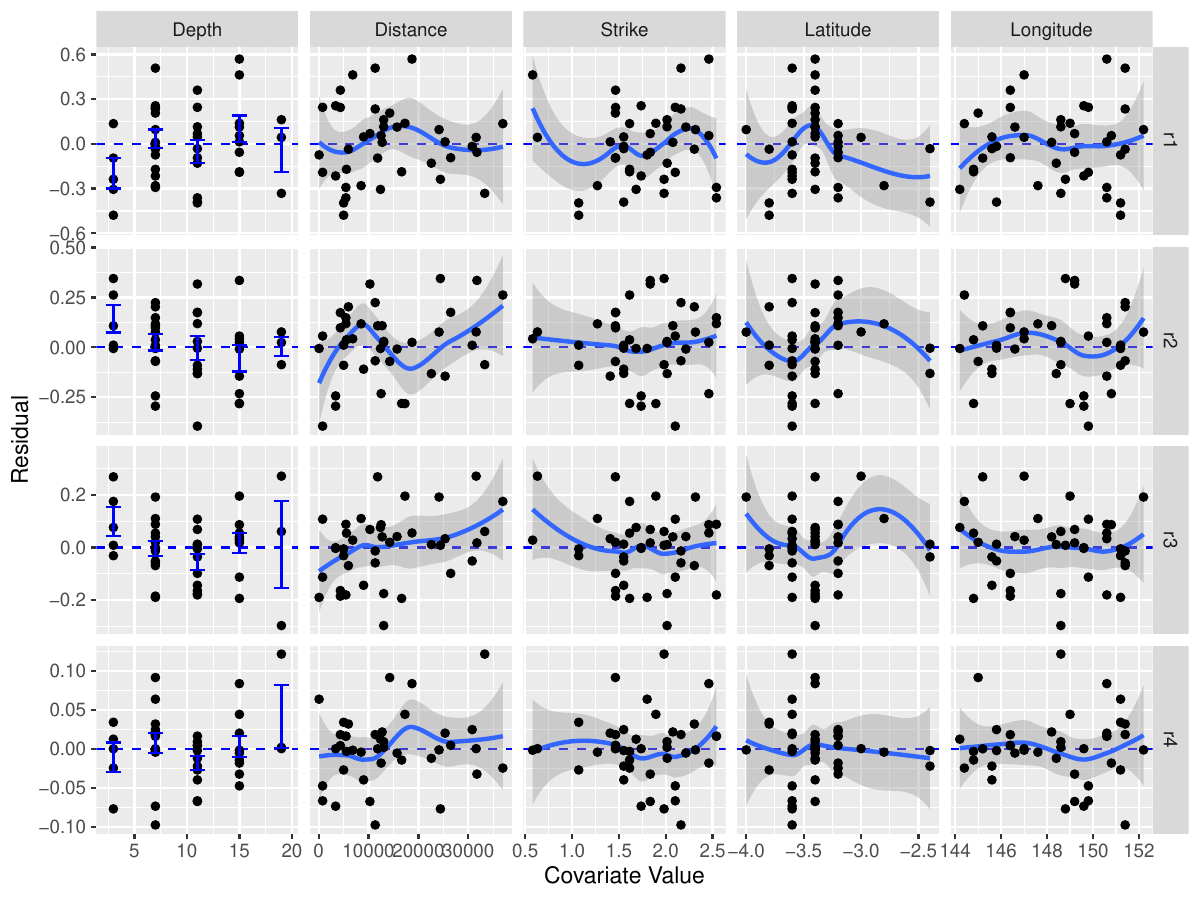}
    \caption{Covariate values against rotated residuals in the tangent space at $\hat{\gamma}_{01}$. The directions r1, r2, r3, r4, correspond to the directions $\hat{\gamma}_{02}$, $\hat{\gamma}_{03}$, $\hat{\gamma}_{04}$, $\hat{\gamma}_{05}$, respectively.
    The mean and 2 standard errors are shown in blue for each discrete distance.
    Smooth lines: local polynomial regression of covariate value against residual value.
    Grey envelopes: two standard errors}
    \label{fig:earthq_rresids2}
\end{figure}

\begin{figure}
    \centering
    \includegraphics[width=\linewidth-2cm]{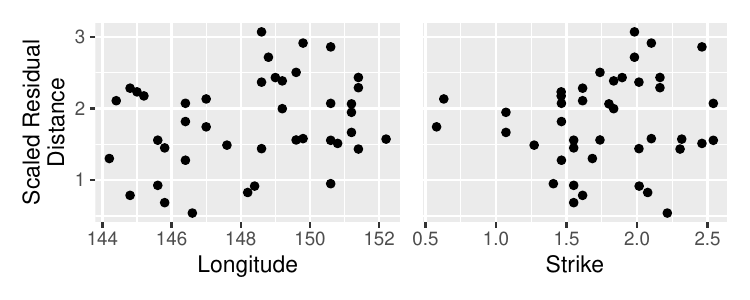}
    \caption{Residual distance, scaled inversely by the estimated scale of each orientation direction, against the covariates Longitude and Strike.}
    \label{fig:earthq_srdist}
\end{figure}

The predicted moment tensors are shown in Figure \ref{fig:earthq_predictions}.
The rotated residuals (parallel-transported residuals; see \citet{Jupp1988}) to the estimated $\gamma_{01}$ and plotted against the estimated directions $\gamma_{02}, \ldots, \gamma_{05}$ (Figure \ref{fig:earthq_rresids}) show the ellipsoidal nature of the scaled von Mises--Fisher.
The rotated residuals plotted against covariate values do not reveal any structure (Figure \ref{fig:earthq_rresids2}).
Residual distance, scaled according to the scales of the scaled von Mises--Fisher orientations, also do not reveal association with covariate values (Figure \ref{fig:earthq_srdist}).

\end{document}